 \documentclass{wiley-article}
 \usepackage[numbers]{natbib}
\usepackage{siunitx}
\usepackage{lineno}
\usepackage[utf8]{inputenc}
\usepackage{bm}
\usepackage{amsmath}
\usepackage{xcolor}
\usepackage{graphicx}
\usepackage{subcaption}
\usepackage{float} 
\usepackage{amssymb}
\usepackage{soul}
\usepackage{longtable}
\usepackage{mathtools}
\usepackage{amsmath}
\usepackage{hyperref}

\definecolor{nika}{rgb}{0.5,0,0.5}

\definecolor{anur}{rgb}{0.0,0.0,1.0}

\newcommand{\RomanNumeralCaps}[1]{\MakeUppercase{\romannumeral #1}}

\DeclarePairedDelimiter{\ceil}{\lceil}{\rceil}

\numberwithin{equation}{section}

\papertype{Original Article}
\paperfield{Journal Section}

\title{Topology Optimization with Tetra-kai-decahedra and Spheroidal Masks}

\author[1\authfn{1}]{Nikhil Singh}
\author[2\authfn{2}]{Anupam Saxena}

\affil[1]{Mechanical Engineering Department, Indian Institute of Technology Kanpur, Kanpur, Uttar Pradesh, 208016, India}
\affil[2]{Mechanical Engineering Department, Indian Institute of Technology Kanpur, Kanpur, Uttar Pradesh, 208016, India}

\corraddress{Nikhil Singh, Mechanical Engineering Department, Indian Institute of Technology Kanpur, Kanpur, Uttar Pradesh, 208016, India}
\corremail{singhn@iitk.ac.in}

\presentadd[\authfn{2}]{Mechanical Engineering Department, RWTH Aachen University, 52062, Germany}

\runningauthor{Singh and Saxena}

\begin{document}

\begin{frontmatter}
\maketitle

\begin{abstract}
A novel meshing scheme, based on regular tetra-kai- decahedron, also referred to as truncated octahedron, cells is presented for use in spatial topology optimization. A tetra-kai-decahedron mesh ensures face connectivity between elements thereby eliminating singular solutions from the solution space. Various other benefits of implementing the said mesh are also highlighted, and the corresponding finite element is introduced. Material mask overlay strategy or MMOS, a feature based method for topology optimization is extended for use in 3-dimensions (MMOS-3D) via the aforementioned finite element and spheroidal negative masks. Formulation for density computation and sensitivity analysis for gradient based optimization is developed. Examples on traditional structural topology optimization problems are presented with detailed discussion on efficacy of the proposed approach.

\end{abstract}
\end{frontmatter}

	\section{Introduction}
Topology optimization pertains to optimal design of continua, in topology, shape and size, for a given objective. \cite{MartinPhilipBendsoe1988} expressed topology optimization as a material distribution problem over a fixed domain. Since then, the field has grown immensely and a variety of problems including but not limited to
structural compliance minimization, design of compliant mechanisms and large deformation continua problems have been solved \citep{Sigmund1998, Poulsen2002a, sigmund1997design, jog1996stability, sigmund1994design, frecker_et_al_1994, saxena2003honeycomb, saxena2007honeycomb, talischi2009honeycomb, reddy2010solution, nagendra2012systematic, nagendra2021topology}. The field is quite well developed in 2D with topology synthesis methods broadly classifiable into, (a) density based \citep{sigmund1994design, Sigmund1998}, (b) implicit level set
\citep{Allaire2016, Guo2014, VanDijk2013}, (c) feature based \citep{Saxena2008, Saxena2011, Singh2020, guo2014doing}  and other methods. A detailed review is provided in \cite{sigmund_maute_2013}. Topology optimization methods in 3D are still under development with an existing few as direct extensions of aforementioned techniques in 2D. \cite{beckers1997optimisation}, \cite{diaz1997optimal} and \cite{borrvall2001large} employ density based method with hexahedral discretization while \cite{Gain2015} use 3D Voronoi-tessellations. \cite{allaire2005structural}, \cite{yamada2010topology}, \cite{van2010level} and \cite{liu20153d} extend the implicit level set method for the same. Feature based methods of topology optimization have also been extended for 3D problems \citep{zhang2018topology,zhang2016new,zhang2017explicit,nguyen2020moving}. \cite{zhang2016new} introduces the MMC approach for three dimensional problems using cuboid like features, while \cite{zhang2017explicit} implement the MMV approach using closed NURBS and Hermite interpolations.

Density methods define an indicator density for each finite element in the design domain as a design variable. The densities vary between 0 (void state of the cell) and 1 (solid state). Determining topology, shape and size of a structure is equivalent to finding the indicator densities. The number of design variables (densities) equals the number of elements and thus mesh refinement leads to increase in number of variables making it inconvenient to be adopted for large problems, especially in 3D. In the initial implementation of the density method that mainly employed rectangular cells (finite elements), it was observed that 2D optimal solutions can be singular and can exhibit point connections \citep{Sigmund1998, Poulsen2002a, sigmund1997design}. \cite{diaz1995checkerboard} and \cite{jog1996stability} argued that this was a consequence of numerical anomalies and proposed the use of higher order finite elements or perimeter constraints. \cite{sigmund1994design} proposed use of filtering, which smears out densities thereby imposing minimum length scale implicitly. However, filtering by itself, leads to intermediate densities in the solution. Dedicated SIMP based methods exist now that can yield solutions close to ‘black and white’ e.g. \cite{guest2004achieving, wang2011projection}. Use of methods like the level set which defines the topology of a structure using the zero level set of a scalar function, $\phi$, over the domain, or feature based which define cell densities as a function of location, size and orientation of geometrical features may also lead to singular solutions with rectangular parametrization. These methods are unlikely to generate checker board patterns but can give point connections at certain locations. Observing that hexagonal tessellation has only edge connections among cell neighbors in 2D, \cite{saxena2003honeycomb, saxena2007honeycomb} proposed their use to discretize the design domain, thus removing the possibility of singular solutions. \cite{talischi2009honeycomb} generated solutions with hexagonal and rectangular elements for common problems solved using the same algorithm. For 3D cases, \cite{beckers1997optimisation} presented a numerical study implementing hexahedral discretization and showed that final structures can exhibit point/edge connections. \cite{Gain2015} implemented density based topology optimization using Voronoi-tessellations which ensure face connections between elements. However, finite element analysis over Voronoi-tessellations is computationally expensive as stiffness matrix for each element has to be evaluated and stored separately. Analogous to hexagonal cells in 2D, \cite{Saxena2011} envisaged the use of a regular tetra-kai-decahedral element mesh in 3D cases as the lattice guarantees face connections throughout.

A truncated octahedron, also known as Kelvin cell, is studied in the past in context of foam structures \citep{thomson1887division}. Kelvin proposed tetra-kai-decahedral cells with slightly curved faces as solution to the Kelvin problem, claiming the shape to be best equal sized structure, in terms of volume, to fill space with minimal surface area \citep{thomson1887division, weaire2009kelvin}. Years later, \cite{weaire1994counter} proposed the Weaire-Phelan structure, made up of two different kinds of cells of the same size and exhibiting only face connections throughout in 3D. Weaire-Phelan structure could also potentially be used in spatial topology optimization. Truncated octahedral lattice structure \footnote{these are constructed by joining bars along edges of a unit cell} is of interest for numerous other reasons, e.g., strength to weight ratio under various loading conditions, and numerical and experimental  \citep{zhu1997analysis, meza2017reexamining, jang2008microstructure, qi2019mechanical} investigations have been conducted to understand its mechanical properties. Such a lattice also provides low pressure drops in flows, exceptional heat transfer, enhanced mixing during flow and high surface area per unit volume, and has been adapted as an open-cell foam in
multiple studies \citep{klostermann2013meshing, iasiello2017developing, sinn2020cfd, kumar2014determination}.

In this paper, we implement regular tetra-kai-decahedra, also referred to as truncated octahedra, to discretize the domain for topology optimization. In addition to ensuring face connectivity throughout, another advantage with a single tetra-kai-decahedra is its ability to capture many deformation modes and axes of symmetry as apposed to a single hexahedral cell. Additionally, tetra-kai-decahedron element mesh provides finite stiffness in more directions locally compared to hexahedral element mesh, thus providing more directions for structural development. Effects pertaining to mesh size is also explored. We further use spheroidal masks within the Material Mask Overlay Strategy (MMOS) \cite{Saxena2008} thus expanding the idea into 3D. With MMOS, the number of design variables, pertaining to the size and location of these geometric features, is an independent choice from the size of mesh — the number of features can be chosen to keep the number of design variables much less than that of finite elements (cells) in the mesh.

In what follows, section \ref{sec:mesh_development} discusses a
computational algorithm not presented elsewhere to our knowledge; to develop a regular mesh using tetra-kai-decahedron elements. Section \ref{sec:FEM} presents details of FEM implementation to solve linear elasticity equations using the aforementioned element. Section \ref{sec:topo_opti} focuses on the implementation of topology optimization using MMOS with spheroidal masks.
Section \ref{sec:sample_prob} presents solutions to some well established problems, followed by a discussion and conclusions drawn in section \ref{sec:discussion} and section \ref{sec:conclusion} respectively.

\section{Mesh Generation with Regular Tetra-kai-Decahedra}
\label{sec:mesh_development}
Generating a primordial regular FEM mesh requires evaluating the connectivity matrix $\boldsymbol{\mathrm{C}}$, node location matrix $\boldsymbol{\mathrm{NX}}$, and unit cell information. The connectivity matrix associates elements to nodes, that is, given an element number, $\boldsymbol{\mathrm{C}}$ provides the associated node numbers. The node location matrix stores spatial coordinates of each node with respect to a defined coordinate axes. The unit cell information corresponds to the physical shape of each unit cell, that is, the solid shape of the element. For example, an element with 6 nodes can either be a triangular prism or a pentagonal pyramid. We use a regular tetra-kai-decahedron, also known as truncated octahedron, to develop the mesh for use in 3D topology optimization. The algorithm that follows is confined to meshing cuboidal regions which is often the case in topology optimization. A natural extension of the algorithm for a generic domain can be obtained by combining it with a bounding box algorithm \citep{barequet2001efficiently}. The geometric description of the unit tetra-kai-decahedron is presented in section \ref{sec:unitcel} followed by development of the connectivity matrix discussed in section \ref{sec:connectivity}  and determination of node location matrix, in section \ref{sec:node_location}.

\subsection{Unit Cell}
\label{sec:unitcel}
\begin{figure}[htb]
	\centering
	\includegraphics[width=0.4\textwidth]{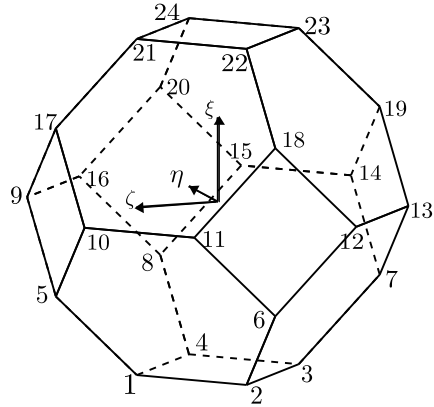}
	\caption{Truncated octahedron unit cell in a local coordinate system. Dashed lines represent hidden faces}
	\label{fig:unit_cell}
\end{figure}

Tetra-kai-decahedron is a convex solid with 24 vertices and 14 faces. Among these faces, six are quadrilaterals and the rest eight are hexagons. Truncated octahedron is a special case of tetra-kai-dechedron where all edges are equal in length (Fig. \ref{fig:unit_cell}). Truncated octahedra form a lattice and hence can be used to discretize the $\mathbb{R}^3$ space. In its lattice structure, each truncated octahedron has 14 such neighbors each sharing a face with the parent octahedron leading to face connectivity throughout between cells. Lack of solely point and/or edge connections makes this lattice ideal to be adopted as a mesh for topology optimization as it inherently removes the possibility of singular solutions.
Just like in conventional FEM, a master truncated octahedron element is established in sec. \ref{sec:FEM}. Developing iso-parametric mapping and connectivity matrix requires setting local node numbering for the element, presented in Fig. \ref{fig:unit_cell}. The same numbering scheme is used for both, physical and parent elements for convenience. Note that, the numbering scheme is independent of the element’s orientation.
This scheme is not unique and can be changed as per convenience and correspondingly, changes need to be made throughout the meshing algorithm. Coordinates of each node in the local axes (Fig. \ref{fig:unit_cell}) with its origin at the centroid of the cell, for an edge length of $a = \sqrt{2}$, are given in Table \ref{tab:1}. These coordinates are permutations of $(0,\pm 1, \pm 2)$.

\begin{table}[h]
	\centering
	\caption{Node locations of a truncated octahedron element for edge length $a = \sqrt{2}$ in local coordinates (Fig. \ref{fig:unit_cell}).} \label{tab:1}
	\begin{tabular}{|c|c||c|c|}
		\hline
		\textbf{Local Node \#} & \textbf{Coordinate} $(\eta, \zeta, \xi)$ & \textbf{Local Node \#} & \textbf{Coordinate} $(\eta, \zeta, \xi)$\\ \hline
		1 & $(0,1,-2)$ & 13 & $(-1,-2,0)$\\ 2 & $(-1,0,-2)$ & 14 & $(1,-2,0)$ \\ 3 & $(0,-1,-2)$ & 15 & $(2,-1,0)$ \\ 4 & $(1,0,-2)$ & 16 & $(2,1,0)$\\ \hline
		5 & $(0,2,-1)$ & 17 & $(0,2,1)$\\ 6 & $(-2,0,-1)$ & 18 & $(-2,0,1)$\\ 7 & $(0,-2,-1)$ & 19 & $(0,-2,1)$\\ 8 & $(2,0,-1)$ & 20 & $(2,0,1)$ \\ \hline
		9 & $(1,2,0)$ & 21 & $(0,1,2)$\\ 10 & $(-1,2,0)$ & 22 & $(-1,0,2)$ \\ 11 & $(-2,1,0)$ & 23 & $(0,-1,2)$ \\ 12 & $(-2,-1,0)$ & 24 & $(1,0,2)$ \\ \hline
	\end{tabular}
\end{table}

\subsection{Element Connectivity}
\label{sec:connectivity}
The connectivity matrix, $\boldsymbol{\mathrm{C}}$ takes an element number ($en$) and a local node number ($ln$) as inputs and provides the associated global node number ($gn$), that is, $\boldsymbol{\mathrm{C}}(en, ln) = gn$. Developing the connectivity matrix employs four primary steps, namely:
\begin{enumerate}
	\item Unit cell map and mesh visualization
	\item Element-Point connectivity
	\item Point-Node connectivity
	\item Node selection
\end{enumerate}
Unit cell map and mesh visualization (section \ref{sec:mesh_visual}) develops a map between a physical element and a fictitious element (Fig. \ref{fig:unit_cell_map}) in an abstract space called the mesh space. Correspondingly, visual representation of the physical mesh is obtained in the mesh space. Element-Point connectivity (section \ref{sec:element-point}) develops the connectivity matrix for the obtained mesh representation, followed by Point-Node connectivity (section \ref{sec:point-node}) and Node selection (section \ref{sec:node-selection}) wherein the reverse map is developed to obtain connectivity in the physical space. These steps are discussed in detail next. An overview of the algorithm is given in Fig. \ref{fig:mesh_algo}.

\subsubsection{Unit cell map and mesh visualization}
\label{sec:mesh_visual}
\begin{figure}[!htb]
	\begin{minipage}{0.35\textwidth}
		\centering
		\includegraphics[width=.9\linewidth]{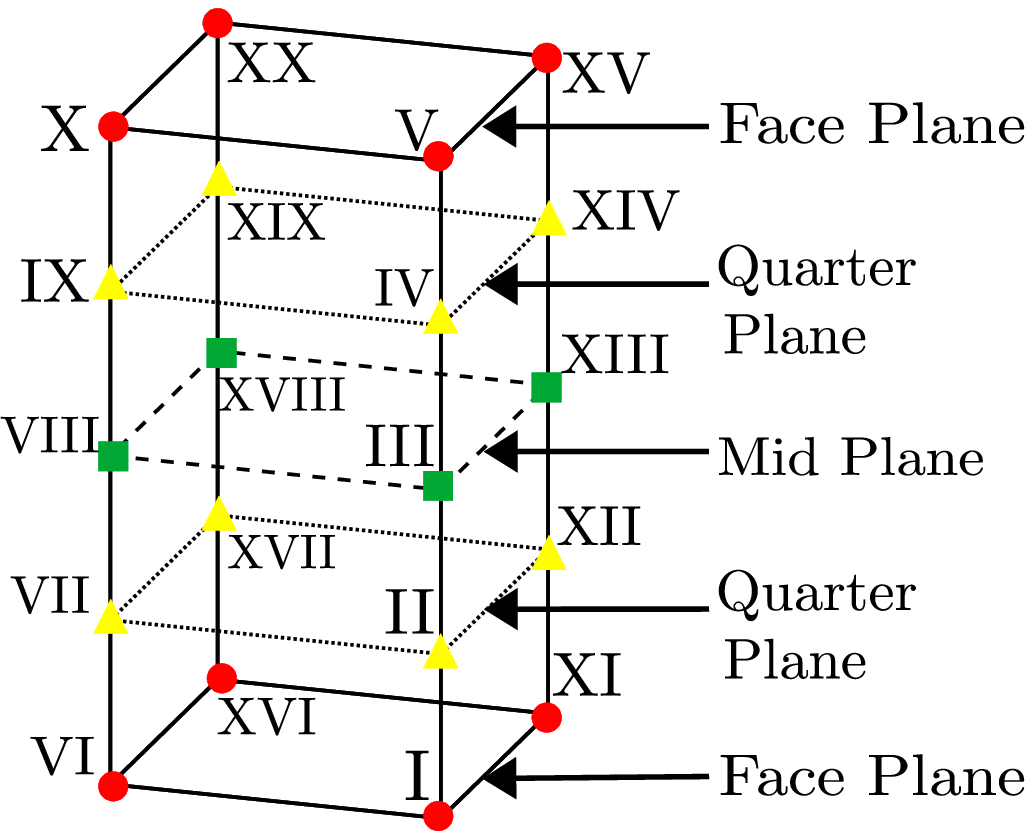}
		\caption{Local Point numbering of unit cuboid element in the mesh space}\label{fig:cuboid_cell}
	\end{minipage}\hfill
	\begin{minipage}{0.6\textwidth}
		\centering
		\includegraphics[width=.9\linewidth]{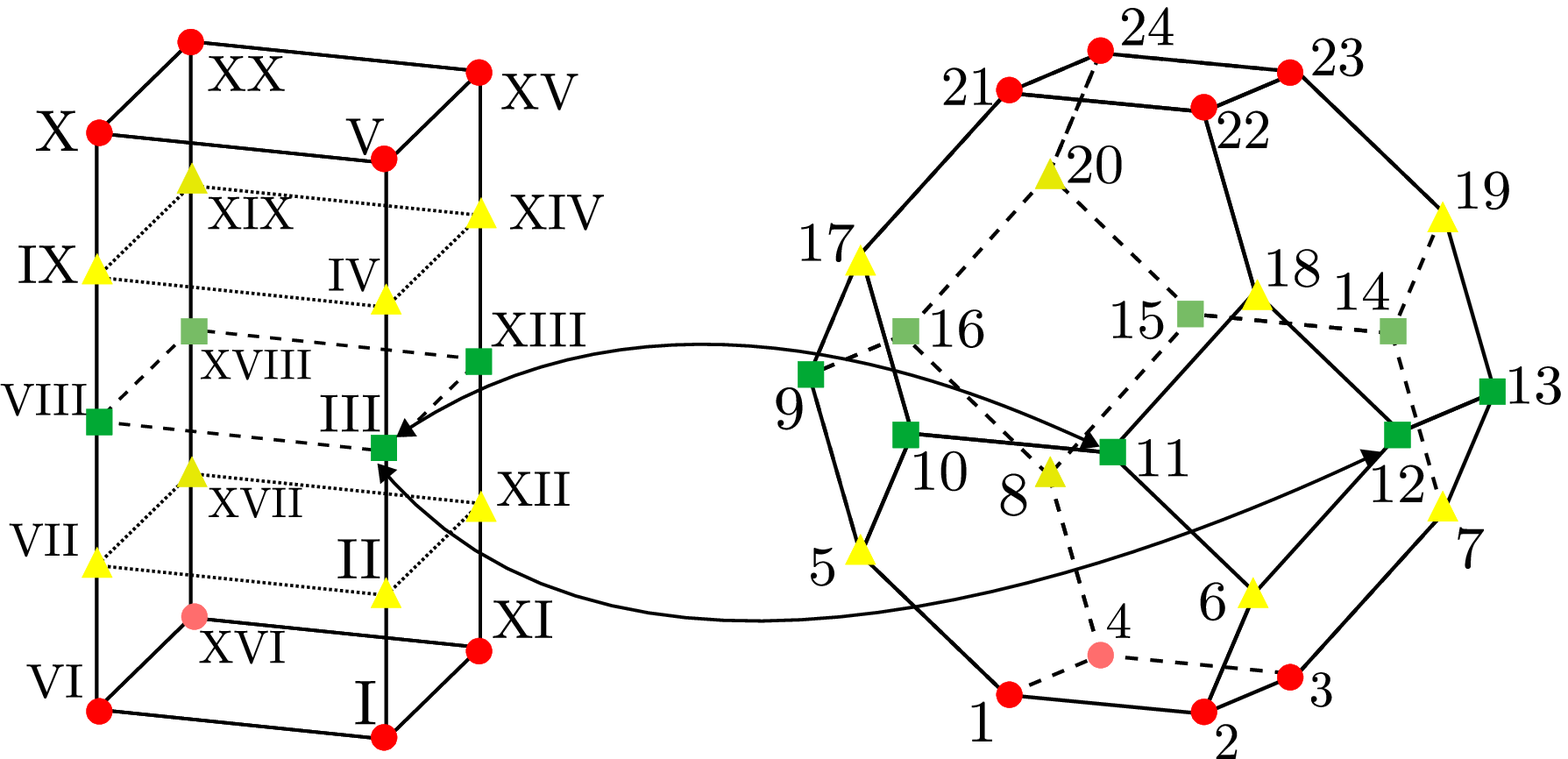}
		\caption{Point-Node map for unit cell in mesh and physical spaces respectively}\label{fig:unit_cell_map}
	\end{minipage}
\end{figure}

We develop a virtual representation of the physical mesh in an abstract space, called the mesh space. A truncated octahedron element in the physical space is represented by a cuboid cell in the mesh space (Fig. \ref{fig:unit_cell_map}). For convenience, within the section, a truncated octahedron element in the physical space and a cuboid cell in the mesh space are referred as \textit{element} and \textit{cell} respectively. Similarly, \textit{nodes} refer to as vertices/nodes of an element and \textit{points} to locations in the mesh space to which the nodes get mapped to. Local point numbering (Roman numerals) for the unit cell (Fig. \ref{fig:cuboid_cell}) is established and thereafter map between a single element and the corresponding cell is given in Table \ref{tab:2}. Every node maps to a unique point, however, not every point corresponds to a unique node. Hence, a map from the physical to mesh space is straight forward but the reverse map is not trivial and is discussed in detail in section \ref{sec:node-selection}. A graphical representation of the element to cell map is presented in Fig. \ref{fig:unit_cell_map}. The bottom and top square faces of an element, created by nodes $\{1,2,3,4\}$ and $\{21,22,23,24\}$ map to the bottom and top face of the cell created by points \{\RomanNumeralCaps{6}, \RomanNumeralCaps{1}, \RomanNumeralCaps{11}, \RomanNumeralCaps{16}\} and \{\RomanNumeralCaps{10}, \RomanNumeralCaps{5}, \RomanNumeralCaps{15}, \RomanNumeralCaps{20}\} respectively. For the remaining square faces, two of the diagonally opposite nodes map to a single point on the mid plane, represented by dashed lines in Fig. \ref{fig:cuboid_cell}. That is, nodes $9$ and $10$ on the square face $\{5,10,17,9\}$ map to point \RomanNumeralCaps{8}, while nodes $5$ and $17$ map to points \RomanNumeralCaps{7} and \RomanNumeralCaps{9} respectively. Hence, this square face in the physical space is degenerated and represented by part of a vertical edge in the mesh space. Similarly, nodes $11$ and $12$ map to point \RomanNumeralCaps{3}, nodes $13$ and $14$ to point \RomanNumeralCaps{13} and nodes $15$ and $16$ to point \RomanNumeralCaps{18} leading to corresponding square faces being mapped to parts of respective vertical edges of the cuboid.

\begin{figure}[h]
	\centering
	\subcaptionbox{Type A (\textit{Full face}) neighbours}{\includegraphics[width=0.2\textwidth]{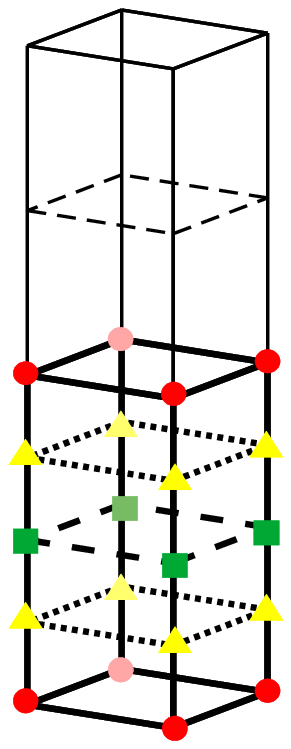}}
	\label{fig:Cell-neighbours_a}%
	\hspace{0.7cm}
	\subcaptionbox{Type B (\textit{Half face}) neighbours}{\includegraphics[width=0.2\textwidth]{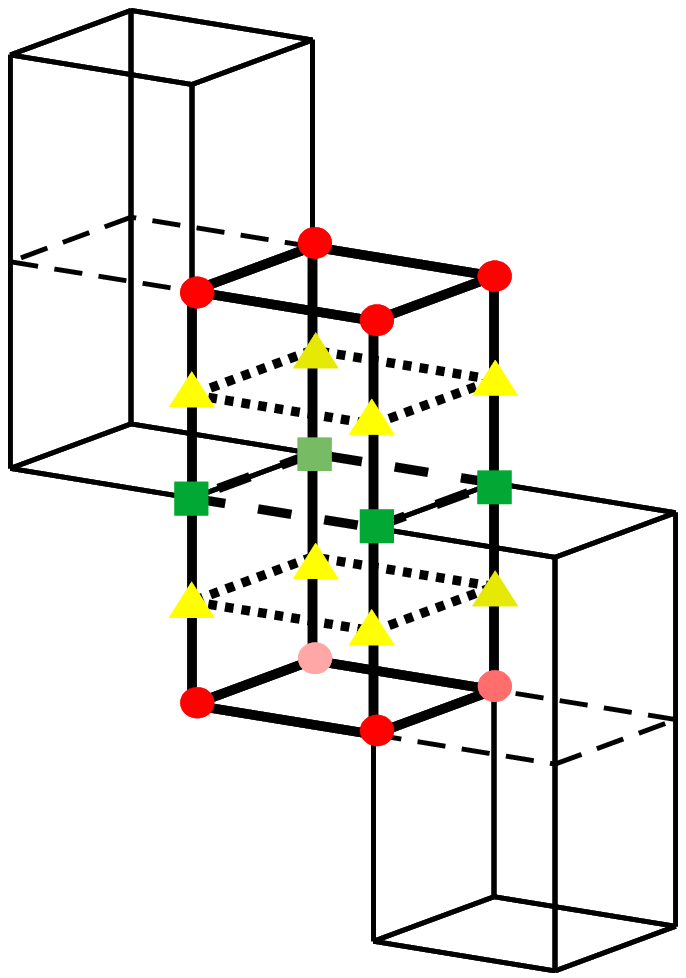}}
	\label{fig:Cell-neighbours_b}%
	\hspace{0.7cm}
	\subcaptionbox{Type C (\textit{Edge}) neighbours}{\includegraphics[width=0.2\textwidth]{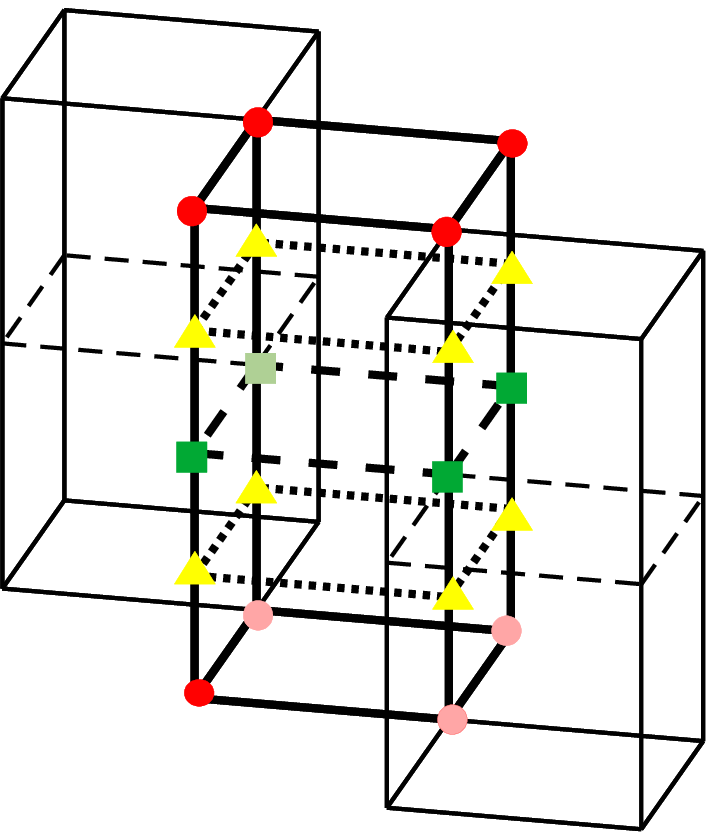}}
	\label{fig:Cell-neighbours_c}%
	\hspace{0.7cm}
	\subcaptionbox{Type D (\textit{Point}) neighbours}{\includegraphics[width=0.2\textwidth]{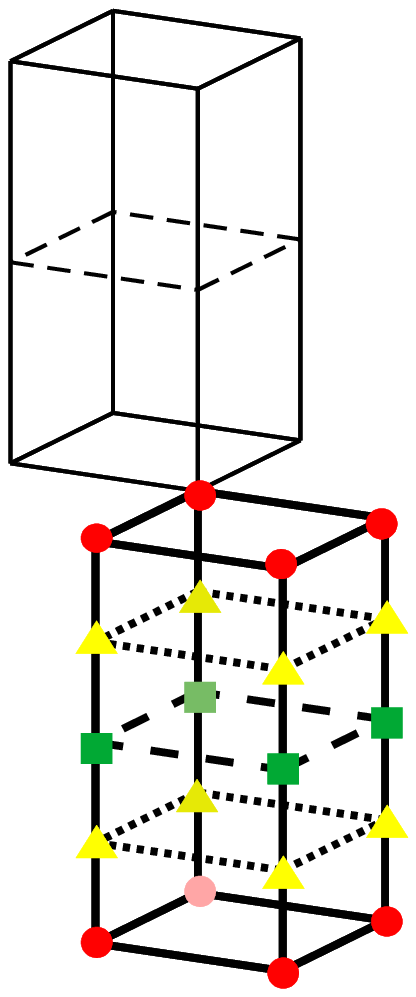}}
	\label{fig:Cell-neighbours_d}%
	\caption{Unit cell (thick solid line) and different categories of cell neighbors (thin solid line) in the mesh space.}
	\label{fig:Cell-neighbours}  
\end{figure}

Points on a cell are classified into 3 categories: \textit{Face-Plane Points} (FPP), \textit{Mid-Plane Points} (MPP) and \textit{Quarter-Plane Points} (QPP). FPPs are points on the top and bottom face of the cuboid, highlighted by red circles in Fig. \ref{fig:cuboid_cell}, MPPs are points on the mid plane (green squares) and QPPs are those which lie on the planes between the top/bottom face and the mid plane, depicted by yellow triangles. Nodes corresponding to each category of points are highlighted using the same shape and color in Fig. \ref{fig:unit_cell_map}. Note that each MPP corresponds to 2 nodes.
\begin{table}[h]
	\centering
	\caption{Point-node mapping between truncated octahedral element (physical space) and cuboid cell (mesh space), and node selection procedure}
	\begin{tabular}{|c|c|c||c|c|c|}
		\hline
		\textbf{Local Node \#} & \textbf{Local Point \#} & \textbf{ Node 1/2} & \textbf{Local Node \#} & \textbf{Local Point \#} & \textbf{Node 1/2}\\ \hline
		1 & \RomanNumeralCaps{6} & 2& 13 & \RomanNumeralCaps{13} & 1\\
		2 & \RomanNumeralCaps{1} & 2 & 14 & \RomanNumeralCaps{13} & 2\\
		3 & \RomanNumeralCaps{11} & 1 & 15 & \RomanNumeralCaps{18} & 2 \\
		4 & \RomanNumeralCaps{16} & 1 & 16 & \RomanNumeralCaps{18} & 1 \\
		5 & \RomanNumeralCaps{7} & 1 & 17 & \RomanNumeralCaps{9} & 1 \\
		6 & \RomanNumeralCaps{2} & 1 & 18 & \RomanNumeralCaps{4} & 1 \\
		\hline
		7 & \RomanNumeralCaps{12} & 1 & 19 & \RomanNumeralCaps{14} & 1 \\
		8 & \RomanNumeralCaps{17} & 1 & 20 & \RomanNumeralCaps{19} & 1 \\
		9 & \RomanNumeralCaps{8} & 2 & 21 & \RomanNumeralCaps{10} & 2 \\
		10 & \RomanNumeralCaps{8} & 1 & 22 & \RomanNumeralCaps{5} & 2 \\
		11 & \RomanNumeralCaps{3} & 1 & 23 & \RomanNumeralCaps{15} & 1 \\
		12 & \RomanNumeralCaps{3} & 2 & 24 & \RomanNumeralCaps{20} & 1 \\ \hline         
	\end{tabular}
	\label{tab:2}
\end{table}
Next, we determine the arrangement of neighboring cells in the mesh space. Knowledge that (a) all neighbors of an element share a face connection and (b) each face is connected to a single neighbor, is sufficient to describe neighboring cell arrangement in the mesh space. For convenience, neighbors of a cell are categorized into 3 types:
\begin{enumerate}
	\item Type A or \textit{Full face neighbors}: They share the top or bottom face of the cuboid unit cell corresponding to two of the six square faces in the truncated octahedron element. A cell can have up to two Type A neighbors (Fig. \ref{fig:Cell-neighbours}a).
	\item Type B or \textit{Half face neighbors}: They share half of each vertical face of the cuboid unit cell corresponding to all eight hexagonal faces in the truncated octahedron element. Two Type B neighbors of a cell are shown in Fig. \ref{fig:Cell-neighbours}b. A cell can have up to eight Type B neighbors.
	\item Type C or \textit{Edge neighbors}: They share a vertical edge of the cuboid unit cell corresponding to the remaining four square faces of the truncated octahedron (Fig. \ref{fig:Cell-neighbours}c).
	\item One also encounters Type D or \textit{point neighbors} (Fig. \ref{fig:Cell-neighbours}d), cases which are discussed later in the section.
\end{enumerate}
From the local cell arrangement, one observes that (a) FPPs of a cell are MPPs for its Type B neighbors (Fig. \ref{fig:Cell-neighbours}b), (b) Type C neighbors share FPPs (Fig. \ref{fig:Cell-neighbours}c) even though they do not share the corresponding nodes and (c) QPPs for one cell are QPPs for the neighboring cells (Fig. \ref{fig:Cell-neighbours}b amd \ref{fig:Cell-neighbours}c). Because MPPs must always correspond to two nodes as discussed earlier, and FPPs must at least map to one node for each cell, in a mesh, an FPP maps to two nodes if the cell containing the FPP has Type B, Type C or both Type B and Type C neighbors sharing that FPP.
\begin{figure}[h]
	\centering
	\subcaptionbox{Reoriented cuboid cell }{\includegraphics[width=0.35\textwidth]{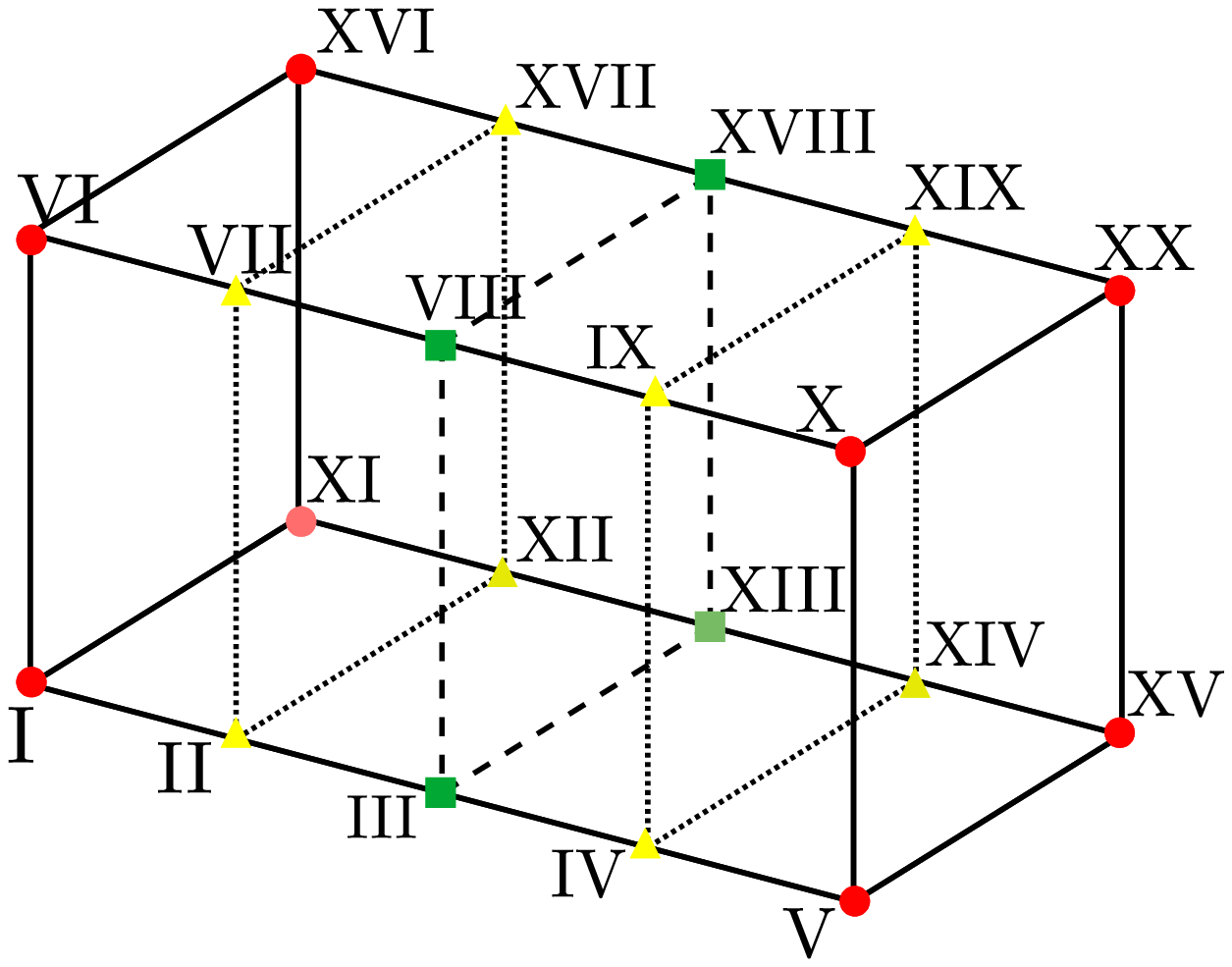}}
	\label{fig:reoriented_brick_element}%
	\hspace{1cm}
	\subcaptionbox{Corresponding reoriented truncated octahedron element with new axes }{\includegraphics[width=0.3\textwidth]{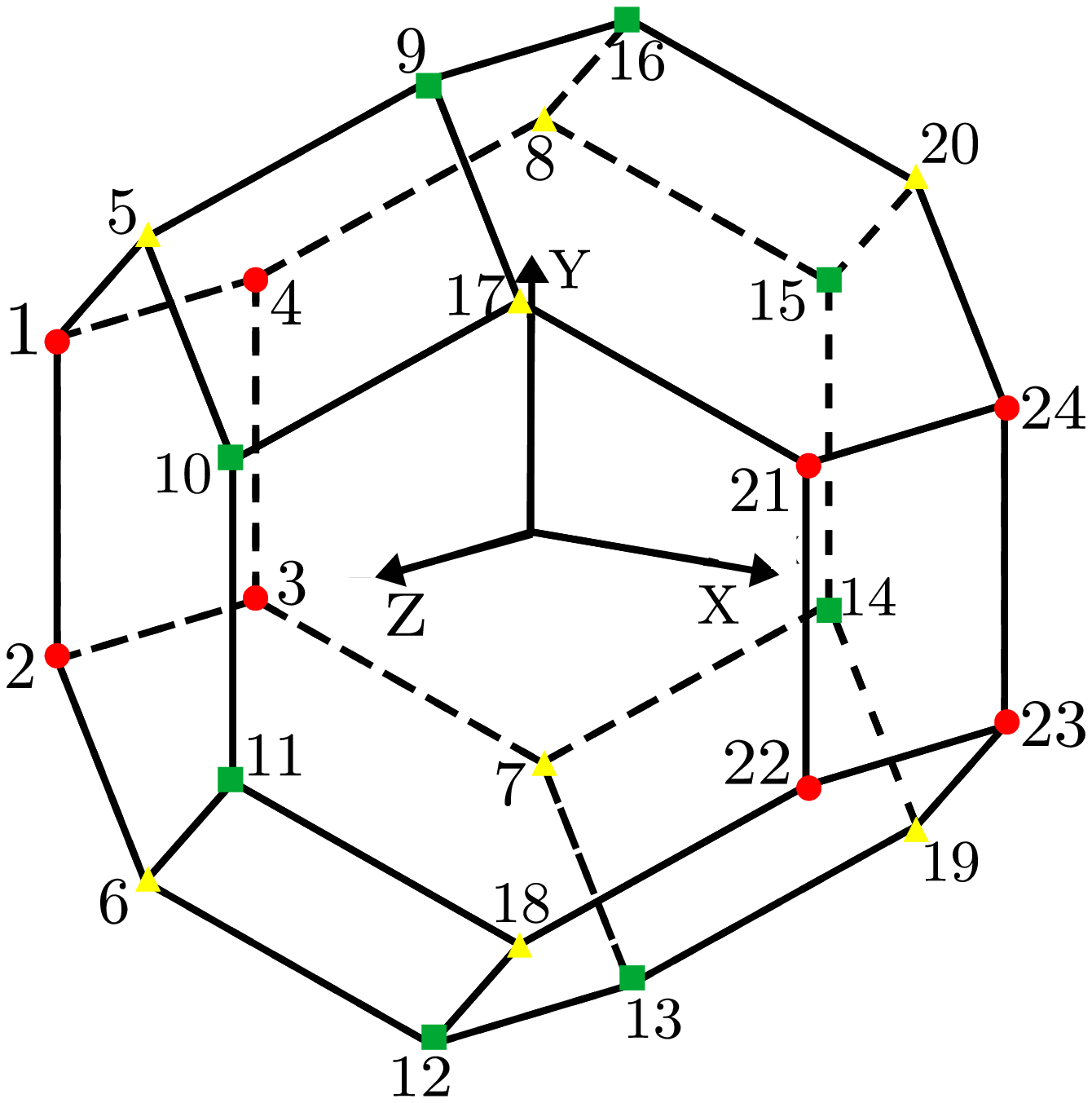}}
	\label{fig:reoriented_unit_element} 
	\caption{Reoriented cell in mesh space and its corresponding element orientation in physical space with a new set of axes.}
	\label{fig:reoriented_elements}
\end{figure}
For ease in mesh visualization, the cuboid cell is reoriented such that its length lies along the horizontal as shown in Fig. \ref{fig:reoriented_elements}a. The corresponding orientation of the physical element is presented in Fig. \ref{fig:reoriented_elements}b. A new set of axes is introduced (Fig. \ref{fig:reoriented_elements}b) such that position vector, $\boldsymbol{\mathrm{X}}$, of a local node in the new orientation is given by $\boldsymbol{\mathrm{X}} = \boldsymbol{\mathrm{R}}_2\boldsymbol{\mathrm{R}}_1 \boldsymbol{\mathrm{X}}_0$, where $\boldsymbol{\mathrm{X}}_0$ is the vector of co-ordinates of the local node in Table \ref{tab:1} and, $\boldsymbol{\mathrm{R}}_1$ and $\boldsymbol{\mathrm{R}}_2$ are orthogonal matrices given as,
\begin{eqnarray}\nonumber
\boldsymbol{\mathrm{R}}_1 = \begin{bmatrix}
\frac{1}{\sqrt{2}} & -\frac{1}{\sqrt{2}} & 0 \\
\frac{1}{\sqrt{2}} & \frac{1}{\sqrt{2}} & 0 \\
0 & 0 & 1
\end{bmatrix} ~\text{and}~
\boldsymbol{\mathrm{R}}_2 = \begin{bmatrix}
0 & 0 & 1 \\ 0 & 1 & 0 \\ -1 & 0 & 0
\end{bmatrix}.
\end{eqnarray}
\begin{figure}[h]
	\centering
	\subcaptionbox{Regular cell arrangement of cuboid cells}{\includegraphics[width=0.4\textwidth]{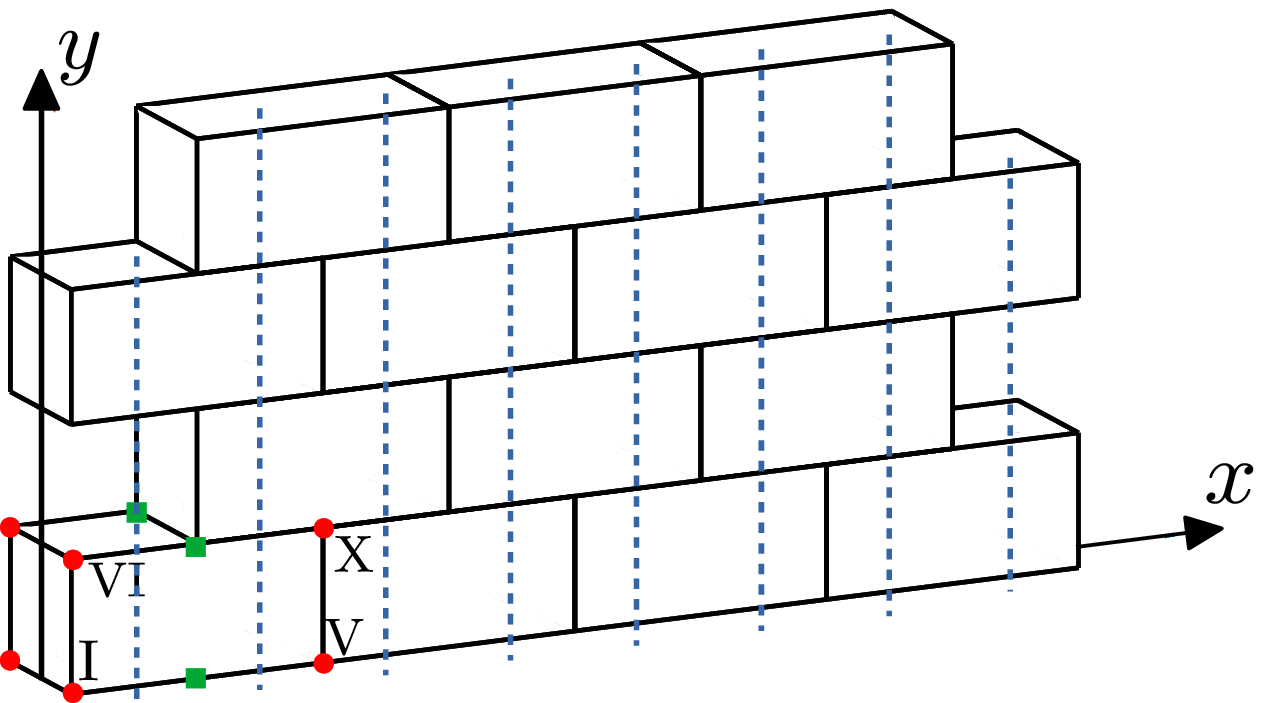}}
	\label{fig:Regular_arrangement}%
	\hspace{1cm}
	\subcaptionbox{Conjugate cell arrangement of cuboid cells}{\includegraphics[width=0.4\textwidth]{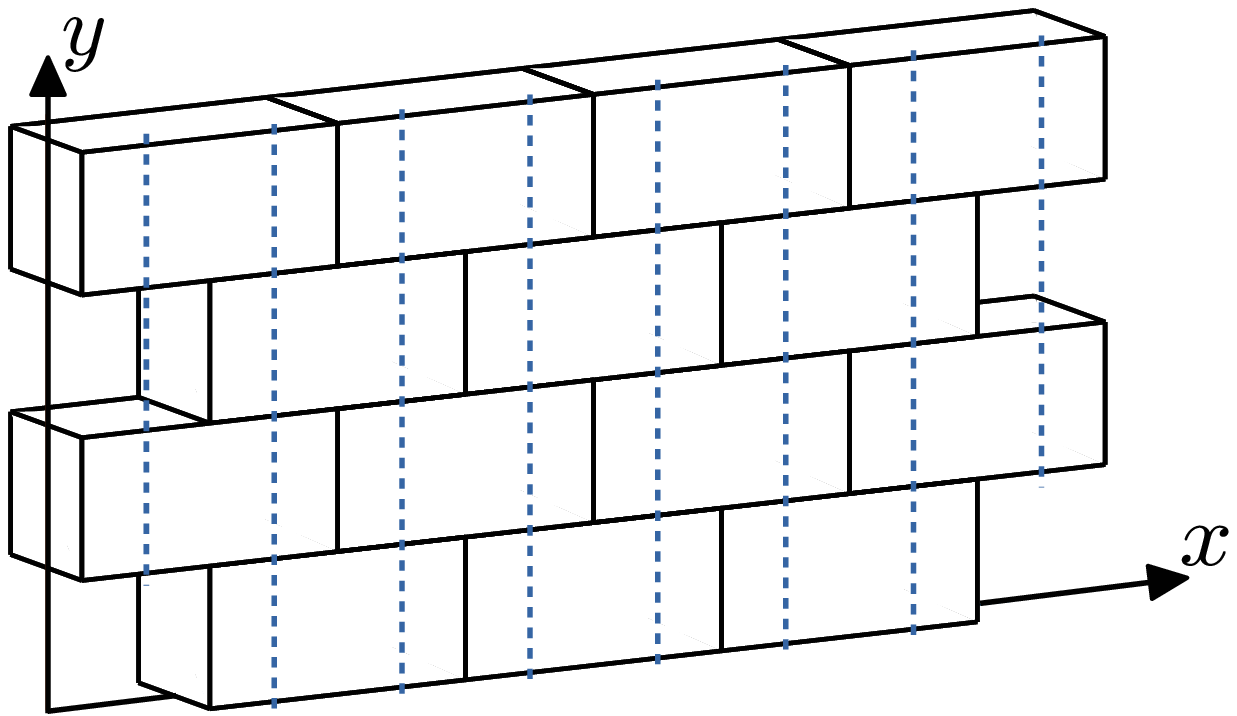}}
	\label{fig:Conjugate_arrangement} 
	\vspace{0.5cm}
	\subcaptionbox{Juxtaposed layers of Regular and Conjugate cell arrangement}{\includegraphics[width=0.4\textwidth]{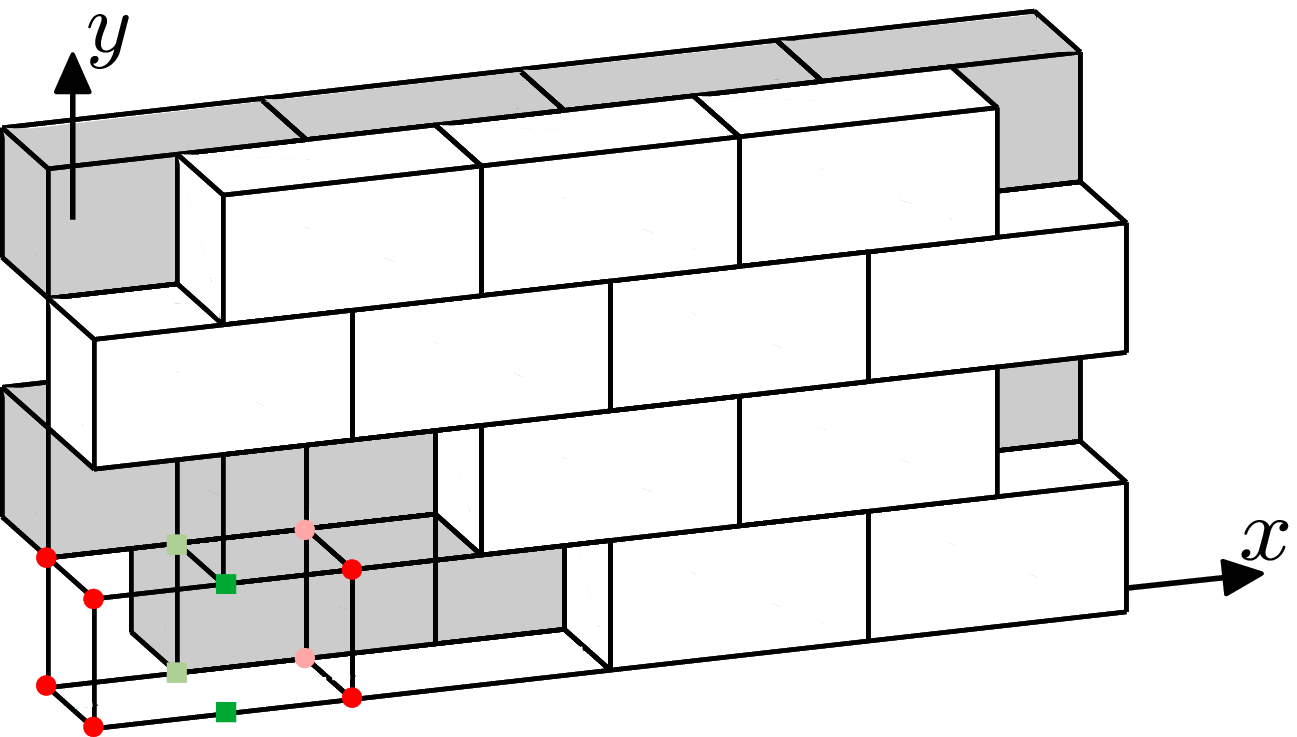}}
	\label{fig:Cell_arrangement_stack} 
	\hspace{1cm}
	\subcaptionbox{Point grid for the mesh representation in \ref{fig:Cell-Arrangement}c}{\includegraphics[width=0.4\textwidth]{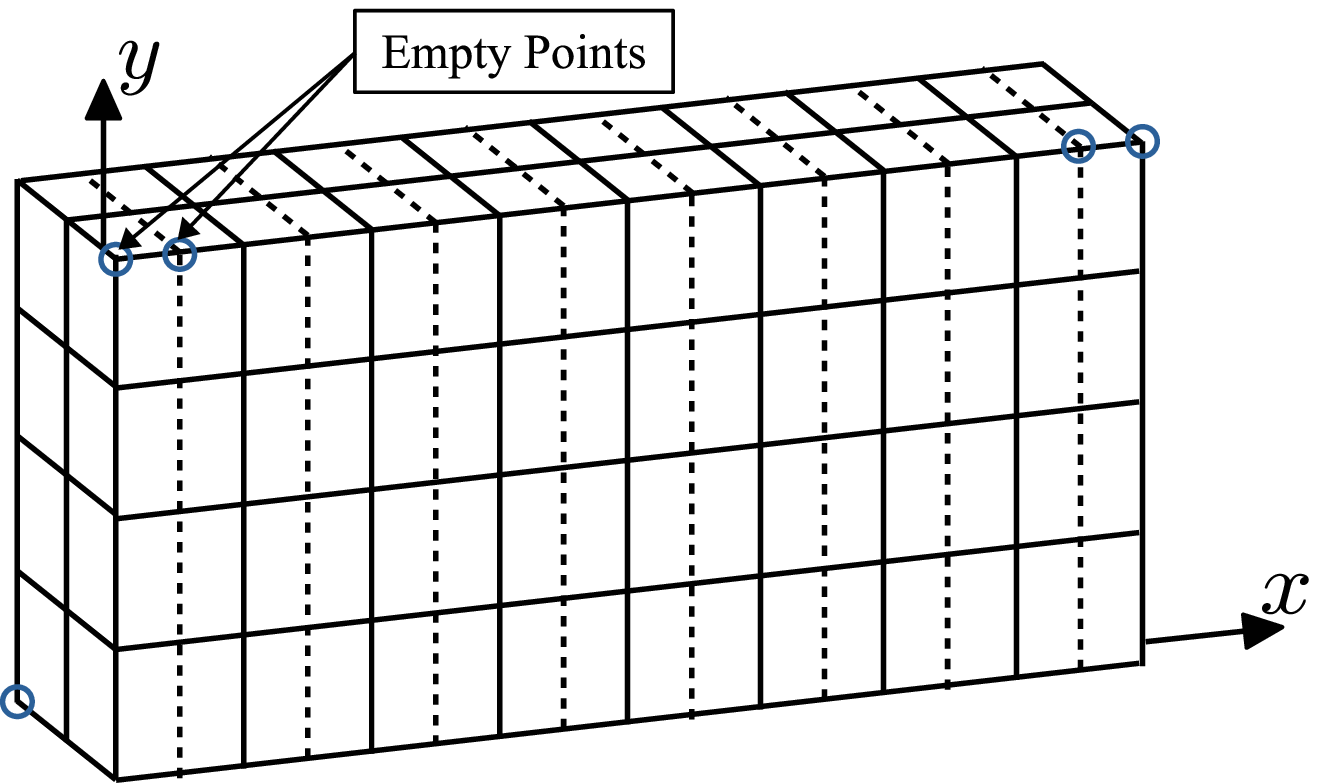}}
	\label{fig:point-grid}
	\caption{Cell arrangement in mesh space.}
	\label{fig:Cell-Arrangement}
\end{figure}
In the new orientation, positive X-axis passes through centroid of the square face $\{21,22,23,24\}$ (Fig. \ref{fig:reoriented_elements}b), the positive Y and Z-axis through the mid point of the edge $\{9-16\}$ and $\{10-11\}$ respectively. Face of the cell \{\RomanNumeralCaps{1}, \RomanNumeralCaps{5}, \RomanNumeralCaps{10}, \RomanNumeralCaps{6}\} is considered the front face while the face \{\RomanNumeralCaps{11}, \RomanNumeralCaps{16}, \RomanNumeralCaps{20}, \RomanNumeralCaps{15}\} is treated as the back face. In a lattice, \textit{what holds for one, holds for all}. Thus, repetition of the local cell arrangement developed above gives the virtual representation of the physical mesh in the mesh space. This leads to a staggered brick arrangement of cells (Fig. \ref{fig:Cell-Arrangement}c). Within a plane, the two observed arrangements are named Regular (Fig. \ref{fig:Cell-Arrangement}a) and Conjugate (Fig. \ref{fig:Cell-Arrangement}b) cell arrangements. The vertical dotted lines in Fig. \ref{fig:Cell-Arrangement}a and \ref{fig:Cell-Arrangement}b lie on the plane of the front faces of cells and intersect with the horizontal edges at locations of QPPs. Similar lines can also be drawn for back faces. Stacking regular cell and conjugate cell arrangement one after the other provides visual representation of the physical mesh in the mesh space. Proper alignment for stacking is presented in Fig. \ref{fig:Cell-Arrangement}c where the conjugate cell arrangement is highlighted in gray and some cells in the regular cell arrangement are made transparent for easier understanding. Note that the co-ordinate axes in Fig. \ref{fig:Cell-Arrangement} is independent of the axes in Fig. \ref{fig:reoriented_elements}. As a consequence of cell stacking, a new kind of cell neighbor is observed, categorized as Type D or \textit{Point neighbor} (Fig. \ref{fig:Cell-neighbours}d). Type D neighbors share a single FPP but do not share the corresponding node, therefore are not neighbors in the physical space. A cell can have a maximum of eight Type D neighbors.

\subsubsection{Element-Point connectivity}
\label{sec:element-point}
Having established a visual representation of the physical mesh in the mesh space, we develop connectivity for the mesh space, that is, connectivity between cells and points called the \textit{element-point connectivity}. To this end, we develop a \textit{point grid} using imaginary vertical lines and horizontal cell edges (Fig. \ref{fig:Cell-Arrangement}d). Vertical dotted lines intersect with horizontal lines at locations of QPPs while intersection of continuous lines locate FPPs and MPPs. To ensure that all planes of the point grid are rectangular grids, extra points, encircled by blue circles in Fig. \ref{fig:Cell-Arrangement}d, are introduced in the first and last planes of the abstract mesh. These extra points are referred to as empty points as they do not map to any node and are introduced only for convenience. Using the above setup of point grid, element-point connections can be established. For an $Nx \times Ny \times Nz$ mesh, the total number of cells $TE$, is 
\begin{eqnarray}
TE = (2Nx - 1)\times Ny \times \ceil[\bigg]{\frac{Nz}{2}} + \left[(2Nx - 1)\times \ceil[\bigg]{\frac{Ny}{2}} + Nx \times (Ny|2)\right]\times (Nz|2).
\end{eqnarray}
$\ceil{x}$ represents the greatest integer less than $x$ and $x|2$ is the remainder (0 or 1) when $x$ is divided by 2. $Nx$ is the maximum number of elements along the $x$ axis and, $Ny$ and $Nz$ are those along $y$ and $z$ axis respectively. The total number of points, $TP = (4Nx + 1) \times (Ny + 1) \times (Nz + 1)$. Note that, point grid of a mesh may or may not have empty points depending on the mesh size, for example, point grid of a $4 \times 4 \times 2 $ mesh as in Fig. \ref{fig:Cell-Arrangement} will have empty points but point grid for a $5 \times 5 \times 3$ mesh will not. Connectivity matrix for mesh space is stored as $\boldsymbol{\mathrm{EP}}(cn,lp) = gp$, where $\boldsymbol{\mathrm{EP}}$ is the element-point connectivity matrix, $cn$ is the cell number, $lp$ is the local point number and $gp$ is the global point number. As each cell maps to a single unique element, cell number, $cn$, and element number, $en$, are one and the same, and hence will be used interchangeably. The algorithm implemented to assign points their respective global point numbers is of little importance. One may implement any numbering scheme as long as $\boldsymbol{\mathrm{EP}}$ defines connectivity in the mesh space. In $\boldsymbol{\mathrm{EP}}$, points in column $\{1,5,6,10,11,15,16,20\}$ (Fig. \ref{fig:reoriented_elements}a) are FPPs for their corresponding cells. Similarly, points in column $\{ 3,8,13,18 \}$ and $\{ 2,4,7,9,12,14,17,19 \}$ are MPPs and QPPs respectively.

\subsubsection{Point-Node connectivity}
\label{sec:point-node}
\begin{figure}
	\centering
	\includegraphics[width=0.6\textwidth]{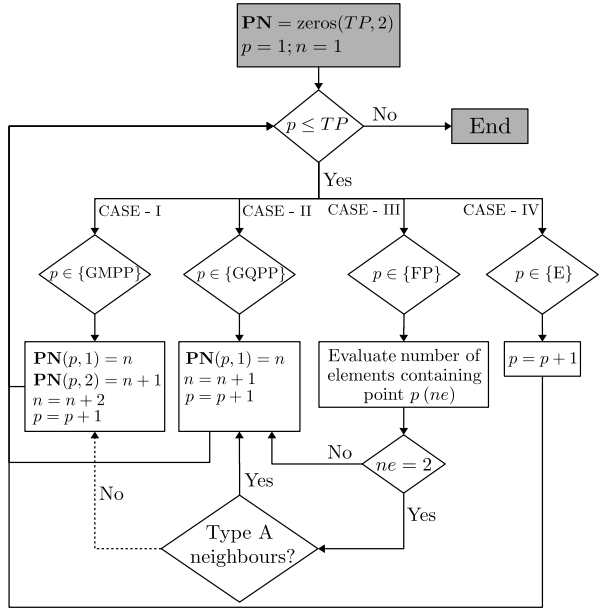}
	\caption{Algorithm for developing Point-Node connectivity. $p$ is point number, $n$ is node number and $ne$ is the number of elements containing point $p$}
	\label{fig:point-node}
\end{figure}

This section establishes relation between global point number and global node number. We construct 4 sets, specifically, \{GFPP\} (\textit{Global Face Plane Point}), \{GMPP\}(\textit{Global Mid Plane Point}), \{GQPP\}(\textit{Global Quarter Plane Points}) and \{E\}(\textit{Empty}). \{GFPP\}, \{GMPP\} and \{GQPP\} are respectively union over FPPs, MPPs and QPPs of all cells, and \{E\} is the set of all empty points. \{GFPP\} and \{GMPP\} have overlapping points because FPP of a cell is MPP of its Type B neighbor (\ref{fig:Cell-neighbours}b). The point-node connectivity can be established using the following rules, with \{P\} as a set of all points,
\begin{enumerate}
	\item \{P\} $\in$ \{GMPP\} maps to two nodes.
	\item \{P\} $\in$ \{GQPP\} maps to one node.
	\item \{P\} $\in$ \{FP\} = \{\{GFPP\} $-$ \{GFPP\}$\cap$\{GMPP\}\}, that is, points which are FPPs but not MPPs, map to 
	\begin{enumerate}
		\item one node if the point is not shared between elements or is shared by Type A neighbors.
		\item two nodes if the point is shared between elements which are not Type A neighbors.
	\end{enumerate}
	\item \{P\} $\in$ \{E\} does not map to any node.
\end{enumerate}
The algorithm for developing point-node connectivity is depicted in Fig. \ref{fig:point-node}. The fact that all points in \{FP\} are either members of a single cell or shared between at most 2 cells is incorporated into the algorithm. Point-node connectivity $\boldsymbol{\mathrm{PN}}$, takes global point number as input and gives the associated nodes. If a point maps to only one node, one of the outputs is automatically zero (0).

\subsubsection{Node-Selection}
\label{sec:node-selection}
\begin{figure}
	\centering
	\includegraphics[width=0.55\textwidth]{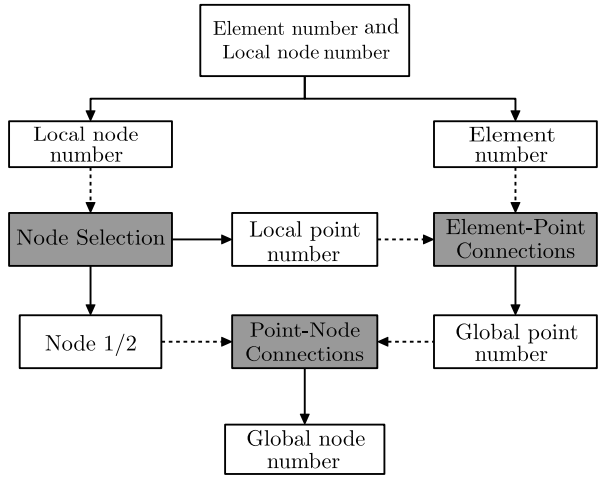}
	\caption{Overview of Connectivity matrix generation algorithm. Steps are highlighted using gray boxes. Solid and dotted arrows represent output and input of a step respectively.}
	\label{fig:mesh_algo}
\end{figure}

Obtaining global node number from element number and local node number involves the following sequence of steps:
\begin{enumerate}
	\item Insert local node number into unit cell map in Table \ref{tab:2} to get the associated local point number.
	\item Input the element number and local point number into the element-point connections to obtain the corresponding global point number.
	\item Use the point-node connectivity to obtain the global node/s for the global point obtained in the previous step.
\end{enumerate}
Complications arise when the global point maps to two global nodes. In such cases a choice between the 2 node numbers needs to be made. The choice is dependent on the local information, that is, local point and local node number of the associated node. To this end, the node selection strategy is presented in Table \ref{tab:2}, where node 1 and node 2 refer to the node number in the 1\textsuperscript{st} and 2\textsuperscript{nd} column of point-node connections, the $\boldsymbol{\mathrm{PN}}$ matrix, respectively. The strategy assumes that an FPP maps to 2 nodes. In case an FPP maps to only 1 node, the choice is straightforward. Local nodes corresponding to FPPs on the front plane, i.e. local points \{\RomanNumeralCaps{1}, \RomanNumeralCaps{5}, \RomanNumeralCaps{6} ,\RomanNumeralCaps{10}\} (Fig. \ref{fig:reoriented_elements}a) map to node 2 while local nodes associated to FPPs on the back plane, i.e. local points \{\RomanNumeralCaps{11}, \RomanNumeralCaps{15}, \RomanNumeralCaps{16}, \RomanNumeralCaps{20}\} map to node 1 (Table \ref{tab:2}). Local nodes associated with QPPs map to node 1 as QPPs always correspond to a single node. In cases where the local node maps to an MPP the node selection procedure is established by looking at the neighboring elements. For example, in Fig. \ref{fig:reoriented_elements}b, local node 13 attaches to local node 4 of its neighbor which shares the hexagonal face $\{12,13,19,23,22,18\}$, implying that local node 13 of the primary element and local node 4 of the neighboring element correspond to the same node in the mesh. Hence, they should have the same output from point-node connections and node-selection procedure. As local node 4 maps to node 1, so should local node 13. \\
A pictorial representation of the algorithm to achieve the connectivity matrix is provided in Fig. \ref{fig:mesh_algo}. The algorithm takes element number and local node number as input and gives the corresponding global node number as output. The unit cell map and choice of node are grouped under Node selection.

\subsection{Node Locations and Scaling}
\label{sec:node_location}
\begin{figure}[h]
	\centering
	\subcaptionbox{Element arrangement in mesh space. Shaded regions represent shared faces }{\includegraphics[width=0.4\textwidth]{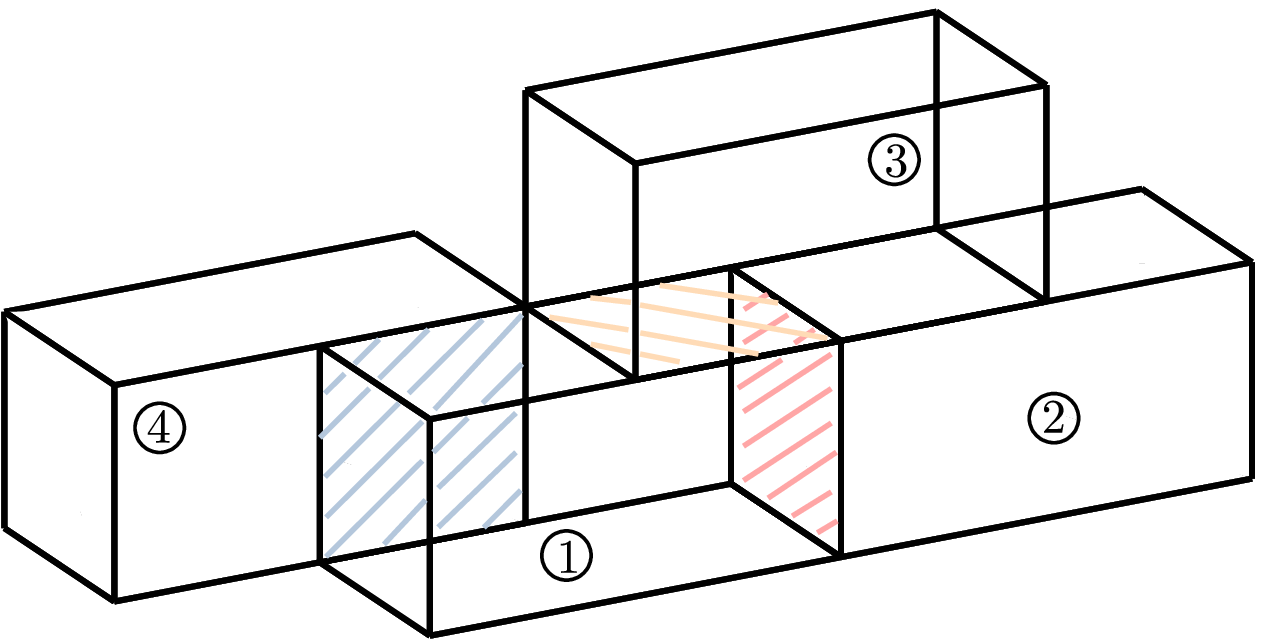}}
	\label{fig:NL_schem_brick}%
	\hspace{1cm}
	\subcaptionbox{Physical manifestation of element arrangement in Fig. \ref{fig:NL_schem}a }{\includegraphics[width=0.4\textwidth]{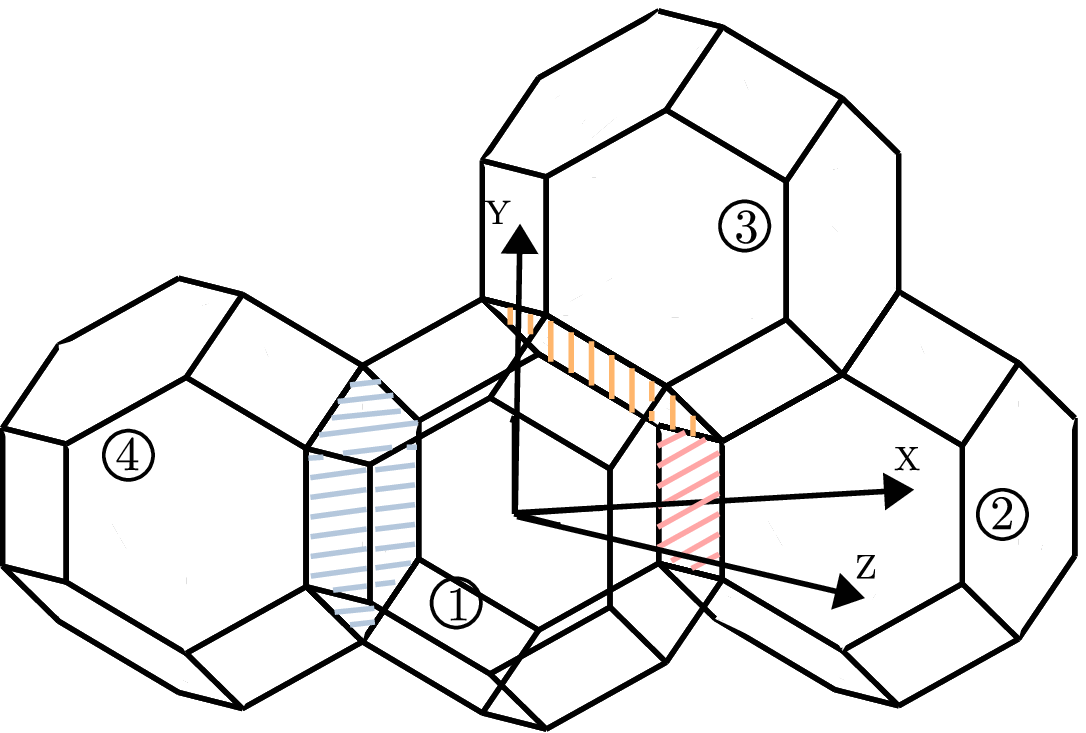}}
	\label{fig:NL_schem_physical} 
	\caption{Truncated octahedron with its neighbors in the physical space, and the corresponding cuboids in the mesh space. Corresponding shared faces are shaded.  }
	\label{fig:NL_schem}
\end{figure}

For an element of edge length $a = \sqrt{2}$, location of the local nodes with respect to the element centroid for the orientation in Fig. \ref{fig:reoriented_elements}b can be obtained using Table \ref{tab:1} and the orthogonal matrices $\boldsymbol{\mathrm{R}}_1$ and $\boldsymbol{\mathrm{R}}_2$ presented in section \ref{sec:connectivity}. Hence, for an element of edge length $l$, the position vector of nodes is given by $\boldsymbol{\mathrm{x}} = \left(\frac{l}{a}\right) \boldsymbol{\mathrm{X}}$, where $\boldsymbol{\mathrm{X}}$ is position vector of the considered node for edge length $a$. As the connectivity matrix and local node locations with respect to cell centroids are known, locating element centroids will provide enough information to determine node locations. Evaluating location of element centroids requires calculating the vector separating an element's centroid from its neighboring cell's centroid. As the mesh visualization is conducted in the mesh space, we look at cell neighbors in that space and determine the physical vector separating element centroids for different neighbors. Fig. \ref{fig:NL_schem}a shows a typical element with three of its neighbors in the mesh space. Fig. \ref{fig:NL_schem}b shows the corresponding element arrangement in the physical space. From symmetry and that elements are placed identically with respect to each other in a lattice, evaluating the vector separating cell centroids of element 1 and its neighbors provides sufficient information to determine vectors separating an element's centroid from all its neighbors. Let $\boldsymbol{\mathrm{x}}^i$ be the centroid of element $i$. For an element oriented as in Fig. \ref{fig:reoriented_elements}b, from geometrical analysis, centroids of element 2 and 1 are separated by $\{2\sqrt{2}l~ 0 ~ 0\}^T$, i.e., $\boldsymbol{\mathrm{x}}^2 - \boldsymbol{\mathrm{x}}^1 = \{2\sqrt{2}l~ 0 ~ 0\}^T$. Similarly, $\boldsymbol{\mathrm{x}}^3 - \boldsymbol{\mathrm{x}}^1 = \{\sqrt{2}l~ 2l ~ 0\}^T$ and $\boldsymbol{\mathrm{x}}^4 - \boldsymbol{\mathrm{x}}^1 = \{-\sqrt{2}l ~ 0 ~ -2l\}^T$. The face shared between elements in the physical space are shaded and their corresponding representations are shaded by the same color in the mesh space. As all elements in the mesh are same, node locations are not needed to conduct the FE analysis but can be useful when imposing boundary conditions.

\section{FEM using truncated octahedron}
\label{sec:FEM}
\begin{figure}[h]
	\centering
	\includegraphics[width=0.5\textwidth]{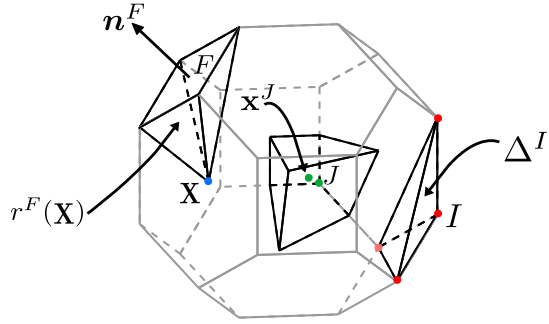}
	\caption{\textit{Nodal volume contribution} and \textit{point to face volume} for truncated octahedron}
	\label{fig:shape_function}
\end{figure}

FEM for linear elasticity problems is well established. The solution procedure for small deformation is reflected as a system of linear equations represented by,
\begin{eqnarray}
\boldsymbol{\mathrm{K}}\boldsymbol{\mathrm{u}} = \boldsymbol{\mathrm{F}}
\end{eqnarray}
where $\boldsymbol{\mathrm{K}}$ is the global stiffness matrix, $\boldsymbol{\mathrm{u}}$ is the unknown nodal displacement vector and $\boldsymbol{\mathrm{F}}$ is the external force vector. $\boldsymbol{\mathrm{F}}$ is evaluated from the forces applied to the structure while the global stiffness matrix is obtained by assembling element stiffness matrices, $\boldsymbol{\mathrm{K}}_0$, where 
\begin{eqnarray}
\boldsymbol{\mathrm{K}}_0 = \int_{\Omega_e} \boldsymbol{\mathrm{B}}^T \boldsymbol{\mathrm{D}} \boldsymbol{\mathrm{B}} ~d \Omega_e.
\label{eqn:elem_stiff}
\end{eqnarray} 
$\boldsymbol{\mathrm{B}}$ is the shape function derivative matrix, $\boldsymbol{\mathrm{D}}$ is the elasticity matrix based on material property and $\Omega_e$ is volume of the element. Evaluating $\boldsymbol{\mathrm{K}}_0$ requires establishing shape functions, their derivatives and a numerical integration technique for the elements used to discretize the domain. In literature, there are multiple methods to develop linear shape functions for convex polyhedral solids such as the mean value coordinates method \citep{Floater2005}, harmonic shape functions \citep{Remacle2012a} and the generalized version of the Wachspress shape functions \citep{Kraus2012}. Mean value coordinates are confined to solids with triangular faces. Harmonic shape functions are numerical solutions to the Laplace's equation over the element with Dirichlet boundary conditions specified on element surface. An advantage of harmonic shape functions is that they can be applied to concave polyhedrals. The generalized Wachspress shape functions are analytical for convex polyhedrons which were first introduced as barycentric coordinates in \cite{Warren1996}. \cite{Kraus2012} adopted and implemented them to solve a non-linear elasticity problem with convex polyhedron elements. Studies implementing iso-geometric analysis using NURBS based shape function have also been conducted in topology optimization studies \citep{hou2017explicit, gao2020comprehensive}. Herein, we use the generalized Wachspress shape functions, given as:
\begin{eqnarray}
\label{eqn:shape_function}
N^I(\boldsymbol{\mathrm{X}}) = \frac{k^I s^I(\boldsymbol{\mathrm{X}})}{\sum_{J = 1}^{Nn} k^J s^J(\boldsymbol{\mathrm{X}})}
\end{eqnarray}
where $N^I(\boldsymbol{\mathrm{X}})$ is the shape function associated with node $I$ at a point $\boldsymbol{\mathrm{X}}$ inside the element, $Nn$ is the number of vertices/nodes in the element, $k^I$ is the \textit{nodal volume contribution} given by 
\begin{eqnarray}
\label{eqn:kisi}
k^I = \kappa^I \Delta^I ~~ \text{and} ~~ s^I = \prod_{F \notin \Gamma^I} r^F(\boldsymbol{\mathrm{X}}).
\end{eqnarray}
$\kappa^I = (1/n^I)^3$ where $n^I$ is the number of nodes connected to node $I$ via an edge. For instance, $n^I = 3$ for truncated octahedron. $\Delta^I$ is the volume of convex hull created by node $I$ (Fig. \ref{fig:shape_function}) and its edge-connected neighbors, $\Gamma^I$ is the set of faces containing the node $I$ and $r^F(\boldsymbol{\mathrm{X}})$ is \textit{point to face volume}, that is, volume of the convex hull created by the point $\boldsymbol{\mathrm{X}}$ and vertices of face $F$ (Fig. \ref{fig:shape_function}). 
For a truncated octahedron element, $k^I = k^J ~\forall I,J \in \mathbb{N}$ where $\mathbb{N}$ is the set of nodes in the element. Thus, $N^I(\boldsymbol{\mathrm{X}})$ in Eqn. \ref{eqn:shape_function} is simplified to
\begin{eqnarray}
N^I (\boldsymbol{\mathrm{X}}) = \frac{s^I(\boldsymbol{\mathrm{X}})}{\sum_{J = 1}^{Nn} s^J (\boldsymbol{\mathrm{X}})}
\label{eqn:shape_fun}
\end{eqnarray}
and $r^F$ in Eqn. \ref{eqn:kisi} is given as
\begin{eqnarray}
r^F(\boldsymbol{\mathrm{X}}) = \frac{1}{3}B^F h^F(\boldsymbol{\mathrm{X}})
\end{eqnarray}
where $B^F$ is the area of face $F$ and $h^F(\boldsymbol{\mathrm{X}})$ the perpendicular distance between point $\boldsymbol{\mathrm{X}}$ and the face $F$.
\\
Shape functions derivatives with respect to $\boldsymbol{\mathrm{X}}$ are given as
\begin{eqnarray}
\frac{\partial N^I (\boldsymbol{\mathrm{X}})}{\partial \boldsymbol{\mathrm{X}}} = \frac{1}{\sum_{J = 1}^{Nn} s^J (\boldsymbol{\mathrm{X}})} \frac{\partial s^I (\boldsymbol{\mathrm{X}})}{\partial \boldsymbol{\mathrm{X}}} - \frac{N^I}{\sum_{J = 1}^{Nn} s^J (\boldsymbol{\mathrm{X}})} \sum_{K =1}^{Nn}  \frac{\partial s^K (\boldsymbol{\mathrm{X}})}{\partial \boldsymbol{\mathrm{X}}} 
\end{eqnarray}
where 
\begin{eqnarray}
\frac{\partial s^I (\boldsymbol{\mathrm{X}})}{\partial \boldsymbol{\mathrm{X}}} = s^I(\boldsymbol{\mathrm{X}}) \sum_{F \notin \Gamma^I}  \frac{1 }{r^F (\boldsymbol{\mathrm{X}})} \frac{\partial r^F (\boldsymbol{\mathrm{X}})}{\partial \boldsymbol{\mathrm{X}} }
\end{eqnarray}
with
\begin{eqnarray}
\frac{\partial r^F}{\partial \boldsymbol{\mathrm{X}}} = - \frac{B^F}{3} \boldsymbol{n}^F
\end{eqnarray}
where $\boldsymbol{n}^F$ is the outward normal to the face $F$, as shown in Fig. \ref{fig:shape_function}. 
\\

Numerical integration approximates an integral over a domain as the weighted sum of integrand value at different points within the domain. The traditional method for numerical integration over a convex element is to divide the element into tetrahedrons and implement either a single point or three point Gauss-Quadrature technique over each tetrahedron \citep{Kraus2012, Martin2008}. Tetrahedrons in the aforementioned method are created by joining a face's centroid, nodes of an edge on the face and the element centroid. For a truncated octahedron element, this leads to 72 tetrahedrons and therefore, for a three point Gauss-Quadrature method, one works with 216 integration points. This being a relatively large number, we adopt the integration technique in \cite{Rashid2006}. The number of integration points therein equals the number of vertices/nodes in the element, leading to 24 integration points for our case. This method divides an element into subregions where each subregion is associated with a node. Integration points and weights are the centroids and volumes of the subregions respectively. Subregion associated with a node is the convex hull of centroid of faces containing the node, mid point of edges connecting the node, the node itself and the centroid of the element. Subregion for node $J$, of a truncated octahedron element is depicted in Fig. \ref{fig:shape_function}. As all nodes of a truncated octahedron are identically placed with respect to the cell centroid, thus, volume of all subregions are identical. Thus, all integration weights, $w^J$, are $1/24$ of the element volume, that is, $w^J = \sqrt{2}a^3/3$ where $a$ is the edge length. Location of all integration points $\boldsymbol{\mathrm{x}}^J$, can be evaluated using nodal coordinates.
\\
As all elements in the mesh are truncated octahedrons we define a master element of the same shape in a parametric space. The master element has an edge length of $a = \sqrt{2}$ and has the same orientation with respect to natural coordinates $(\xi,\eta,\zeta)$ as the physical element in Fig. \ref{fig:reoriented_elements}b has with respect to physical coordinates. The same local node numbering as in Fig. \ref{fig:reoriented_elements}b is adopted for convenience. A linear iso-parametric mapping between the master and the physical element is given as
\begin{eqnarray}
\boldsymbol{\mathrm{X}} = \sum_{I = 1}^{Nn} N^I(\xi,\eta,\zeta) \boldsymbol{\mathrm{X}}_I
\end{eqnarray}
where, $\boldsymbol{\mathrm{X}}$ is the location of point in the physical space corresponding to the point $(\xi,\eta,\zeta)$ in the parametric space, and $\boldsymbol{\mathrm{X}}_I$ are the node locations in the physical space. $N^I(\xi,\eta,\zeta)$ are evaluated using Eqn. \ref{eqn:shape_fun}.
\\
We implement the Neo-Hooken material model for which components of the elasticity tensor, $\boldsymbol{\mathrm{c}}$, are given as
\begin{eqnarray}
c_{pqrs} = \frac{\lambda}{J} \delta_{pq}\delta_{rs} -\frac{2}{J}(\lambda \ln J - \mu) \delta_{pr}\delta_{qs}
\end{eqnarray} 
where $\lambda$ and $\mu$ are Lame's constants, $\delta_{ij} = 1$ for $i=j$ and $0$ otherwise, and $J$ is jacobian for the deformation that is determinant of the deformation gradient. $J = 1$ for linear elastic problems. The elasticity matrix $\boldsymbol{\mathrm{D}}$ in Eqn. \ref{eqn:elem_stiff}. is the Voigt notation of the elasticity tensor $\boldsymbol{\mathrm{c}}$. We implement $\lambda = \mu = 10$ for all examples presented.

\section{Topology optimization with MMOS}
\label{sec:topo_opti}
Topology optimization for solid structures is often presented as a material distribution problem, that is, given the design domain, external forces, boundary conditions on domain surfaces and mass/volume of material available, the aim is to determine the optimal distribution of material within the domain to minimize a given objective function. In a numerical setting, a topology optimization problem can be expressed as,
\begin{flalign}\label{eqn:2}
&~& min. &~~~~ \mathbb{I}(\rho_i) &\\ \nonumber
&\text{subject to} & \bm{\mathrm{R}}(\rho_i,\bm{\mathrm{u}})  &= \bm{0} &\\ \nonumber
&\text{such that}& \sum_{i=1}^{ne} \rho_i v_i &\leq V^* = v_f \sum_{i=1}^{ne} v_i & \\ \nonumber
& ~ & 0 \leq \rho_i & \leq 1 &
\end{flalign}
where $\rho_i$ is the density for element $i$, $\mathbb{I}(\rho_i)$ is the objective function, $\bm{\mathrm{R}}(\rho_i,\bm{\mathrm{u}})$ is the state equation to be satisfied for any intermediate continuum, $v_i$ is the volume of element $i$, $v_f$ is the volume fraction available for the design and $ne$ is the total number of elements in the FE discretization of the domain. For small deformation problems with mechanical loads, $\bm{\mathrm{R}}(\rho_i,\bm{\mathrm{u}}) = \bm{0}$ represents the force balance equation $\bm{\mathrm{K}}(\rho_i)\bm{\mathrm{u}} = \bm{\mathrm{F}}$, where $\bm{\mathrm{K}}(\rho_i)$ is the global stiffness matrix for any intermediate continuum, $\bm{\mathrm{u}}$ is the nodal displacement vector and $\boldsymbol{\mathrm{F}}$ is the external force vector. The objective function and stiffness matrix are expressed as functions of elemental densities. The objective depends on the problem while the relation between stiffness matrix and density depends on the choice of material model. We implement the SIMP material model \citep{bendsoe1999material, Guest2009} for which the elemental stiffness, $\boldsymbol{\mathrm{Ke}}$, for an element with density $\rho_i$ is given as,
\begin{eqnarray}
\label{eqn:SIMP}
\boldsymbol{\mathrm{Ke}} = \{\rho_i^\eta (1-\rho_{min}) + \rho_{min}\}\boldsymbol{\mathrm{K}}_0
\end{eqnarray}
where $\eta$ is the SIMP penalty parameter, $\rho_{min}$ is a small positive number introduced to remove potential singularity of the stiffness matrix and $\boldsymbol{\mathrm{K}}_0$ is the elemental stiffness of a solid cell obtained using Eqn. \ref{eqn:elem_stiff}. Topology optimization methods vary in the procedure to obtain element densities. We implement spheroidal masks within the Material Mask Overlay Strategy (MMOS) to compute these densities. Both, the density field evaluation and sensitivity analysis are discussed in sec. \ref{sec:density_distribution}. Some well established example problems in structural topology optimization are presented in sec. \ref{sec:sample_prob}. 

\subsection{Density distribution and sensitivity analysis}
\label{sec:density_distribution}
\begin{figure}[!htb]
	\begin{minipage}{0.35\textwidth}
		\centering
		\includegraphics[width=0.8\textwidth]{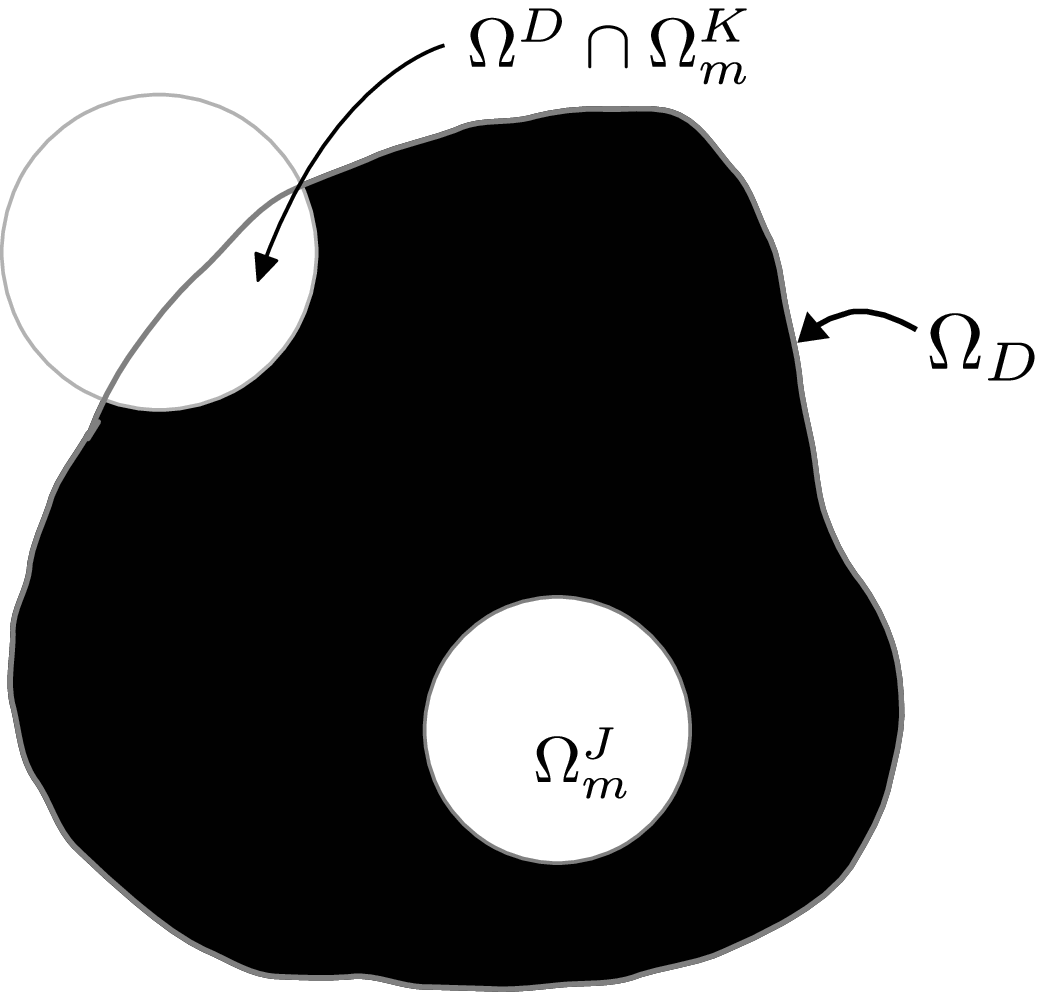}
		\caption{Density distribution with negative masks}
		\label{fig:mask_domain_schematic}
	\end{minipage}\hfill
	\begin{minipage}{0.65\textwidth}
		\centering
		\subcaptionbox{Spheroidal mask }{\includegraphics[width=0.45\textwidth]{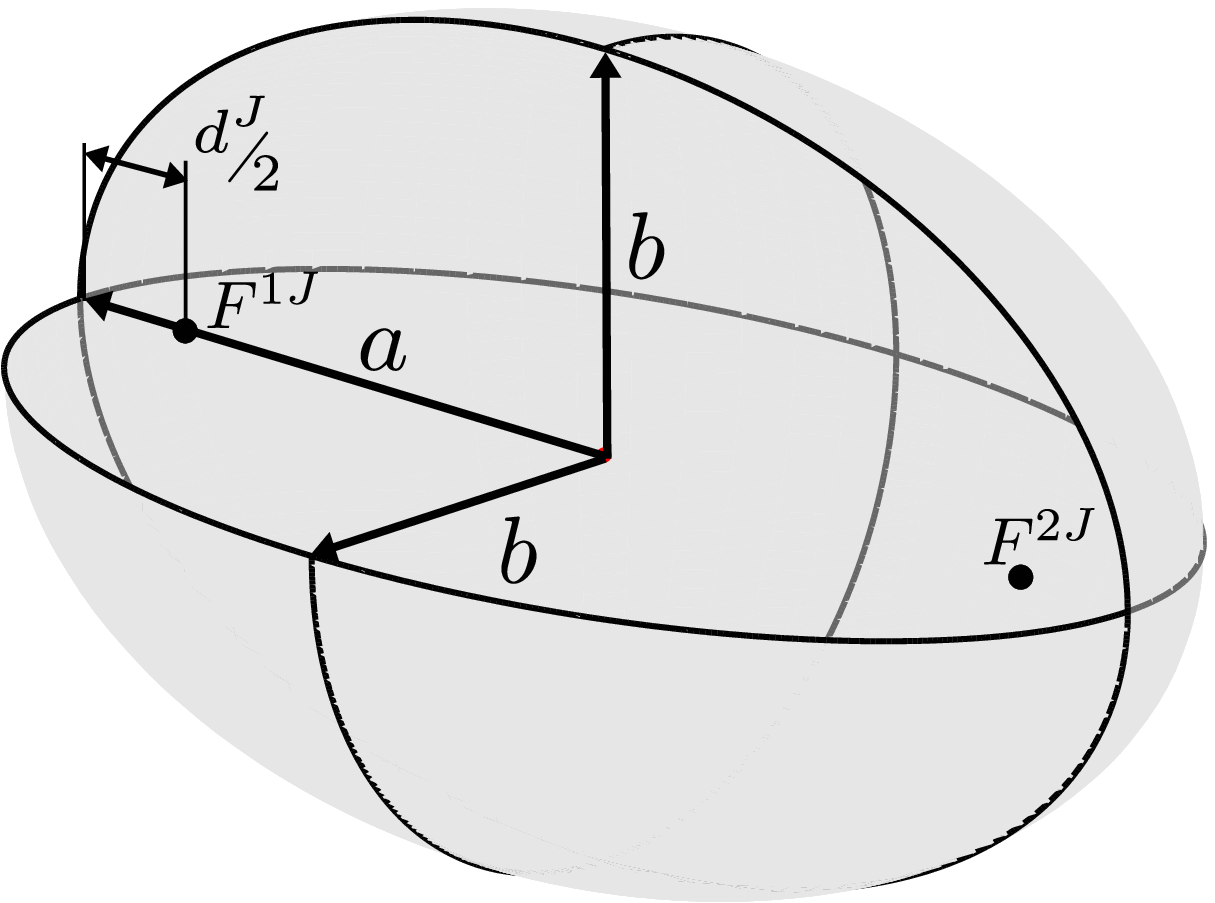}}
		\subcaptionbox{Variation of $h_J(\boldsymbol{\mathrm{x}})$ with $\phi_J(\boldsymbol{\mathrm{x}})$ }{\includegraphics[width=0.45\textwidth]{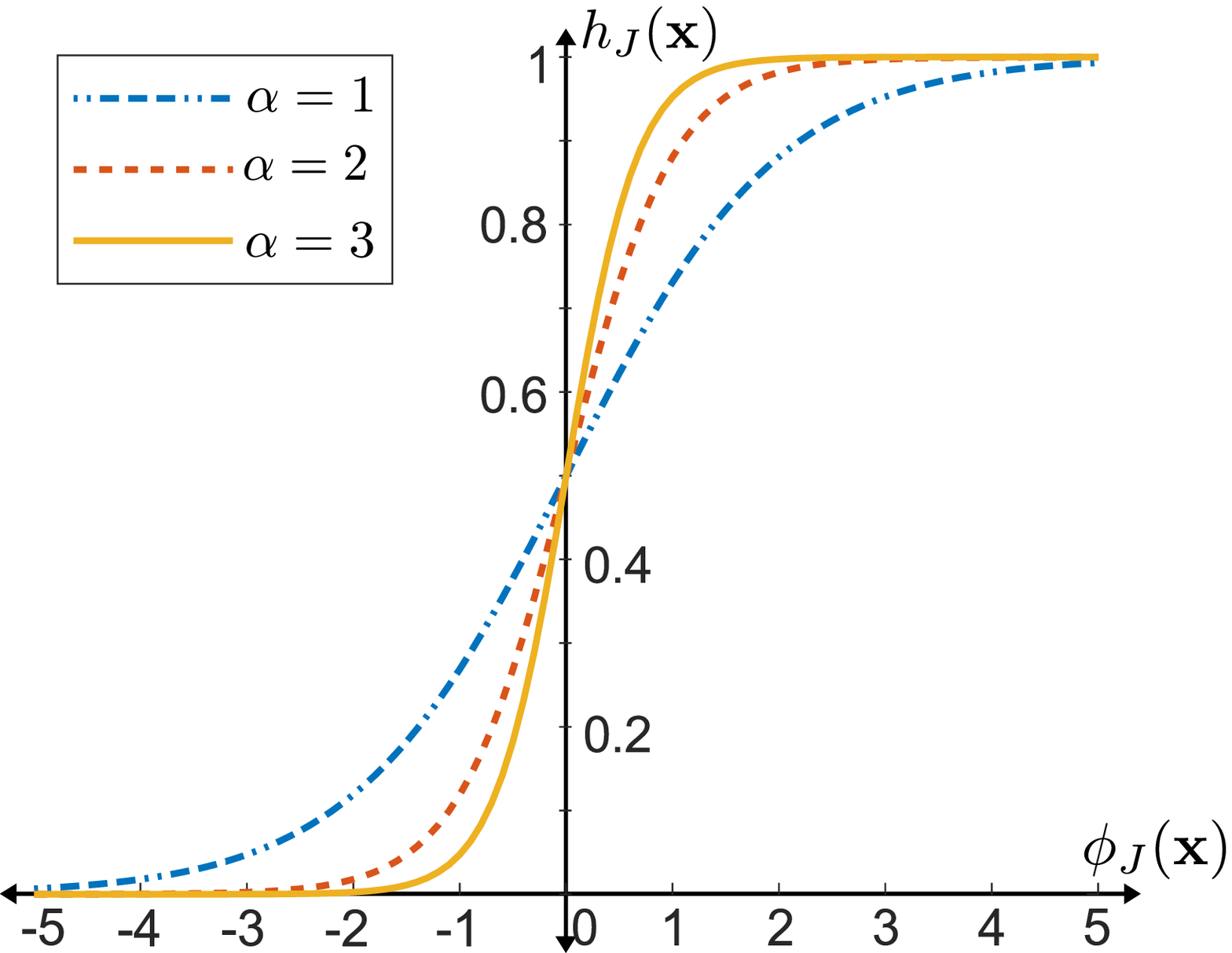}}
		\caption{Schematics for gradient based MMOS}
		\label{fig:mask_schematic}
	\end{minipage}
\end{figure}

Material Mask Overlay Strategy (MMOS) initially proposed in \cite{Saxena2008}, is a feature based method \citep{Saxena2011, Guo2016, Hoang2017} for topology optimization. Such methods express the density distribution over the domain as a function of shape and location of a collection of geometrical features which in MMOS are referred to as \textit{masks}. These are of two types, (a) positive and (b) negative. Positive masks add material while negative masks remove material from regions enclosed within. When working with negative masks (Fig. \ref{fig:mask_domain_schematic}), the domain is initially taken to be filled and masks are then systematically maneuvered, sized and oriented to remove material and thus determine the optimal topology.

Contribution of a negative mask to the density distribution can be expressed by a function $H(\boldsymbol{\mathrm{x}})$ where $H(\boldsymbol{\mathrm{x}}) = 0$ for $\boldsymbol{\mathrm{x}} \in \Omega_D \cap \Omega_m$ and $H(\boldsymbol{\mathrm{x}}) = 1$ for $\boldsymbol{\mathrm{x}} \in \Omega_D / \Omega_m$. $\Omega_D$ is the design domain while $\Omega_m$ is the region enclosed within the mask. Density distribution over $\Omega_D$ is achieved by multiplying the contribution of each mask, that is,
\begin{eqnarray}
\rho(\boldsymbol{\mathrm{x}}) = \prod_{J=1}^{TM} H_J(\boldsymbol{\mathrm{x}})
\end{eqnarray}
where $H_J(\boldsymbol{\mathrm{x}})$ is the contribution from the $J^{th}$ mask and $TM$ is the total number of masks. As $H_J(\boldsymbol{\mathrm{x}})$ is a discontinuous function, to implement gradient based optimization, it is replaced by its differentiable approximation
\begin{eqnarray}
\label{eqn:fj}
h_J(\boldsymbol{\mathrm{x}}) = \frac{1}{1 + \exp^{-\alpha \phi_J(\boldsymbol{\mathrm{x}})}}
\end{eqnarray}
where $\alpha$ is a positive scalar and $\phi_J(\boldsymbol{\mathrm{x}})$ is a scalar function such that $\phi_J(\boldsymbol{\mathrm{x}}) < 0$ for $\boldsymbol{\mathrm{x}} \in \Omega_m^J$, $\phi_J(\boldsymbol{\mathrm{x}}) = 0$ for $\boldsymbol{\mathrm{x}} \in \partial \Omega_m^J$ and  $\phi_J(\boldsymbol{\mathrm{x}}) > 0$ for $\boldsymbol{\mathrm{x}} \in \Omega_D/\Omega_m^J$. $\Omega_m^J$ is the region enclosed within the $J^{th}$ mask with boundary $\partial \Omega_m^J$. The variation of $h_J(\boldsymbol{\mathrm{x}})$ with $\phi_J(\boldsymbol{\mathrm{x}})$ for various values of $\alpha$ is presented in Fig. \ref{fig:mask_schematic}b. Spheroidal masks are implemented herein, which are a special case of ellipsoidal masks with both the minor axes equal. To this end,
\begin{eqnarray}
\phi_J(\boldsymbol{\mathrm{x}}) = |\boldsymbol{\mathrm{x}} - \boldsymbol{\mathrm{F}}^{1J}| + |\boldsymbol{\mathrm{x}} - \boldsymbol{\mathrm{F}}^{2J}| - |\boldsymbol{\mathrm{F}}^{1J} - \boldsymbol{\mathrm{F}}^{2J}| - d^J
\end{eqnarray}
where $\boldsymbol{\mathrm{F}}^{1J}$ and $\boldsymbol{\mathrm{F}}^{2J}$ are position vectors of the focal points of the spheroid and $\frac{d^J}{2}$ is the difference between semi-major axis $a$ and distance of the focal point $F^{1J}$ from center of the spheroid (Fig. \ref{fig:mask_schematic}a). The above function choice is based on the property that sum of distance from the foci of any point on the spheroid is constant, equal to twice the semi-major axis, $a$. Using $\boldsymbol{\mathrm{F}}^{1J}$, $\boldsymbol{\mathrm{F}}^{2J}$ and $d^J$, semi-major axis $a$, and semi-minor axis $b$, for the mask can be evaluated as
\begin{eqnarray}
a = \frac{|\boldsymbol{\mathrm{F}}^{1J} - \boldsymbol{\mathrm{F}}^{2J}| + d^J}{2} ~~\text{and}~~ b = \sqrt{\left(2a - \frac{d^J}{2}\right) \frac{d^J}{2}}.
\end{eqnarray} 
Let
\begin{eqnarray}
\left\{\boldsymbol{\mathrm{F}}^{1J}\right\} = \begin{Bmatrix}
\Psi_1^J \\[6pt] \Psi_2^J \\[6pt] \Psi_3^J
\end{Bmatrix} ~,~
\left\{\boldsymbol{\mathrm{F}}^{2J}\right\} = \begin{Bmatrix}
\Psi_4^J \\[6pt] \Psi_5^J \\[6pt] \Psi_6^J
\end{Bmatrix} ~\text{and}~ 
d^J = \Psi_7^J
\end{eqnarray}
such that location, size and orientation of the $J^{th}$ spheroidal mask are controlled via design variables $\Psi_i^J$, $i = \{1,2,...,7\}$.
For a topology optimization problem that utilizes FEM for function evaluation, the element densities are evaluated as
\begin{eqnarray}
\rho_i \coloneqq \rho(\boldsymbol{\mathrm{x}}^i) = \prod_{J=1}^{TM} H_J(\boldsymbol{\mathrm{x}}^i) \approx \prod_{J=1}^{TM} h_J(\boldsymbol{\mathrm{x}}^i) = \prod_{J=1}^{TM} \left[\frac{1}{1 + \exp^{-\alpha \phi_J(\boldsymbol{\mathrm{x}}^i)}}\right]
\end{eqnarray}
where $\rho_i$ and $\boldsymbol{\mathrm{x}}^i$ are the density and centroid of the $i^{th}$ element respectively. For numerical implementation we allow for negative values of $d^J := \Psi_7^J$ which corresponds to disappearing of mask. Thus, setting the lower limit $d^J < 0$ is recommended as $d^J = 0$ can theoretically have non-zero contribution to the density distribution.

To implement gradient based optimization, derivative of the objective function $\mathbb{I}(\rho_i)$ with respect to design variable, $\Psi_i^J$ is evaluated as follows. Utilizing chain rule gives,
\begin{eqnarray}
\frac{\partial \mathbb{I}(\rho_i)}{\partial \Psi_k^J} = \frac{\partial \mathbb{I}(\rho_i)}{\partial \rho_i} \left(\frac{\partial \rho_i}{\partial \Psi_k^J}\right).
\end{eqnarray}
$\frac{\partial \mathbb{I}(\rho_i)}{\partial \rho_i}$ depends on the objective function while $\frac{\partial \rho_i}{\partial \Psi_k^J}$ can be evaluated independently as
\begin{eqnarray}
\frac{\partial \rho_i}{\partial \Psi_k^J} = \alpha \rho_i \left[1 - h_J(\boldsymbol{\mathrm{x}}^i)\right] \frac{\partial \phi_J(\boldsymbol{\mathrm{x}}^i)}{\partial \Psi_k^J}
\end{eqnarray}
where
\begin{eqnarray}
\frac{\partial \phi_J(\boldsymbol{\mathrm{x}}^i)}{\partial \Psi_k^J} = 
\begin{dcases}
\frac{\Psi_k^J - x_k^i}{|\boldsymbol{\mathrm{x}}^i - \boldsymbol{\mathrm{F}}^{1J}|} - \frac{\Psi_k^J - \Psi_{k+3}^J}{|\boldsymbol{\mathrm{F}}^{1J} - \boldsymbol{\mathrm{F}}^{2J}|} ~~~~ &\text{for}~~ k = 1,2,3\\[6pt]
\frac{\Psi_k^J - x_{k-3}^i}{|\boldsymbol{\mathrm{x}}^i - \boldsymbol{\mathrm{F}}^{1J}|} + \frac{\Psi_{k-3}^J - \Psi_{k}^J}{|\boldsymbol{\mathrm{F}}^{1J} - \boldsymbol{\mathrm{F}}^{2J}|} &\text{for}~~ k = 4,5,6 \\[6pt]
-1 &  \text{for}~~ k = 7
\end{dcases}
\end{eqnarray}
where $x_k^i$ is the $k^{th}$ component of $\boldsymbol{\mathrm{x}}^i$. In case both focal points, or a focal point and an element centroid coincide, denominator in the above expressions tend to 0 causing numerical difficulties. Therefore, a small real number, $\varepsilon$ is added to all denominators outside of the modulus sign.

\subsection{Numerical Implementation}
\label{sec:numerical_imp}
3D topology optimization problems typically have large numbers of degrees of freedom requiring solution to large linear systems for each function evaluation. As direct methods for solving linear systems are computationally expensive, Pre-conditioned Conjugate Gradient (PCG) method is the preferred way to solve the FE equations. PCG is an iterative method for solving linear systems, $\boldsymbol{\mathrm{A}} \boldsymbol{\mathrm{x}} = \boldsymbol{\mathrm{b}}$, for symmetric positive definite $\boldsymbol{\mathrm{A}}$. The method does not require an explicit evaluation of matrix $\boldsymbol{\mathrm{A}}$ but only needs an operator which provides the evaluation $\boldsymbol{\mathrm{A}}\boldsymbol{\mathrm{y}}$ for given vector $\boldsymbol{\mathrm{y}}$. This makes the method memory efficient. For structural mechanics problems, evaluation of $\boldsymbol{\mathrm{A}} \boldsymbol{\mathrm{y}}$ corresponds to evaluating the internal force vector for a given displacement vector. Given the displacement vector $\boldsymbol{\mathrm{u}}$ for all degrees of freedom, we generate a matrix $\boldsymbol{\mathrm{ue}}$ such that the $e^{th}$ column, that is $\boldsymbol{\mathrm{ue}}(:,e)$, is the local displacement vector for element $e$. As all elements in the mesh are the same,
\begin{eqnarray}
\boldsymbol{\mathrm{Fe}} = \boldsymbol{\mathrm{K}}_0 \boldsymbol{\mathrm{ue}} \boldsymbol{\mathrm{Pe}}
\label{eqn:Fe}
\end{eqnarray}
where $\boldsymbol{\mathrm{Fe}}(:,e)$ is the internal force vector for element $e$, $\boldsymbol{\mathrm{K}}_0$ is the elemental stiffness of a solid element evaluated from Eqn. \ref{eqn:elem_stiff} and $\boldsymbol{\mathrm{Pe}}$ is a diagonal matrix such that $\boldsymbol{\mathrm{Pe}}(e,e) = \{\rho_e^\eta (1-\rho_{min}) + \rho_{min}\}$. An internal force assembly is conducted to obtain the internal force for the system.\\

The rate of convergence of PCG depends on the condition number of the matrix. Higher condition number implies worse rate. For a given density distribution, the condition number depends on the ratio of highest and lowest value of the density. In MMOS using negative masks, elements lying inside a mask have density close to $\rho_{min}$ while elements outside the masks approximately have density 1. Hence, the condition number depends on $\rho_{min}$. We implement the pre-conditioner in \cite{wang2007large}. It is a diagonal matrix constructed using elemental densities and diagonal of the elemental stiffness matrix. \cite{wang2007large} show that the pre-conditioner lowers the condition number to the same order as that of a completely solid domain, essentially making it independent of $\rho_{min}$. Effects of pre-conditioning are discussed in section \ref{sec:discussion}. 

\section{Numerical examples}
\label{sec:sample_prob}
Implementing the above mathematical construct, the topology optimization problem in Eqn. \ref{eqn:2} can be re-expressed as 
\begin{flalign}\label{eqn:topo_prob}
&~& min. &~~~~ \mathbb{I}(\rho_i) &\\ \nonumber
&\text{subject to} & \bm{\mathrm{K}}(\rho_i) \bm{\mathrm{u}} &= \bm{\mathrm{F}} &\\ \nonumber
&\text{such that}& \sum_{i=1}^{ne} \rho_i(\boldsymbol{\mathrm{x}}^i , \Psi_k^J) v_i &\leq V^* = v_f \sum_{i=1}^{ne} v_i & \\ \nonumber
& ~ & \Psi_k^{min} \leq \Psi_k^J & \leq \Psi_k^{max} &
\end{flalign}
where $\Psi_k^{min}$ and $\Psi_k^{max}$ are chosen such that they allow for masks to leave the design domain or collapse onto themselves and disappear. We allow focal points of masks to move 20 units beyond the domain limits in all directions. Also, $\Psi_7^J$ is bound such that, $-3 \leq \Psi_7^J \leq 20$.\\

The most often discussed formulation in topology optimization is the compliance minimization problem where the objective function, $\mathbb{I}(\rho_i)$ in Eqn. \ref{eqn:2}, is the strain energy stored in the system, that is,
\begin{eqnarray}
\mathbb{I}(\rho_i) = \frac{1}{2}\boldsymbol{\mathrm{u}}^T\boldsymbol{\mathrm{K}}(\rho_i)\boldsymbol{\mathrm{u}}
\end{eqnarray}
Gradient of the objective with respect to element densities is given as,
\begin{eqnarray}
\frac{\partial \mathbb{I}(\rho_i)}{\partial \rho_i} = -\frac{1}{2} \boldsymbol{\mathrm{u}}^T\frac{\partial \boldsymbol{\mathrm{K}}(\rho_i)}{\partial \rho_i}\boldsymbol{\mathrm{u}}
\end{eqnarray}
where $\frac{\partial \boldsymbol{\mathrm{K}}(\rho_i)}{\partial \rho_i}$ for the SIMP material model can be evaluated using Eqn. \ref{eqn:SIMP}.\\

We present solutions to three compliance minimization problems, namely, (a) cantilever beam, (b) torsion beam and (c) bridge design problem. Schematics describing these problems are given in Figs. \ref{fig:canti_prob_descrition}, \ref{fig:Torsion_prob_descrition} and \ref{fig:UDL_prob_descrition} respectively. The optimization formulation described is non-convex and hence the solution depends on the parameter values and initial guess. For all examples, SIMP penalty parameter, $\eta = 3$ (Eqn. \ref{eqn:SIMP}), $\alpha = 3$ (Eqn. \ref{eqn:fj}) and minimum density, $\rho_{min} = 10^{-4}$ (Eqn. \ref{eqn:SIMP}) are implemented. All problems are solved using the inbuilt MATLAB function \textit{fmincon} \citep{MATLAB:2010}.

\subsection{Cantilever beam problem}
\begin{figure}[h]
	\centering
	\includegraphics[width=0.45\textwidth]{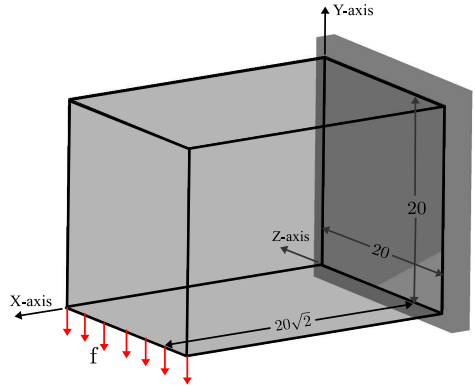}
	\caption{Problem description for cantilever beam.}
	\label{fig:canti_prob_descrition}
\end{figure}
\begin{figure}[h]
	\centering
	\subcaptionbox{$\mathrm{X}$-$\mathrm{Y}$ view }{\includegraphics[width=0.4\textwidth]{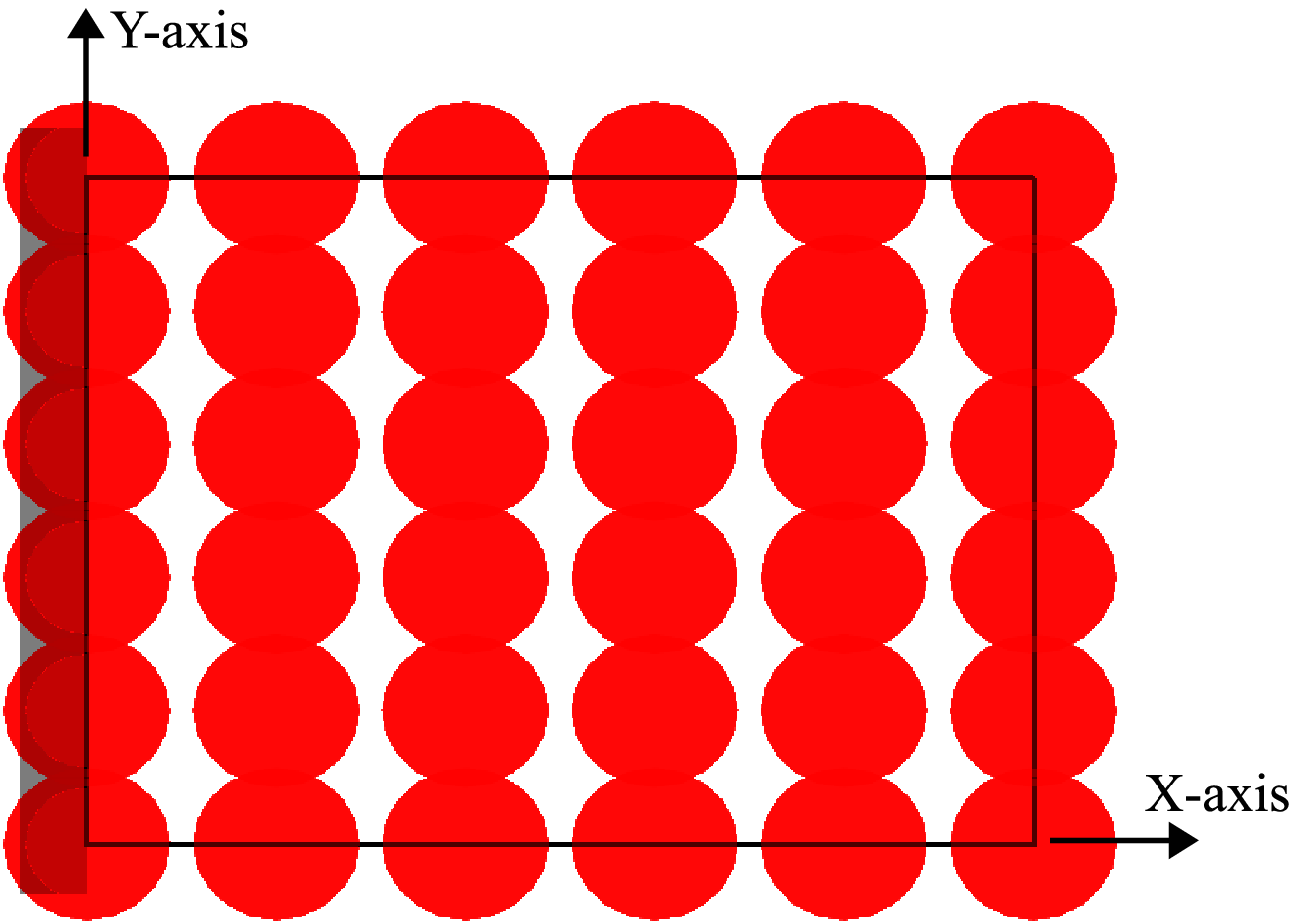}}
	\label{fig:Canti_Initial_guess_xy}%
	\hspace{0.75cm}
	\subcaptionbox{$\mathrm{X}$-$\mathrm{Z}$ view }{\includegraphics[width=0.4\textwidth]{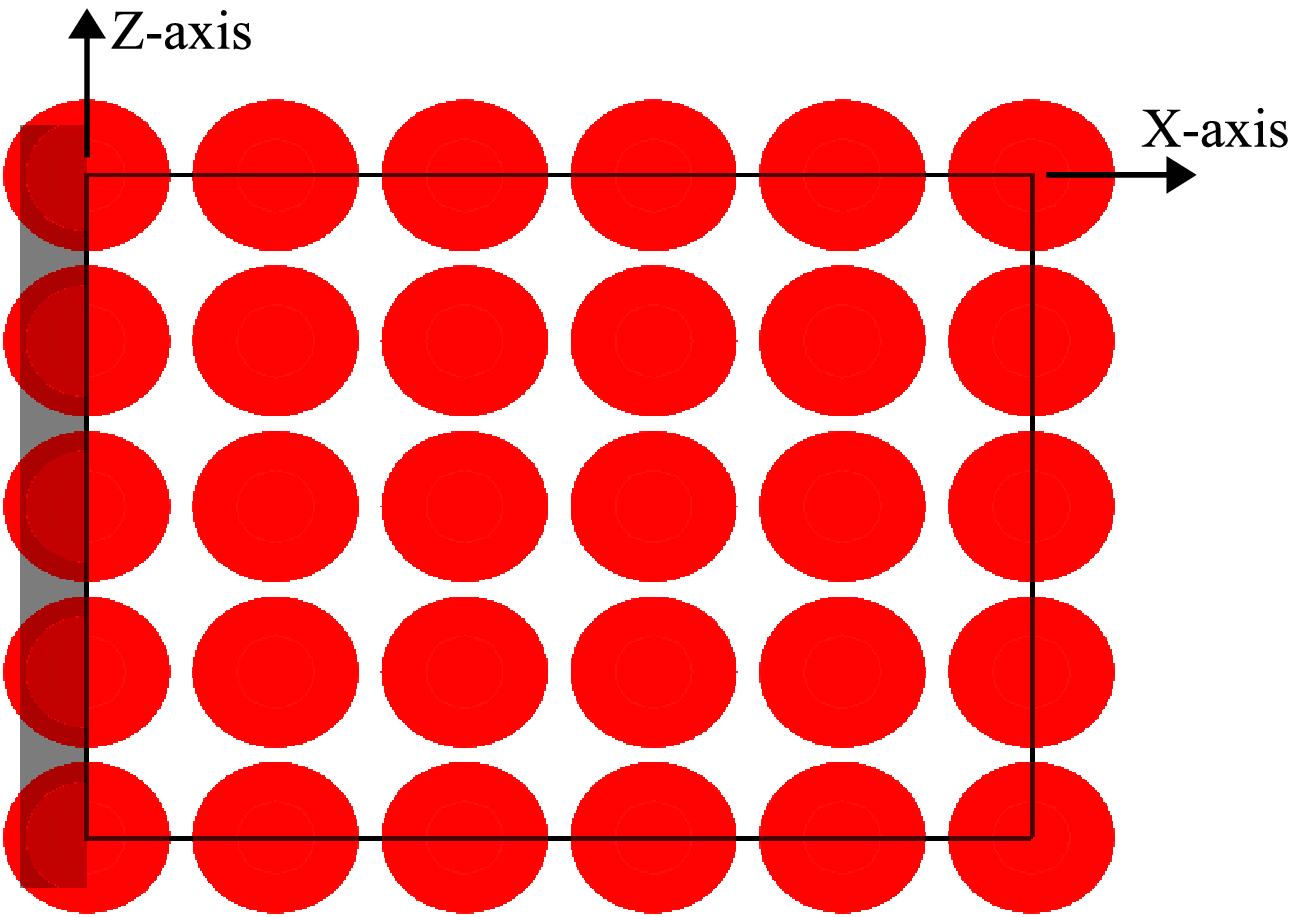}}
	\label{fig:Canti_Initial_guess_xz} 
	\caption{Initial mask arrangement for cantilever beam.}
	\label{fig:Canti_Initial_guess}
\end{figure}

The domain in Fig. \ref{fig:canti_prob_descrition} is discretized using a $41 \times 41 \times 41$ truncated octahedron mesh with an edge length of $a = 0.25$ corresponding to $68,081$ elements and $436,984$ nodes ($1.31$ million degrees of freedom). Elements have the same orientation as presented in Fig. \ref{fig:reoriented_elements}b. Nodes with non-positive $\mathrm{X}$ coordinate are fixed. A vertical downward force of $|\boldsymbol{\mathrm{f}}| = 0.125$ is applied at the local nodes $22$ and $23$ of elements along the loaded edge presented in the Fig. \ref{fig:canti_prob_descrition}. For the initial guess masks are placed such that their centers form a $ 6 \times 6 \times 5$ uniform grid over the domain as shown in Fig. \ref{fig:Canti_Initial_guess}. Initially the foci are placed at a distance of 1 unit from the center in opposite directions along the $\mathrm{X}$ axis and $d^J = 3$. Thus we implement 180 masks corresponding to 1260 design variables. The problem is solved for a volume fraction ($vf$) of 0.15.\\

\begin{figure}[h]
	\centering
	\subcaptionbox{Iteration 20 }{\includegraphics[width=0.4\textwidth]{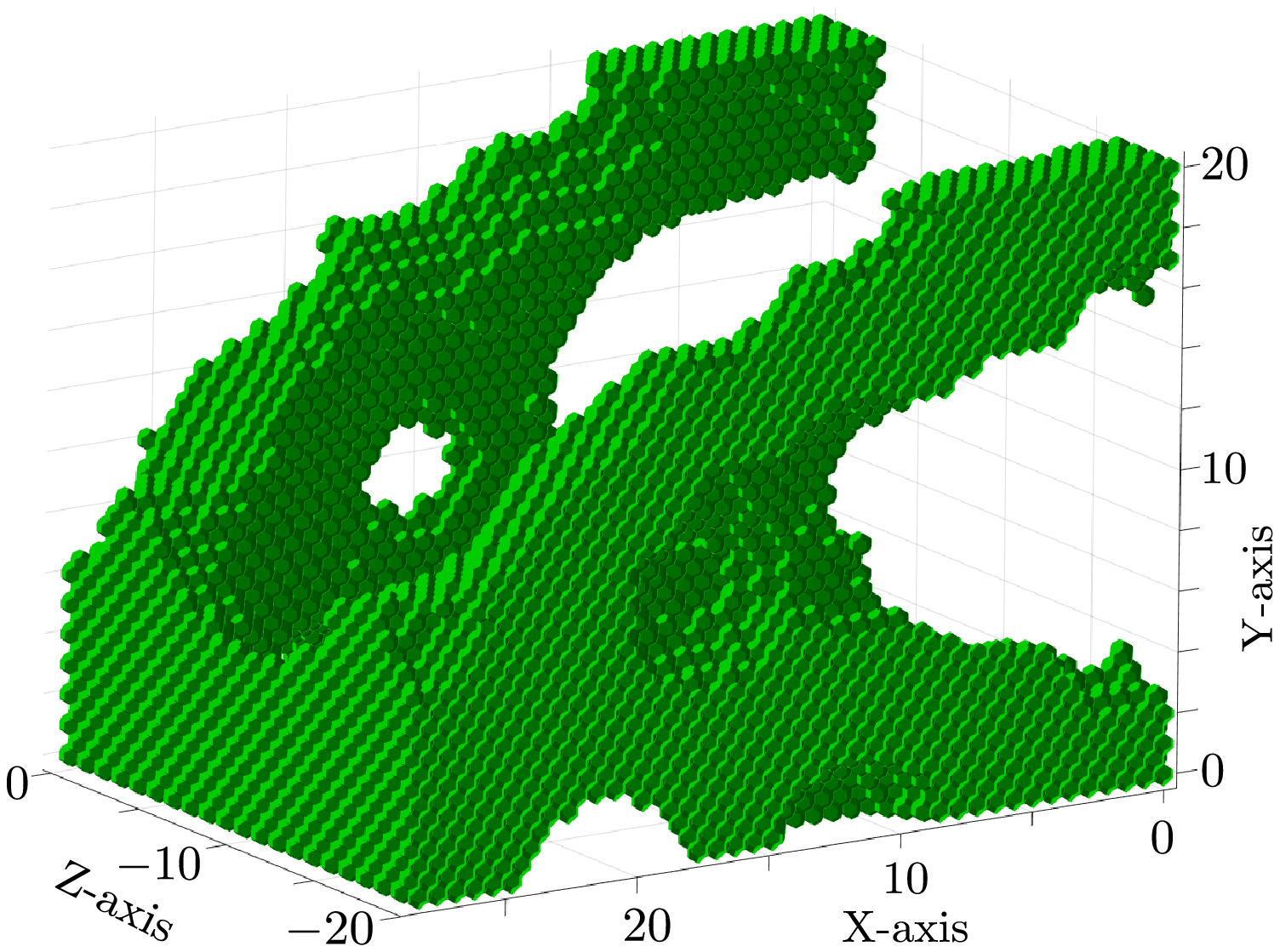}}
	\label{fig:canti_inter_sol_20}%
	\hspace{0.75cm}
	\subcaptionbox{Iteration 50 }{\includegraphics[width=0.4\textwidth]{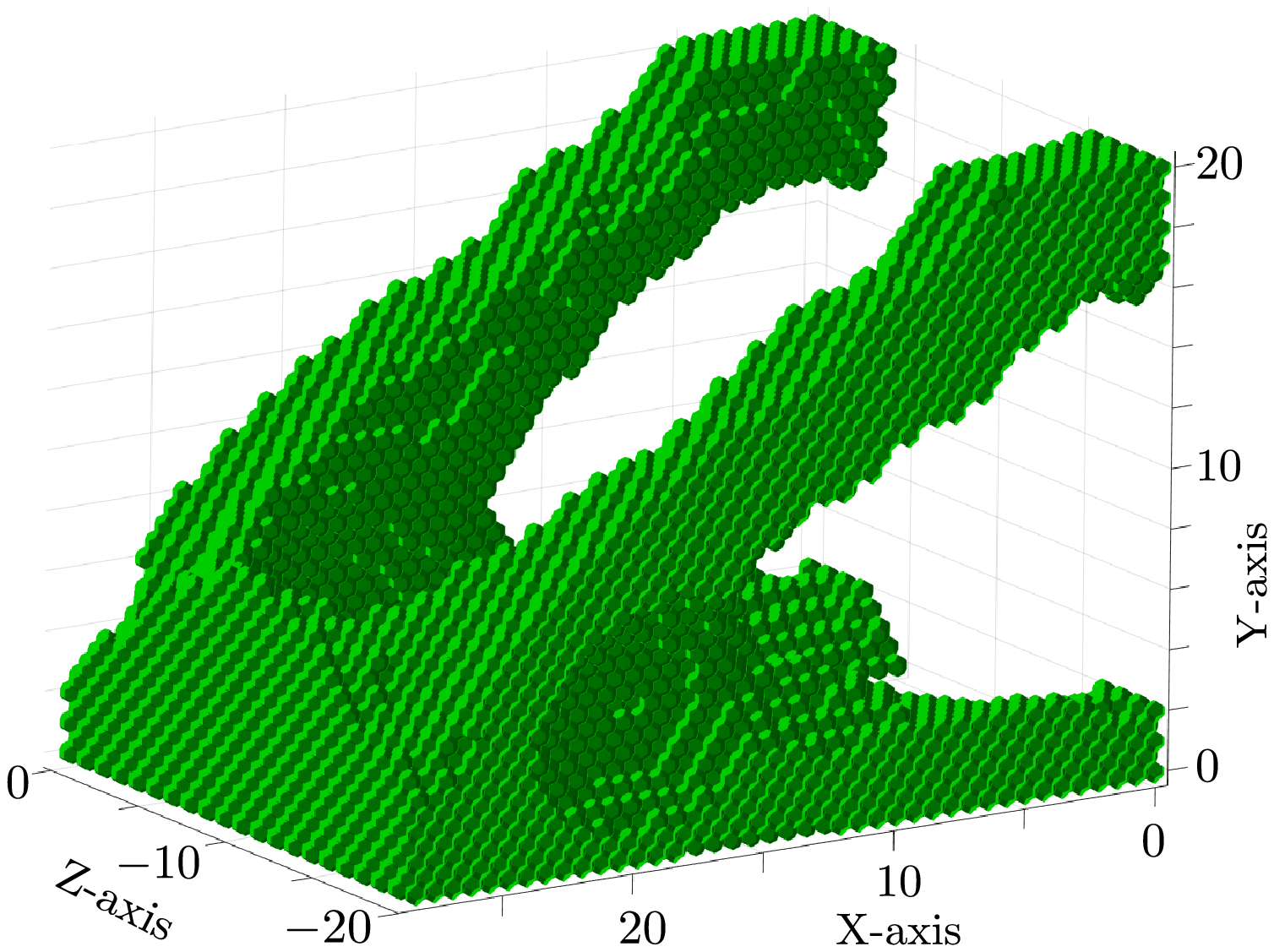}}
	\label{fig:canti_inter_sol_50} 
	\\ \vspace{0.5cm}
	\subcaptionbox{Iteration 100 }{\includegraphics[width=0.4\textwidth]{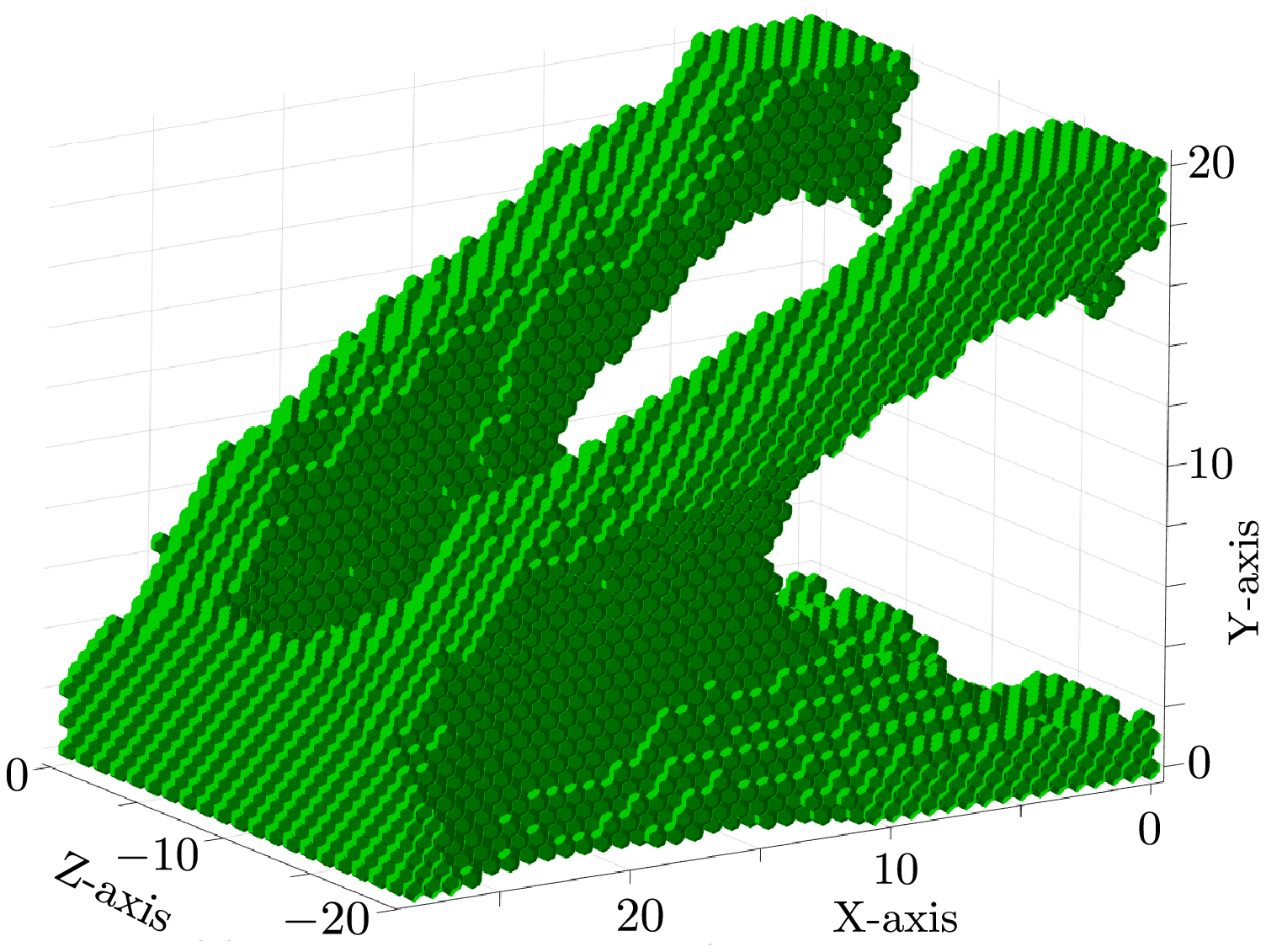}}
	\label{fig:canti_inter_sol_100}
	\hspace{0.75cm}
	\subcaptionbox{Iteration 200 }{\includegraphics[width=0.4\textwidth]{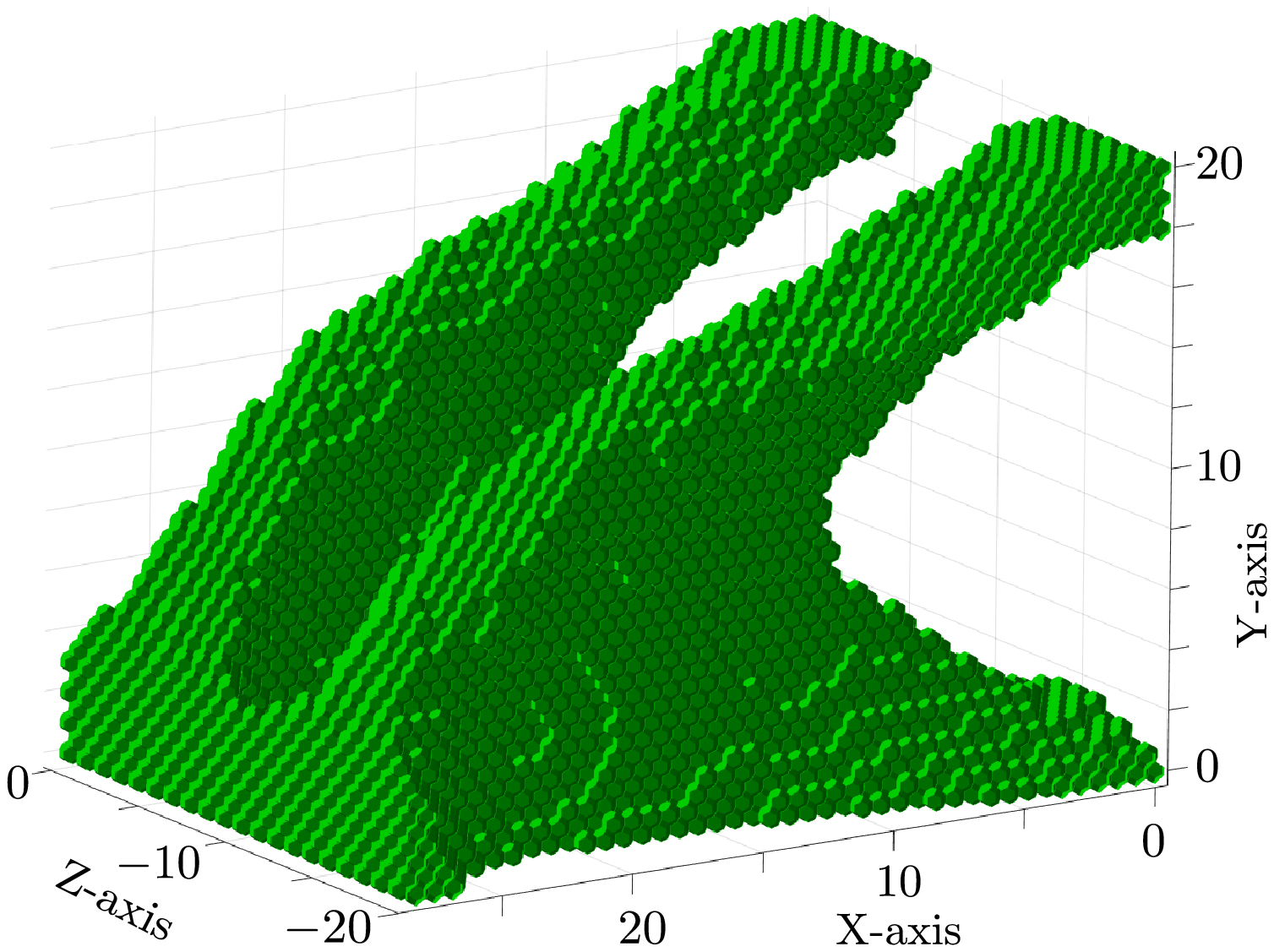}}
	\label{fig:canti_inter_sol_200}
	\caption{Intermediate solutions of cantilever beam at different iteration steps.}
	\label{fig:canti_inter_sol}
\end{figure}

\begin{figure}[h]
	\centering
	\subcaptionbox{Converged solution }{\includegraphics[width=0.4\textwidth]{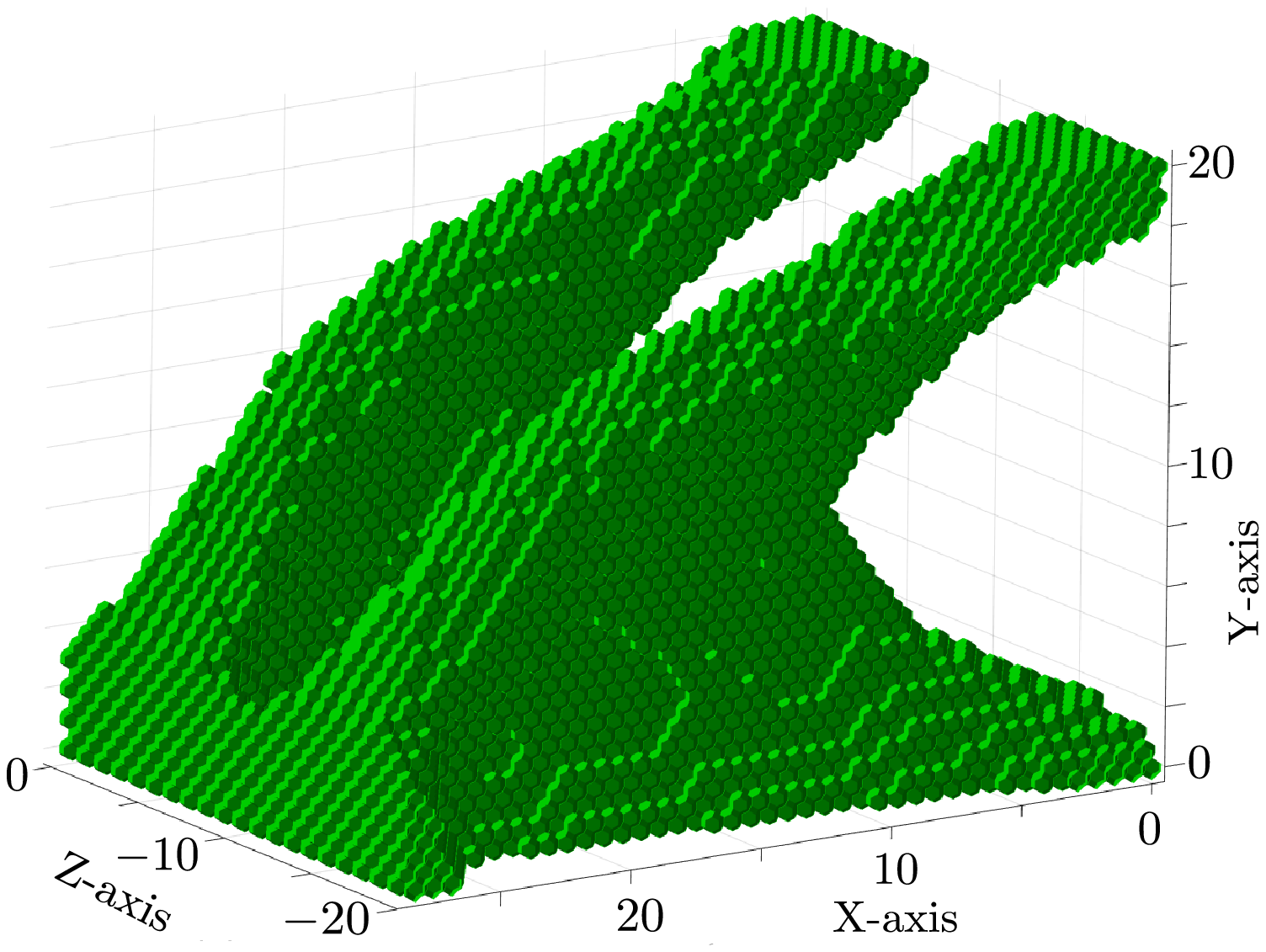}}
	\label{fig:canti_final_sol}%
	\hspace{0.75cm}
	\subcaptionbox{Convergence history }{\includegraphics[width=0.4\textwidth]{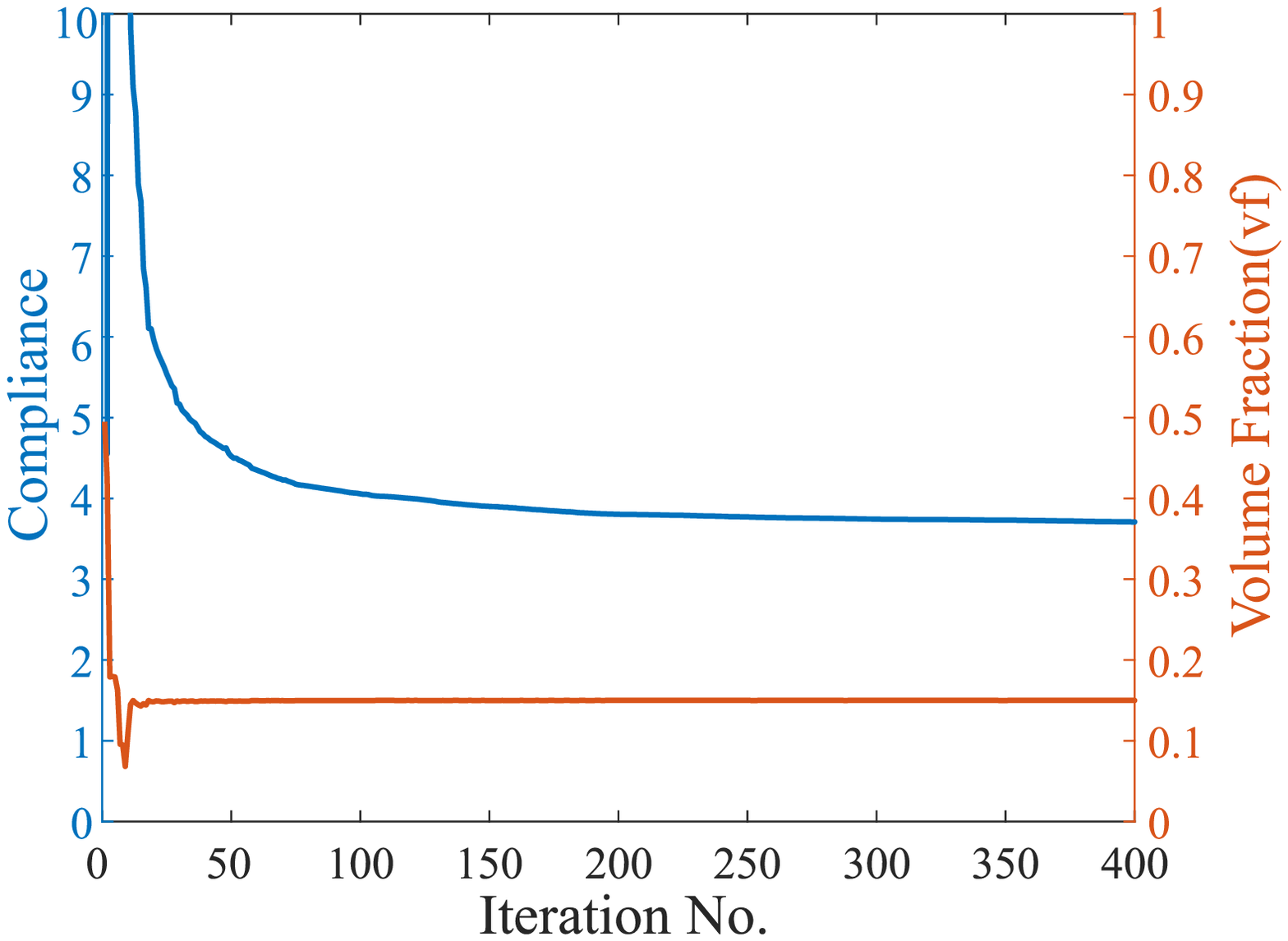}}
	\label{fig:Canti_convergence} 
	\caption{Final solution and convergence history for cantilever beam.}
	\label{fig:canti_final_convergence}
\end{figure}

For solution representation we only plot elements with density greater than 0.2. Fig. \ref{fig:canti_inter_sol} presents intermediate solutions at different optimization iterations. Fig. \ref{fig:canti_final_convergence}a and \ref{fig:canti_final_convergence}b present the final solution obtained after 400 optimization iterations and the convergence history for the objective and constraint respectively. Note that the problem is symmetric about the mid vertical plane $Z = -10$. The solution procedure does not impose this symmetry but the final structure presents the expected symmetry. Convergence history of the objective, and a visual comparison of the solution in iteration 200 and the final solution suggest that convergence was achieved much earlier.

\subsection{Torsional beam problem}
\begin{figure}[h]
	\centering
	\includegraphics[width=0.5\textwidth]{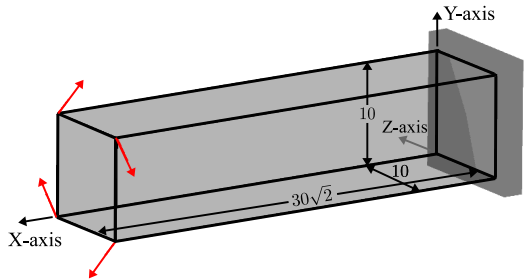}
	\caption{Problem description for torsion beam.}
	\label{fig:Torsion_prob_descrition}
\end{figure}
\begin{figure}[h]
	\centering
	\subcaptionbox{$\mathrm{X}$-$\mathrm{Y}$ view }{\includegraphics[width=0.45\textwidth]{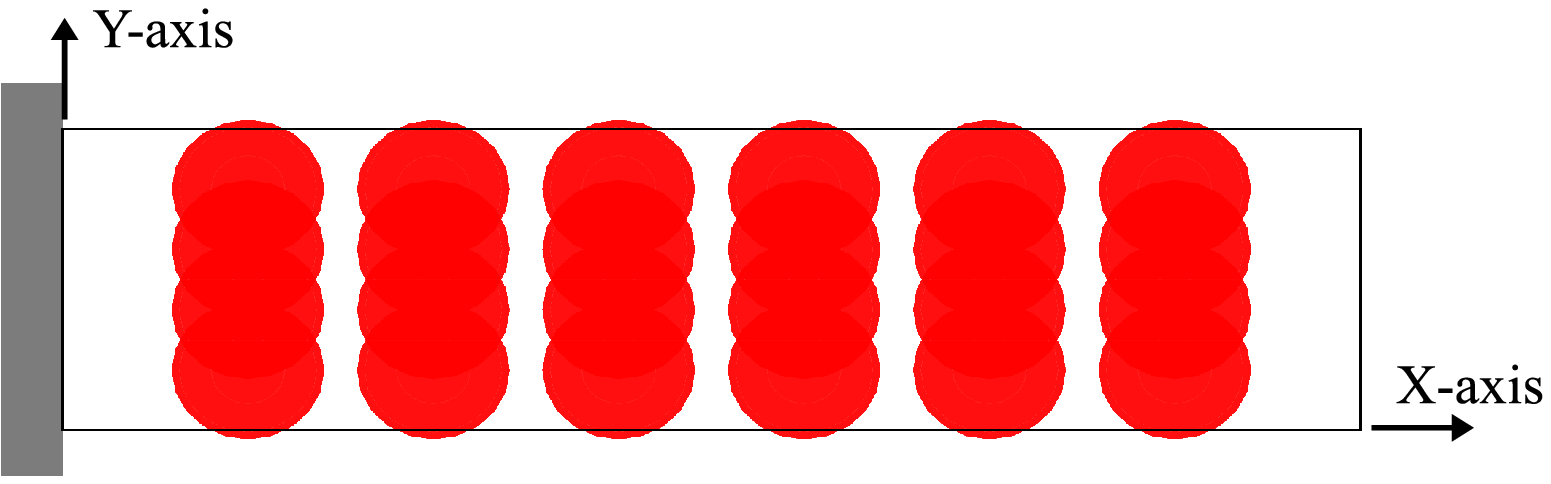}}
	\label{fig:Torsion_Initial_guess_xy}%
	\hspace{0.5cm}
	\subcaptionbox{$\mathrm{X}$-$\mathrm{Z}$ view }{\includegraphics[width=0.45\textwidth]{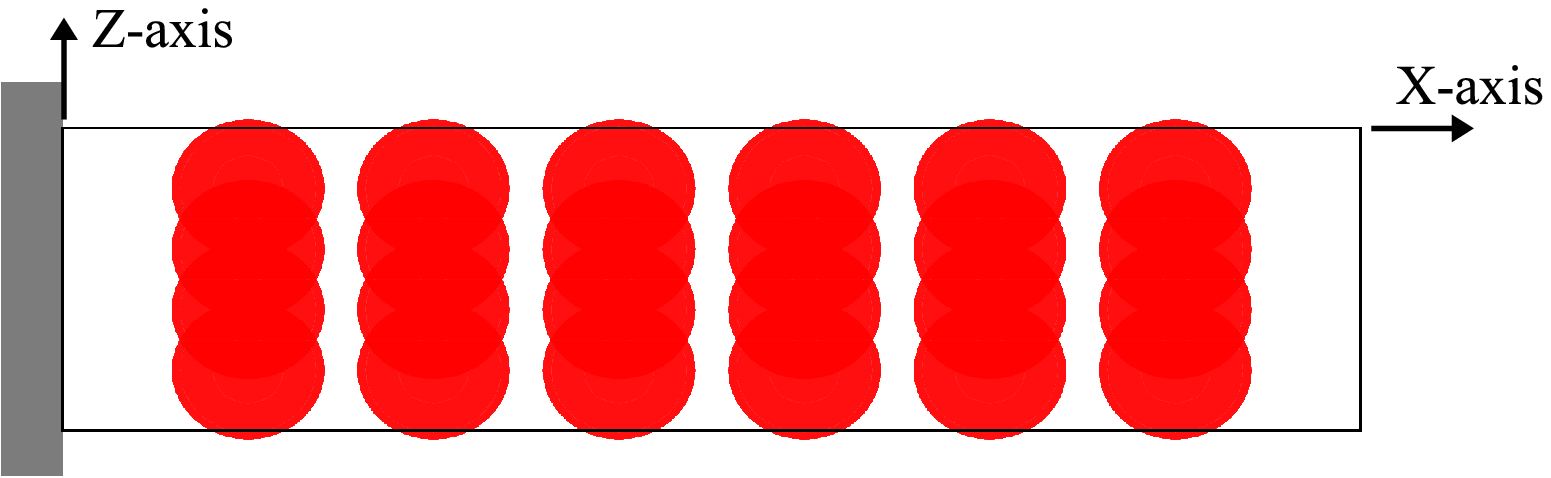}}
	\label{fig:Torsion_Initial_guess_xz} 
	\caption{Initial mask arrangement for torsion beam.}
	\label{fig:Torsion_Initial_guess}
\end{figure}

For torsional beam problem (Fig. \ref{fig:Torsion_prob_descrition}), the domain is discretized using a $61 \times 21 \times 21$ truncated octahedron mesh with an edge length of $0.25$ corresponding to $26,681$ elements and $177,704$ nodes ($533,112$ degrees of freedom). Nodes with non-positive $\mathrm{X}$ coordinate, that is, $\mathrm{X} \leq 0$ are fixed. Loads of magnitude $0.125\sqrt{2}$ each are applied to the local nodes $21,22,23$ and $24$ of the corner elements on the front face, $\mathrm{X} = 30\sqrt{2}$, such that they create a negative moment about the $\mathrm{X}$ axis, as shown in Fig. \ref{fig:Torsion_prob_descrition}. For the initial guess masks of the same size and orientation as in the cantilever beam problem are placed such that their centers form a $ 6 \times 4 \times 4$ uniform grid inside the domain, as shown in Fig. \ref{fig:Torsion_Initial_guess}, thus implementing 96 masks corresponding to 672 design variables. The problem is solved for a volume fraction ($vf$) of 0.15.\\

\begin{figure}[h]
	\centering
	\subcaptionbox{Iteration 20 }{\includegraphics[width=0.4\textwidth]{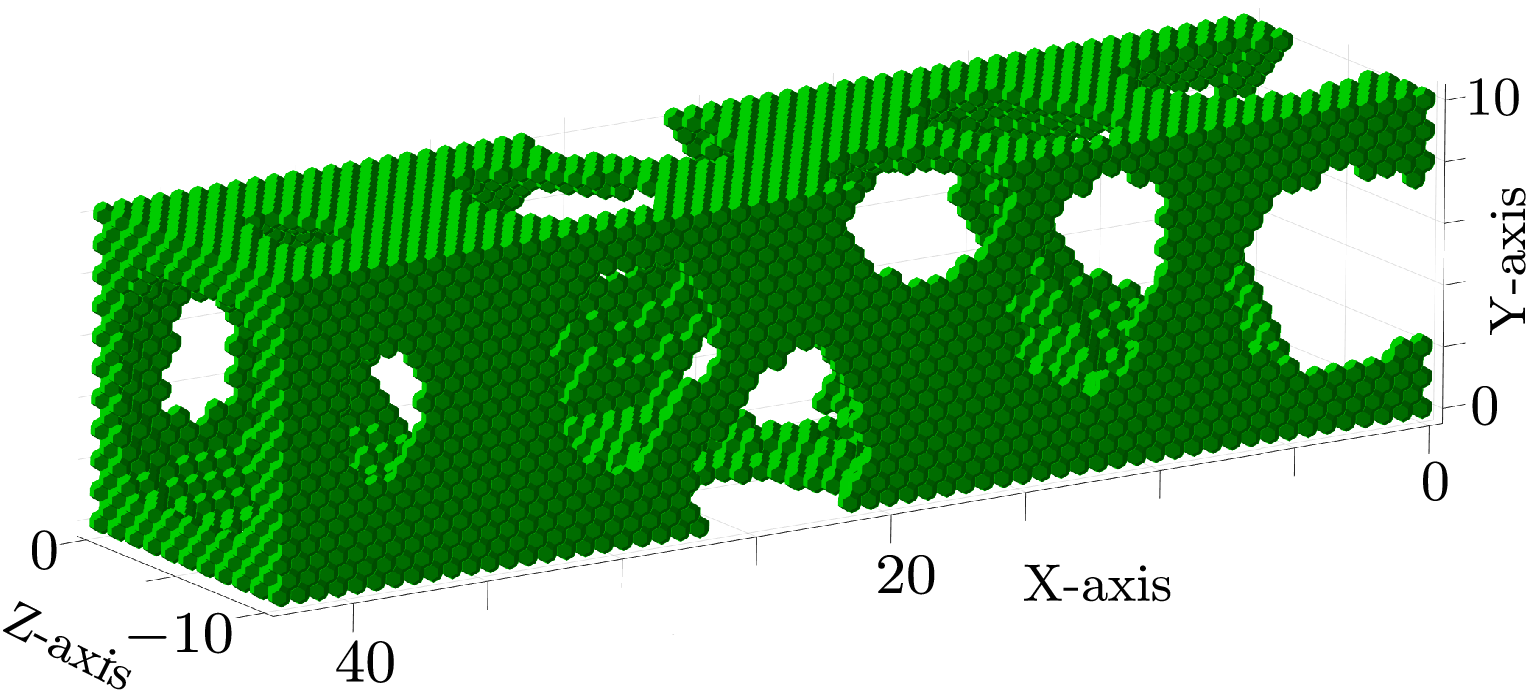}}
	\label{fig:tosrion_inter_sol_20}%
	\hspace{0.75cm}
	\subcaptionbox{Iteration 50 }{\includegraphics[width=0.4\textwidth]{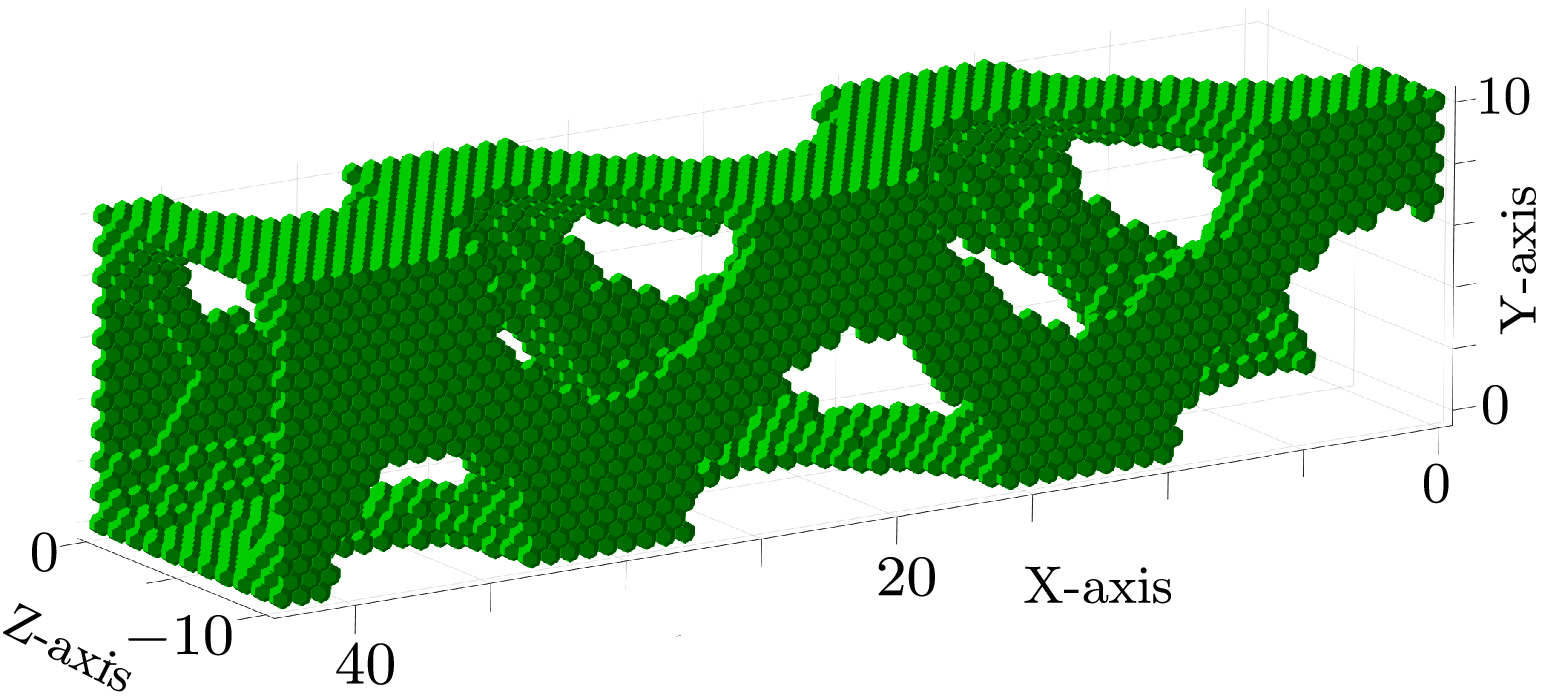}}
	\label{fig:tosrion_inter_sol_50} 
	\\ \vspace{0.5cm}
	\subcaptionbox{Iteration 100 }{\includegraphics[width=0.4\textwidth]{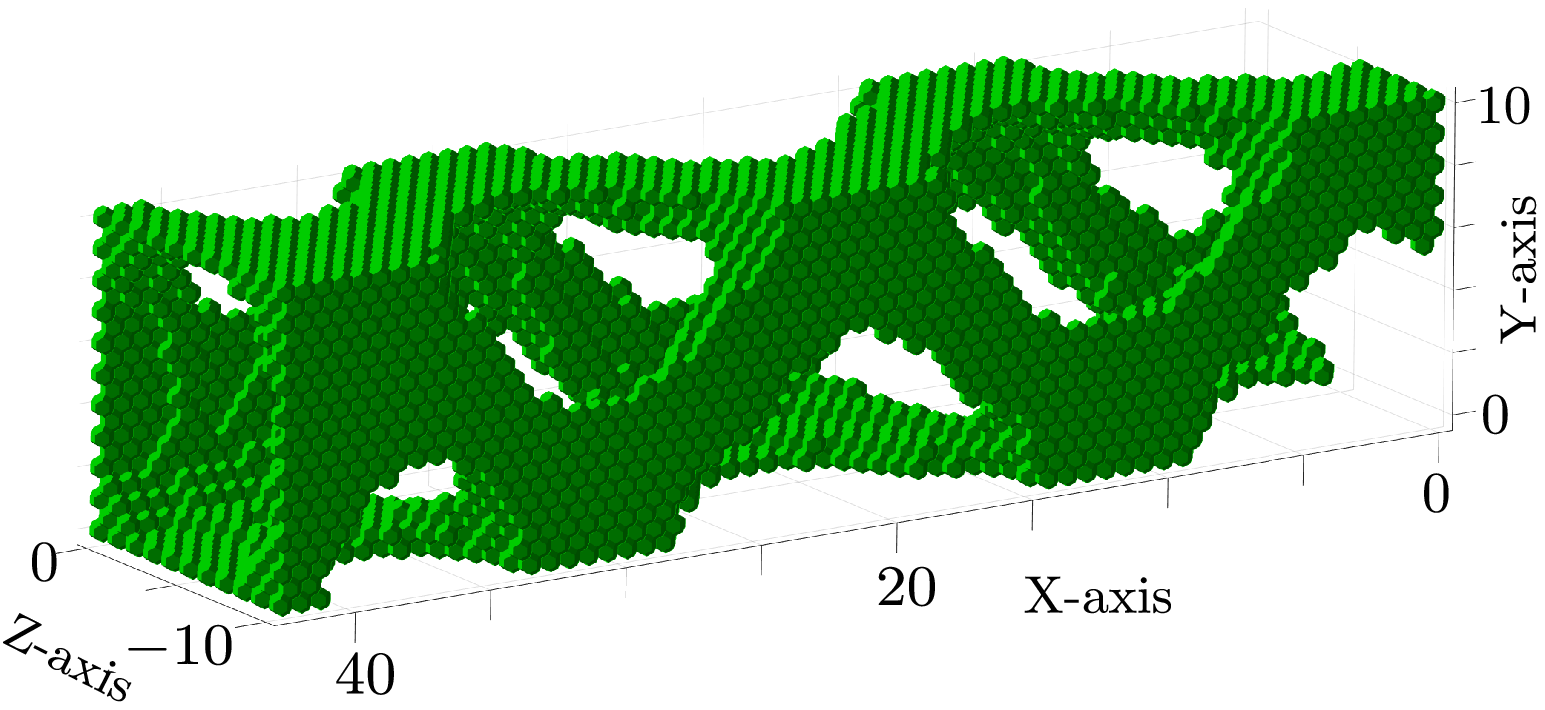}}
	\label{fig:tosrion_inter_sol_100}
	\hspace{0.75cm}
	\subcaptionbox{Iteration 200 }{\includegraphics[width=0.4\textwidth]{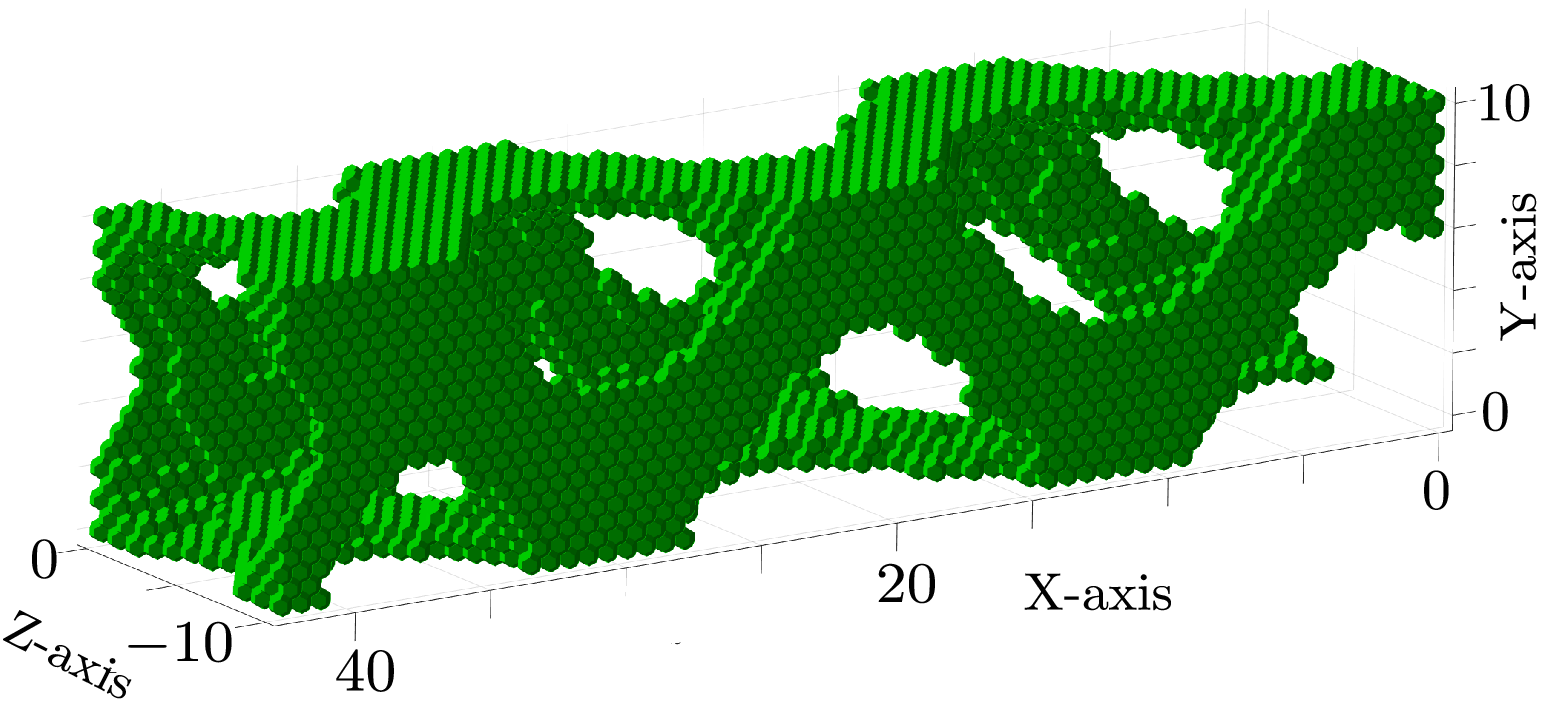}}
	\label{fig:tosrion_inter_sol_200}
	\caption{Intermediate solutions of torsion beam at different iteration steps.}
	\label{fig:tosrion_inter_sol}
\end{figure}

\begin{figure}[h]
	\centering
	\subcaptionbox{Converged solution }{\includegraphics[width=0.45\textwidth]{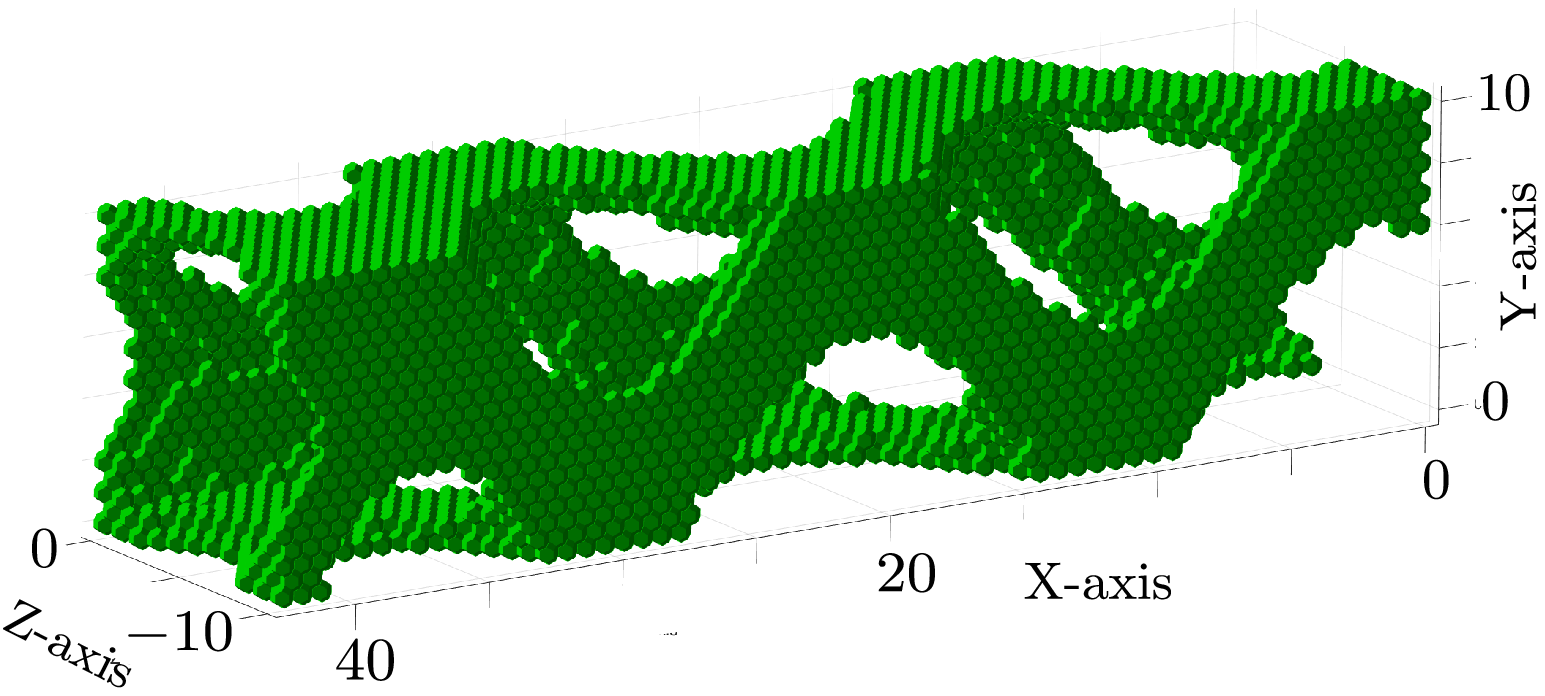}}
	\label{fig:Torsion_Final_sol}%
	\hspace{0.75cm}
	\subcaptionbox{Convergence history }{\includegraphics[width=0.4\textwidth]{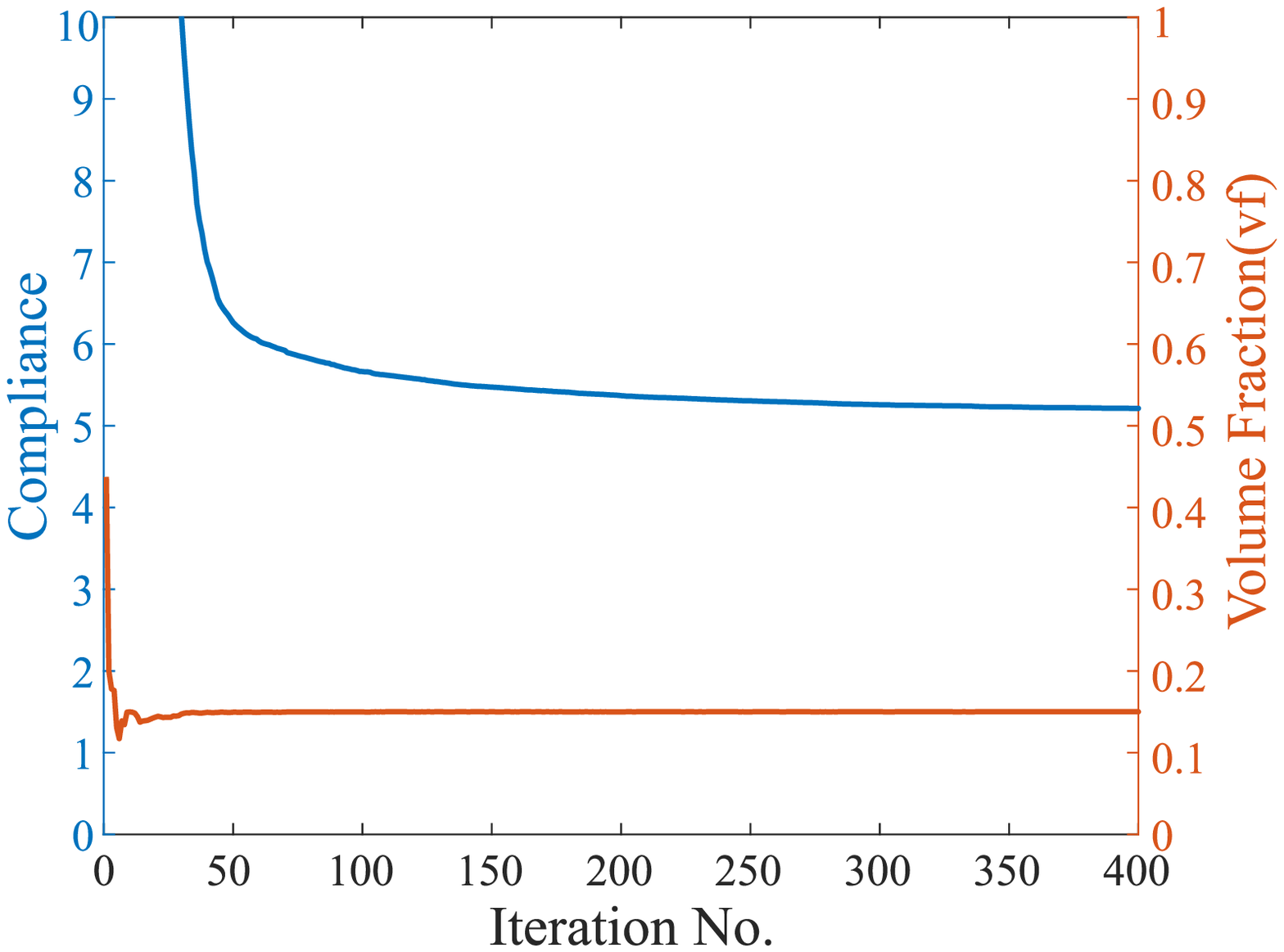}}
	\label{fig:Torsion_Convergence} 
	\caption{Final solution and convergence history for torsion beam.}
	\label{fig:Torsion_Final_Convergence}
\end{figure}

Fig. \ref{fig:tosrion_inter_sol} presents intermediate solutions at different optimization iterations. Fig. \ref{fig:Torsion_Final_Convergence}a and \ref{fig:Torsion_Final_Convergence}b present the final solution obtained after 400 iterations and the convergence history for the objective and constraint respectively. Convergence history suggests that there is not much change in the objective between iteration 100 and 200 but considerable changes in the geometry can be seen among the two solutions while the solutions at iterations 200 and 400 are numerically close and resemble visually suggesting that the solution has converged.

\subsection{Bridge design}
\begin{figure}[h]
	\centering
	\includegraphics[width=0.5\textwidth]{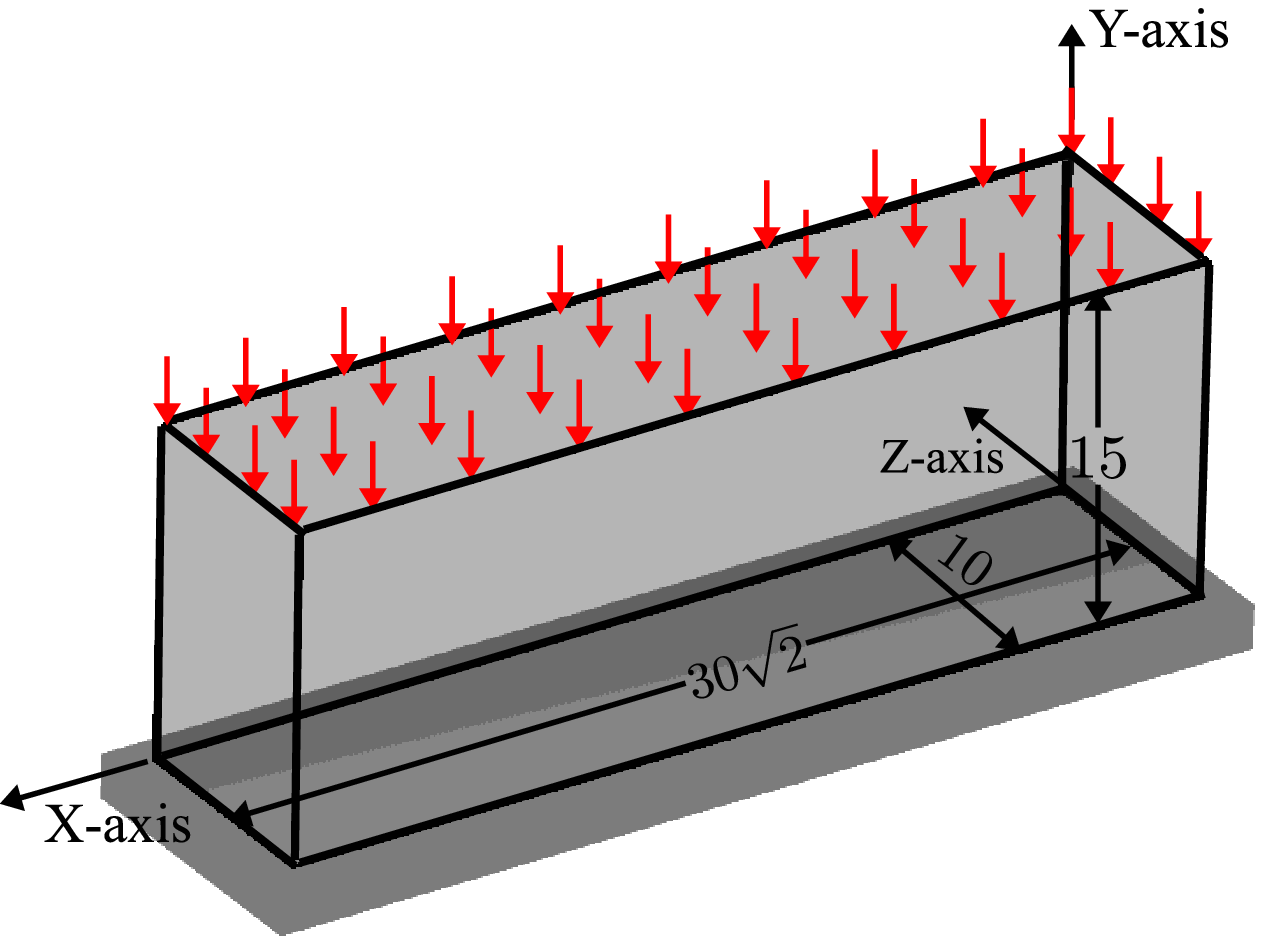}
	\caption{Problem description for bridge design.}
	\label{fig:UDL_prob_descrition}
\end{figure}
\begin{figure}[h]
	\centering
	\subcaptionbox{$\mathrm{X}$-$\mathrm{Y}$ view }{\includegraphics[width=0.45\textwidth]{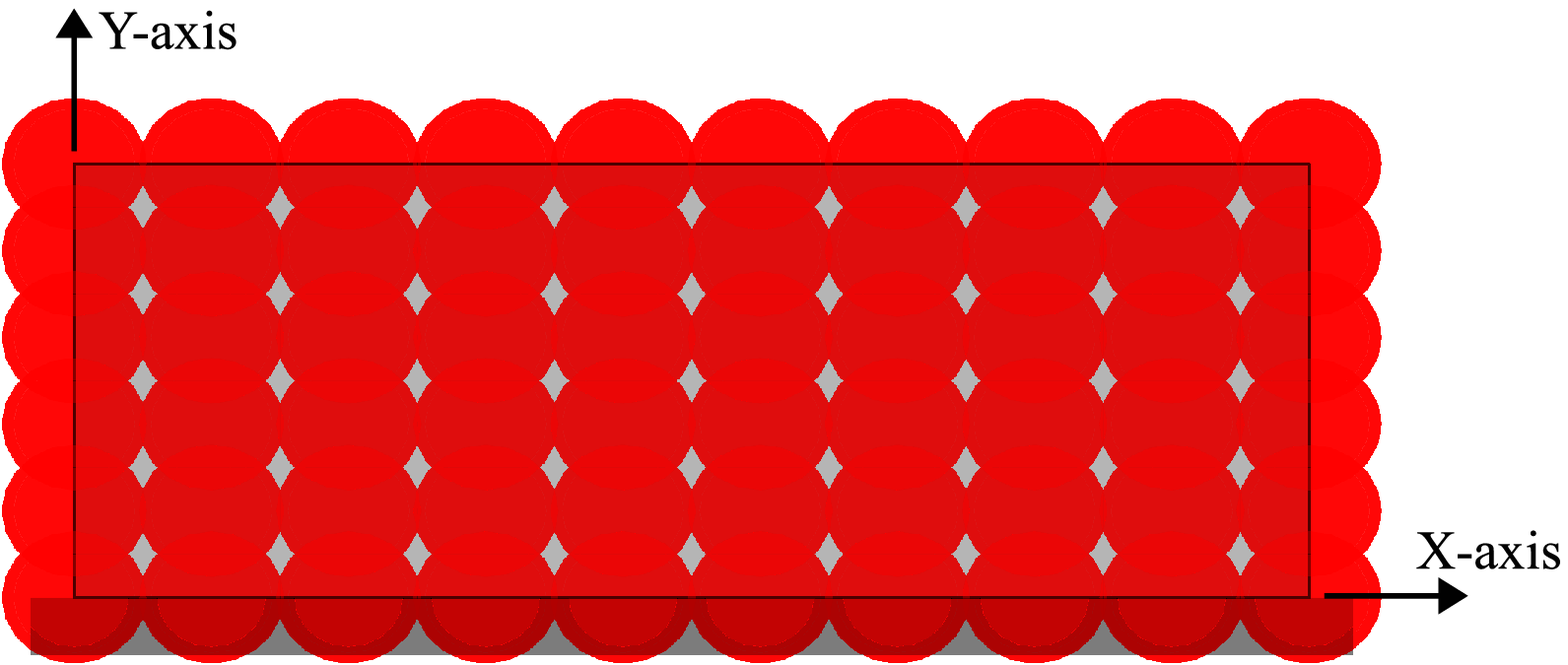}}
	\label{fig:UDL_Initial_guess_xy}%
	\hspace{0.5cm}
	\subcaptionbox{$\mathrm{X}$-$\mathrm{Z}$ view }{\includegraphics[width=0.45\textwidth]{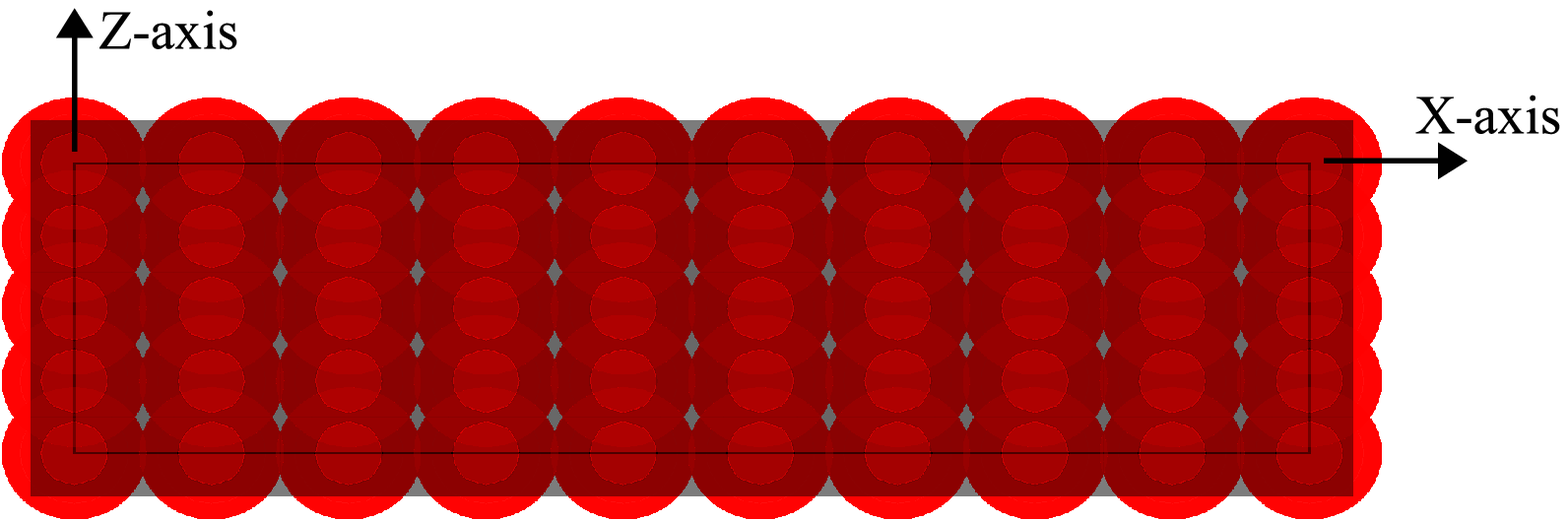}}
	\label{fig:UDL_Initial_guess_xz} 
	\caption{Initial mask arrangement for bridge design problem.}
	\label{fig:UDL_Initial_guess}
\end{figure}

For bridge design (Fig. \ref{fig:Torsion_prob_descrition}), the domain is discretized using a $61 \times 31 \times 21$ truncated octahedron mesh with an edge length of $0.25$ corresponding to $39,386$ elements and $258,624$ nodes ($775,872$ degrees of freedom). Nodes with non-positive $\mathrm{Z}$ coordinate, that is, $\mathrm{Z} \leq 0$ are fixed. Vertical downward loads of magnitude $0.1$, that is, $|\boldsymbol{\mathrm{f}}| = 0.1$ are applied to the local nodes $9$ and $16$ of all elements discretizing the top surface, $\mathrm{Y} = 15$, as shown in Fig. \ref{fig:UDL_prob_descrition}. For the initial guess masks of the same size and orientation as in the cantilever beam problem are placed such that their centers form a $ 10 \times 6 \times 5$ uniform grid over the domain, as shown in Fig. \ref{fig:UDL_Initial_guess}. Thus implementing 300 masks corresponding to 2100 design variables. The problem is solved for a volume fraction ($vf$) of 0.15.\\

\begin{figure}[h]
	\centering
	\subcaptionbox{Iteration 20 }{\includegraphics[width=0.4\textwidth]{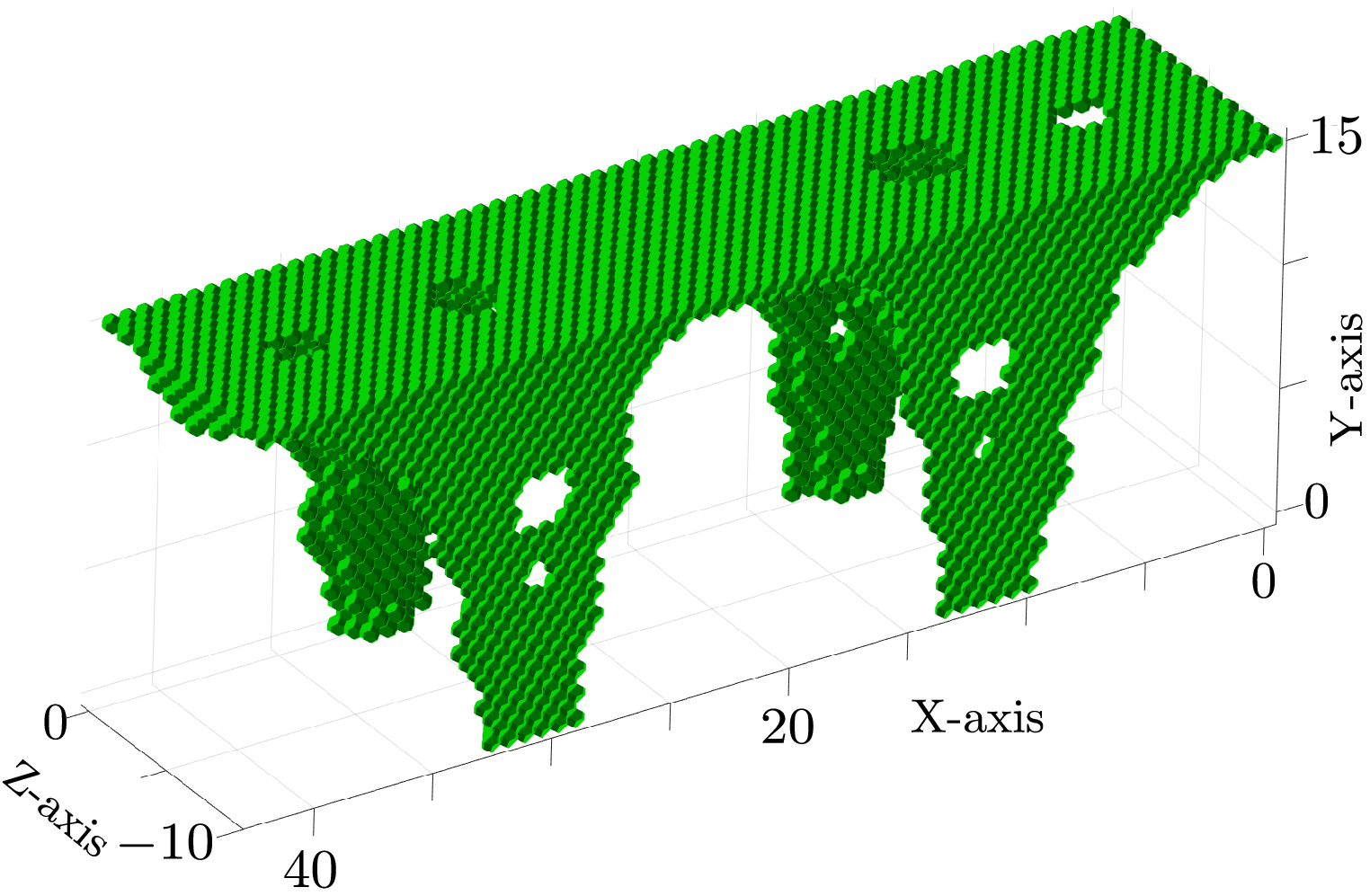}}
	\label{fig:UDL_inter_sol_20}%
	\hspace{0.75cm}
	\subcaptionbox{Iteration 50 }{\includegraphics[width=0.4\textwidth]{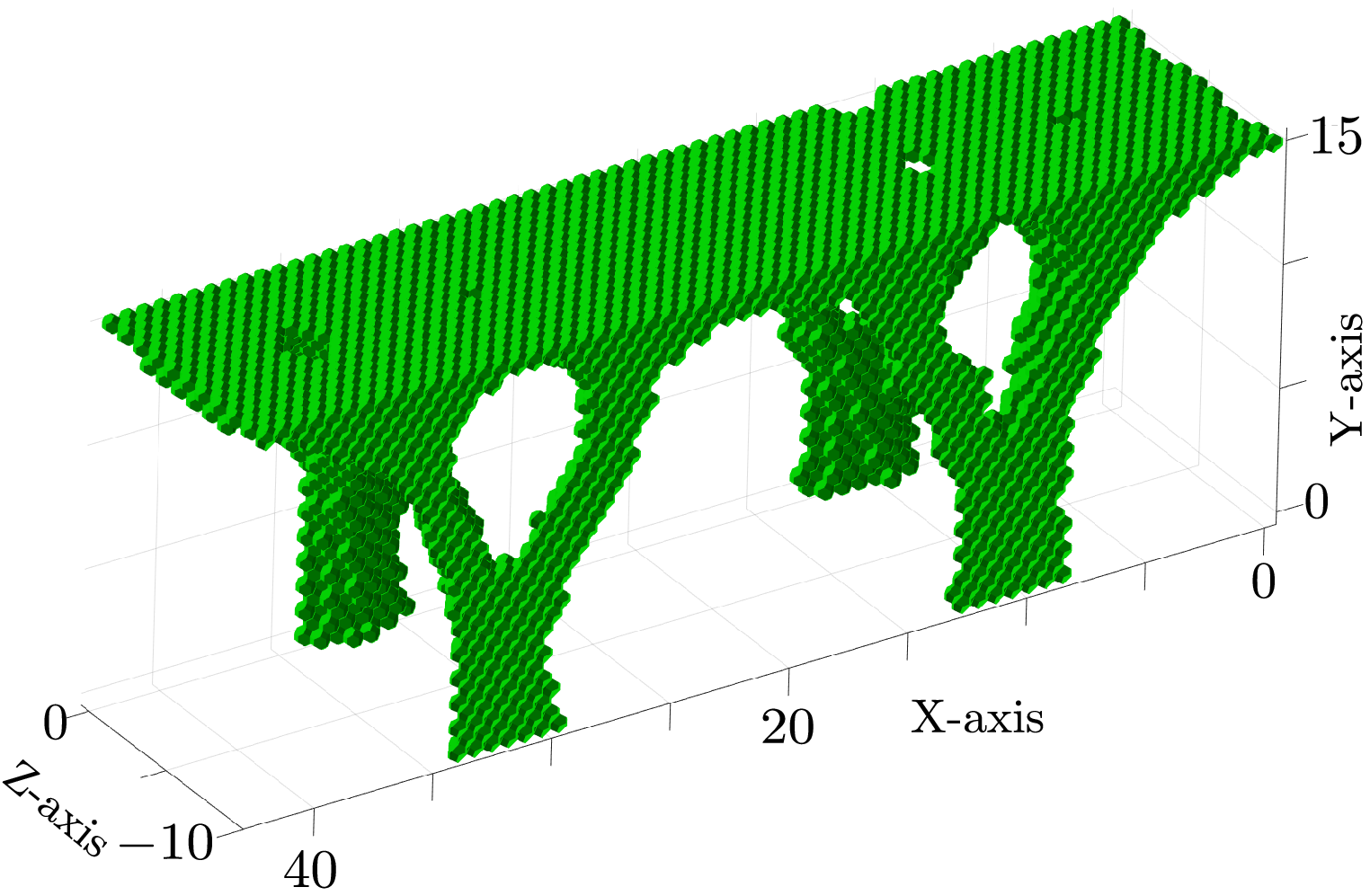}}
	\label{fig:UDL_inter_sol_50} 
	\\ \vspace{0.5cm}
	\subcaptionbox{Iteration 100 }{\includegraphics[width=0.4\textwidth]{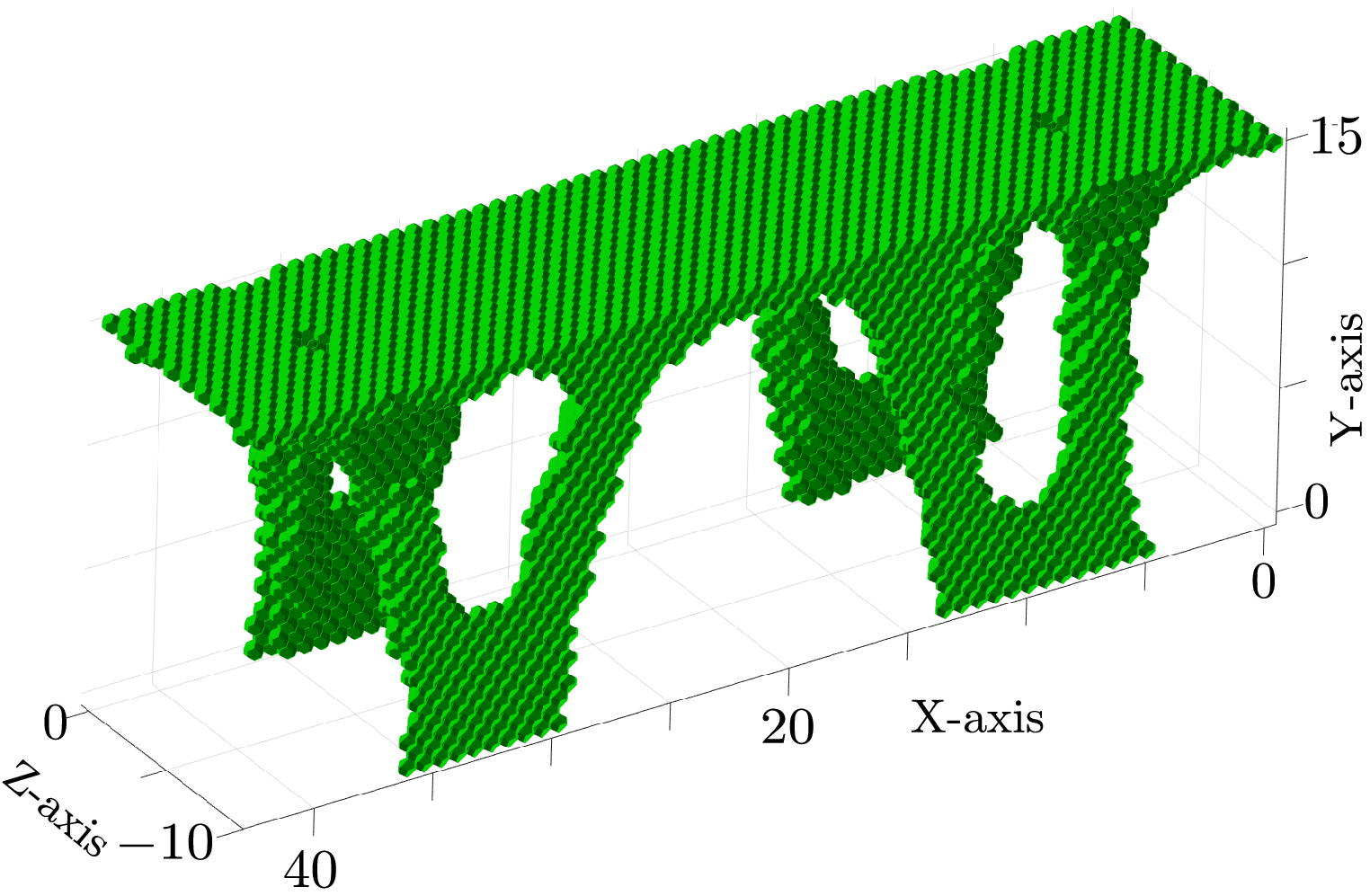}}
	\label{fig:UDL_inter_sol_100}
	\hspace{0.75cm}
	\subcaptionbox{Iteration 200 }{\includegraphics[width=0.4\textwidth]{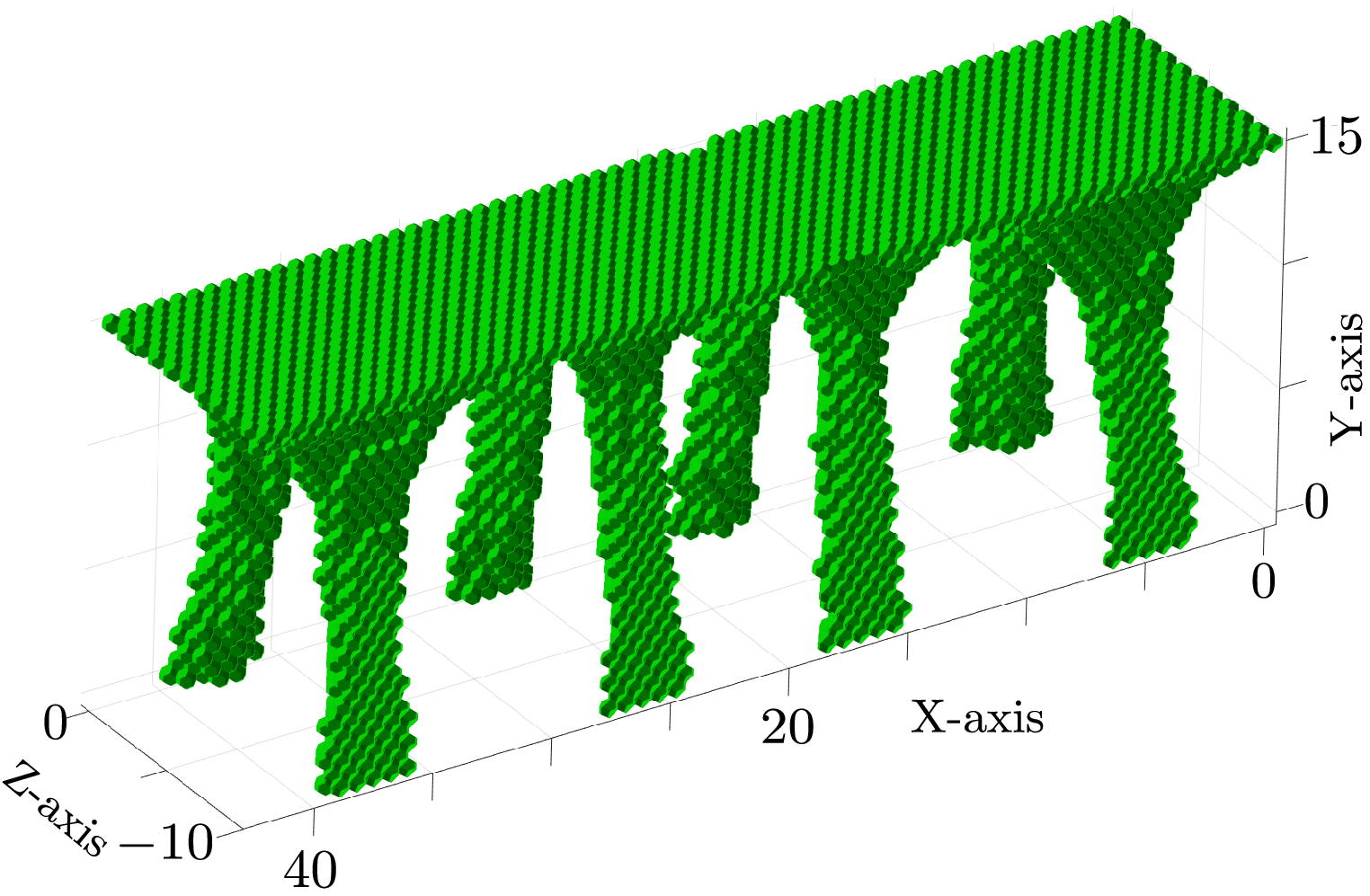}}
	\label{fig:UDL_inter_sol_200}
	\caption{Intermediate solutions of bridge design at different iteration steps.}
	\label{fig:UDL_inter_sol}
\end{figure}

\begin{figure}[h]
	\centering
	\subcaptionbox{Converged solution }{\includegraphics[width=0.45\textwidth]{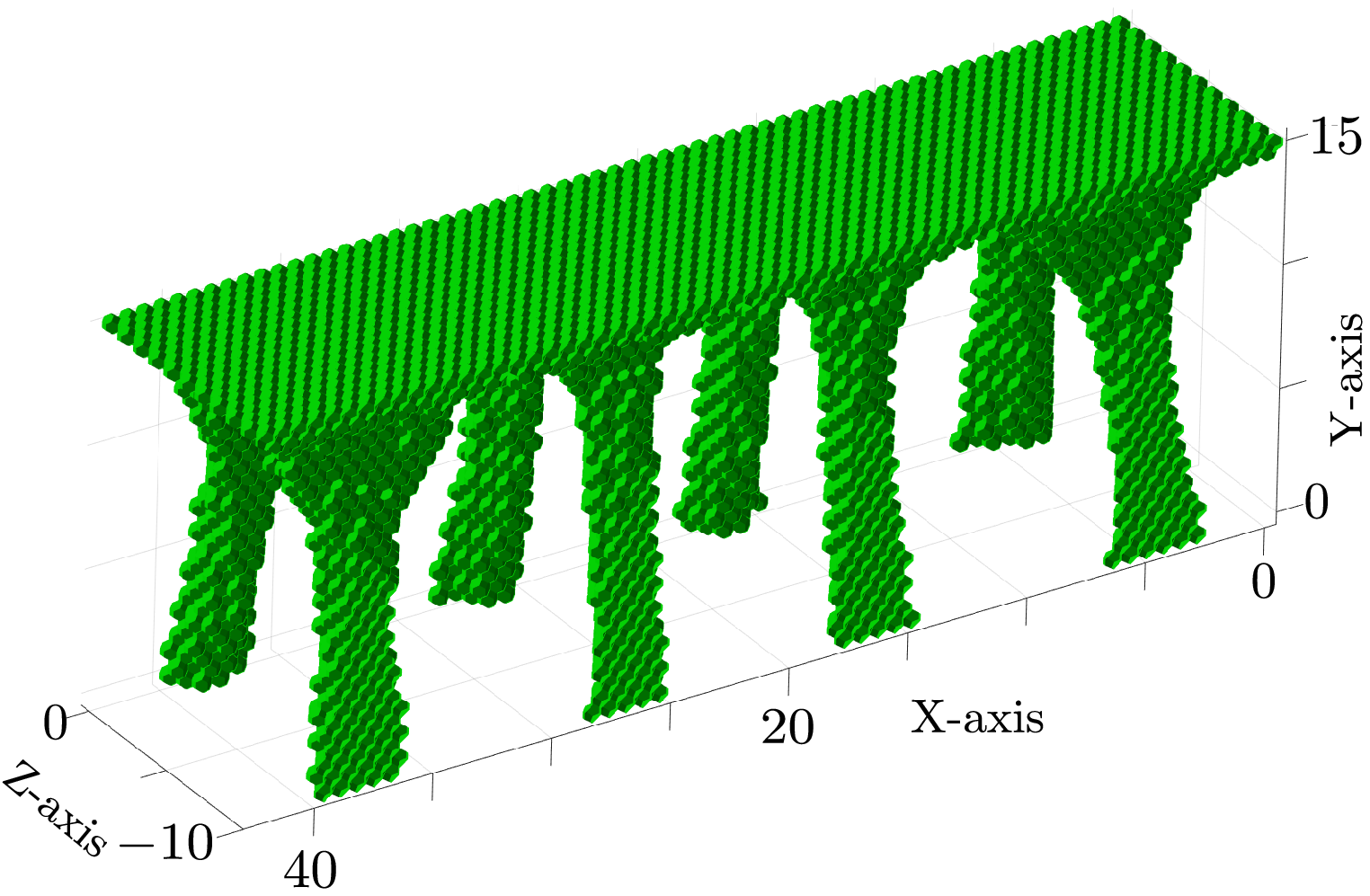}}
	\label{fig:UDL_Final_sol}%
	\hspace{0.75cm}
	\subcaptionbox{Convergence history }{\includegraphics[width=0.4\textwidth]{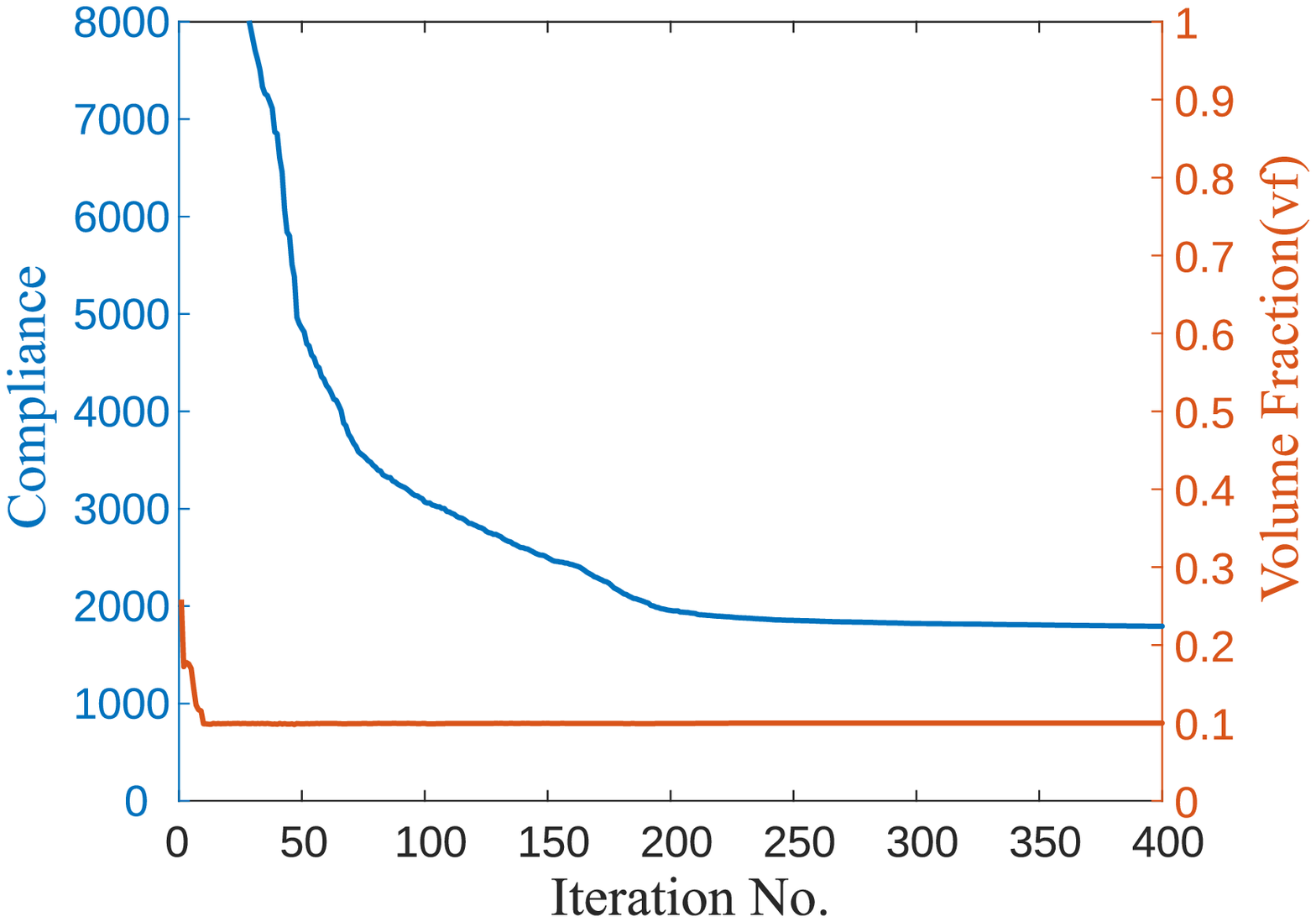}}
	\label{fig:UDL_Convergence} 
	\caption{Final solution and convergence history for bridge design.}
	\label{fig:UDL_Final_Convergence}
\end{figure}

Fig. \ref{fig:UDL_inter_sol} presents intermediate solutions at different optimization iterations. Fig. \ref{fig:UDL_Final_Convergence}a and \ref{fig:UDL_Final_Convergence}b present the final solution obtained after 400 optimization iterations and the convergence history for the objective and constraint respectively. Note that the problem is symmetric about the vertical mid plane, $\mathrm{X} = 15\sqrt{2}$. This symmetry is not explicitly imposed in the solution procedure. Similar to previous examples the density distribution and objective function value in iteration 200 is very close to the final solution suggesting that the solution has converged.

\section{Discussion}
\label{sec:discussion}
We discuss the practicality of implementing a truncated octahedron mesh (sec. \ref{sec:discussion_mesh}), viability of MMOS for 3D topology optimization problems and the effects of pre-conditioning (sec. \ref{sec:discussion_masks}).

\subsection{Truncated Octahedron}
\label{sec:discussion_mesh}
Use of truncated octahedron mesh ensures non-singular solutions, without the use of filtering like suppression methods, by maintaining face connectivity throughout. The above justification for the use of truncated octahedron mesh does not apply to minimum length scale problems as, in that case, singular solutions are not part of the acceptable solution space but the manner in which minimum length scale is formulated can interfere with the optimization process \citep{Zhang2014, Guest2004, Singh2020}. Nevertheless, irrespective of whether minimum length scale is a requirement or otherwise, the proposed method will always yield non-singular solutions. In most cases though, minimum length scale is imposed to satisfy a manufacturing condition. Under such circumstances, an independent justification based on advantages of implementing truncated octahedron mesh over a hexahedral mesh for large systems is provided ahead. The truncated octahedron element captures higher modes of deformation compared to hexahedral mesh as it has higher number of nodes associated with each element. Additionally, the element provides 14 directions of finite stiffness locally permitting that many directions for structural development, unlike hexahedral mesh where direction of finite stiffness are confined to 6. The above justification is based on the idea that the ideal element for topology optimization is one which captures all modes of deformation for better structural analysis and allows for structural development by providing identical stiffness in all directions. The effects of capturing higher mode shapes and higher choices of directions for development can be better studied with density method as it assigns design variables to each element. Further, in case of large grid meshes the ratio of number of nodes, $TN$, to the number of elements, $TE$, for a hexahedral mesh is approximately 1, that is, $\frac{TN}{TE} \approx 1$, while the same ratio in case of truncated octahedron mesh is approximately 6, that is, $\frac{TN}{TE} \approx 6$. These ratios are obtained by the following rationale. In hexahedral meshes, each element contains 8 nodes and each interior node is shared among 8 elements. In truncated octahedron meshes, each element has 24 nodes with each interior node shared by 4 elements. Thus, for meshes with the same number of elements, the truncated octahedron mesh will have approximately 6 times the number of nodes compared to hexahedral meshes. But in most cases, the size of domain and characteristic length of element are the primary variables and not the number of elements. For the same edge length, $a$, the volume of a truncated octahedron is $8\sqrt{2} a^3$ while that of a cube is $a^3$.
Hence, to discretize the same volume, the number of truncated octahedron elements required are approximately $\frac{1}{8\sqrt{2}}$ times the number of hexahedral elements. Thus, for a specified domain and the same characteristic length, the ratio of number of nodes in a truncated octahedron mesh, $TN^{tetra}$, to the number of node in a hexahedral mesh, $TN^{hexa}$, that is, $\frac{TN^{tetra}}{TN^{hexa}} \approx
\frac{3\sqrt{2}}{8} = 0.53$, implying that truncated octahedron mesh reduces the size of stiffness matrix. Other than hexahedral mesh, Voronoi tessellations have also been implemented in topology optimization problems \citep{Gain2015}. Compared to hexahedral and truncated octahedron meshes, Voronoi tessellations can more accurately approximate a variety of design domains, but is computationally expensive to implement. It also leads to a variety of difficulties, requiring storage of the stiffness matrix of each element separately which is undesirable when dealing with large systems. While in regular meshes, such as one implemented here, the stiffness matrix of all solid elements is the same. A drawback of implementing truncated octahedron mesh lies in its inability to perform mesh refinement. That is, unlike a hexahedral element which can be divided into multiple hexahedral elements \citep{shephard1991automatic,yerry1984automatic}, a truncated octahedral element cannot be divided into multiple such elements. Therefore, any change in desired mesh size requires complete reconstruction of the mesh. Considering mesh generation is a relatively inexpensive process for topology optimization problems, the aforementioned drawback is not a major concern. Another drawback of truncated octahedron mesh pertains to imposition of symmetry about different planes. Unlike hexahedral mesh, truncated octahedron mesh do not have the ability to discretize a plane using element faces making it difficult to impose symmetry. Also, the meshing algorithm discussed in section \ref{sec:mesh_development} is confined to a cuboidal domain and hence cannot be used to mesh general domains. A separate study can be conducted for generating truncated octahedron mesh for generalized domain using a bounding box algorithm and incorporating the use of various planes of symmetry. As an iso-parametric truncated octahedron element has been introduced in section \ref{sec:FEM} hence FE analysis can be conducted for any tetra-kai-decahedron element. Also, the algorithms developed for large scale topology optimization \citep{borrvall2001large,wang2007large,aage2021length} using hexahedral elements can be adopted for the aforementioned element. 

\subsection{MMOS with spheroidal masks}
\label{sec:discussion_masks}
In traditional density method for topology optimization the number of design variables equal the number of elements in the mesh. On the other hand, the number of masks used for MMOS is independent of the mesh size and hence the formulation is suitable to be adopted for solving large problems as it allows for less number of design variables. In the examples presented, the cantilever beam problem discretized using approx. $68,000$ elements was solved using just 1,260 design variables. Similarly, the torsion beam problem discretized using 25,000 elements was solved with just 672 variables and the bridge design problem discretized using 40,000 elements was solved using 2,100 variables. Thus MMOS can reduce the number of variables significantly. It seems reasonable to relate the number of masks required to volume of the domain rather than mesh size. From the examples presented, MMOS seems capable of capturing the desired range of shapes and topologies. Also, intermediate cell densities when using MMOS occur only at mask boundaries. Solutions closer to the ideal binary density distribution (black and white solutions) can be achieved by increasing the value of $\alpha$ in Eqn. \ref{eqn:fj}. An increase in $\alpha$ reduces cells with intermediate densities but an increase beyond a certain threshold adversely affects the sensitivity analysis as it increases the magnitude of gradients within a narrow band close to the mask boundaries but reduces their magnitudes everywhere else (Fig. \ref{fig:mask_schematic}b). Consequentially, implementing a high value of $\alpha$ can hinder the optimization process. \cite{Saxena2011} suggests implementing continuation of $\alpha$. Choice of pre-conditioner plays an important role in the efficiency of MMOS, even more so than in traditional density based approach. This is because, in traditional density based approach the initial guess can be provided such that the ratio of the highest and lowest density is low leading to better condition number, but this is not possible in MMOS as both extremes of the density value will always be present for any initial guess. Via numerical experiments it was observed that on applying the pre-conditioner in section \ref{sec:numerical_imp}, number of iterations required to solve the FE system reduced significantly compared to without its implementation. Also, number of iterations remained approximately the same irrespective of the value of $\rho_{min}$. These observations are in agreement with the results reported in \cite{wang2007large}. \citep{zhang2016new, zhang2017explicit, zhang2018topology} discuss the importance of maintaining a well connectedness between the fixed boundary conditions and applied load in the initial guess for feature based methods. Numerical experiments suggest that well connectedness improves the run time by avoiding close to singular matrices which take longer to solve with iterative solvers, but do not grantee better solutions and can converge to a suboptimal local minima. \cite{zhang2017explicit} implement NURBS and Hermite interpolations as features which makes gradient evaluations extensive and involved. The same is true for \cite{zhang2018topology} who use rotation matrices to determine the orientation of a feature. As shown, MMOS with spheroidal masks obtain the gradients in a rather elegant fashion and do not need rotation matrices to determine the orientation of the mask. Implementation of NURBS \citep{zhang2017explicit} or B-splines \citep{Zhang2017} require additional constraints in the problem formulation to prevent self intersection in the features. This adds one constraint per feature, which is undesirable. By fixing the shape of the feature as in MMOS these constraint can be avoided making the formulation straightforward and computationally efficient. It is also important to note that MMOS, by itself, does not inherently guarantee mesh independent solutions. To achieve the latter, adequate length scale measures must be additionally implemented.\\

Structures obtained as a result of the optimization process are defined by discrete elements, hence a boundary smoothening process, irrespective of the 3D tessellation used, needs to be developed before these solutions are manufactured. Boundary smoothening techniques for hexagonal cell structures is proposed in \citep{kumar2015topology, kumar2016synthesis}. Alternatively, iso-geomtric analysis can also be implemented to bridge the gap between structural analysis and CAED modeling \citep{cottrell2009isogeometric}. 

\section{Conclusion}
\label{sec:conclusion}
A novel meshing algorithm for the development of regular tetra-kai-decaherons or truncated octahedrons mesh is established for cuboidal domain. The idea can be extended to mesh any geometry using a bounding box algorithm and will be explored in future works. The FEM uses truncated octahedron cells as elements for solving linear elasticity problem. For the finite element process a truncated octahedron master element is defined and an iso-parametric map between the master element and the physical element is established using analytical linear shape functions. Numerical integration technique introduced in \cite{Rashid2006} is implemented and is found to provide reasonably accurate stiffness matrix evaluation. The MMOS using negative spheroidal masks is successfully implemented and shown to capture acceptable range of shape and topology. The pre-conditioner proposed by \cite{wang2007large} is implemented in conjunction with PCG for matrix inversion. The pre-conditioner is found to be effective in reducing iterations required by the PCG solver effectively reducing run time and computational cost. Elements with intermediate densities are limited and only present on mask boundaries as expected. Implementation of truncated octahedron mesh inherently removes the possibility of singular solutions without the use of filtering methods. The aforementioned mesh is shown to be a viable option numerically, as, for a specified volume and fixed edge length, the stiffness matrix obtained is smaller in comparison to one obtained via hexahedral mesh. The above method for topology optimization is estimated to be well equipped to solve large scale problems because unlike traditional density based methods the number of design variables in MMOS is independent of the number of elements. Also, the implementation of truncated octahedron mesh reduces the size of stiffness matrix which should lead to easier inversions while  maintaining the characteristic length of the discretization.

\section*{Conflict of Interest}
The authors declare that they have no known competing financial interests or personal relationships that could have appeared to influence the work reported in this paper

\bibliography{opti_ref_1}

\end{document}